\documentclass[prd, superscriptaddress,showpacs,nofootinbib,floatfix,include graphics
preprintnumbers]{revtex4-1}
\usepackage{amsmath,amssymb,amsfonts,graphicx,bm,bbm,xcolor,hyperref}
\usepackage{array}
\usepackage{float}
\usepackage{multirow}
\usepackage{subfigure}

\newcolumntype{L}[1]{>{\raggedright\arraybackslash}p{#1}}
\newcolumntype{C}[1]{>{\centering\arraybackslash}p{#1}}
\newcolumntype{R}[1]{>{\raggedleft\arraybackslash}p{#1}}

\newcommand{\be}{\begin{eqnarray}}
\newcommand{\ee}{\end{eqnarray}}
\newcommand{\ba}{\begin{align}}
\newcommand{\ea}{\end{align}}

\newcommand{\Gam}[1]{\Gamma\left(#1\right)}
\newcommand{\Tr}[1]{\text{Tr}\left[#1\right]}
\newcommand{\KeV}{{\rm ~KeV}}

\newcommand{\GeV}{{\rm ~GeV}}
\newcommand{\TeV}{{\rm ~TeV}}
\newcommand{\pb}{{\rm ~pb}}
\newcommand{\fb}{{\rm ~fb}}
\newcommand{\ab}{{\rm ~ab}}
\newcommand{\invfb}{{\rm ~fb^{-1}}}
\newcommand{\invab}{{\rm ~ab^{-1}}}
\newcommand{\CL}{{\rm CL}}
\newcommand{\mN}{m_{N}}

\newcommand{\MWR}{M_{W_R}}
\newcommand{\MZR}{M_{Z_R}}
\newcommand{\mh}{m_{H_{0}^0}}
\newcommand{\mH}{m_{H_{1}^0}}
\newcommand{\mA}{m_{A_{1}^0}}
\newcommand{\mHH}{m_{H_{2}^0}}
\newcommand{\mAA}{m_{A_{2}^0}}
\newcommand{\mHp}{m_{H_1^\pm}}
\newcommand{\mHHH}{m_{H_{3}^0}}

\newcommand{\mHHp}{m_{H_2^\pm}}
\newcommand{\mHppL}{m_{\delta^{\pm\pm}_L}}
\newcommand{\mHppR}{m_{\delta^{\pm\pm}_R}}
\newcommand{\vR}{v_R}
\newcommand{\vL}{v_L}
\newcommand{\vSM}{v_{\rm SM}}
\newcommand{\kp}{k_{+}}
\newcommand{\km}{k_{-}}

\newcommand{\NR}{N_R}
\newcommand{\WR}{W_R}
\newcommand{\ZR}{Z_R}
\newcommand{\jN}{j_N}

\newcommand{\as}{\alpha_s}
\newcommand{\asBar}{\overline{\alpha}}

\newcommand{\MB}{M_{\mathcal{B}}}
\newcommand{\sigmaLO}{\sigma^{\rm LO}}
\newcommand{\sigmaNLO}{\sigma^{\rm NLO}}
\newcommand{\sigmaFO}{\sigma^{\rm FO}}
\newcommand{\sigmaRes}{\sigma^{\rm Res.}}

\newcommand{\kNLO}{K^{\rm NLO}}

\newcommand{\DeltaNRes}{\Delta^{\rm Res.}_N}

\newcommand{\confirm}{\textcolor{black}}
\begin{document}
\preprint{IPPP/16/65,~CPT/16/130}

\title{Neutrino Jets from High-Mass $W_R$ Gauge Bosons in TeV-Scale Left-Right Symmetric Models}

\newcommand{\iiserm}{\affiliation{Department of Physics, Indian Institute of Science Education and Research Mohali (IISER Mohali), Sector 81, SAS Nagar, Manauli 140306, India}}
\newcommand{\ippp}{\affiliation{Institute for Particle Physics Phenomenology {(IPPP)}, Department of Physics, Durham University, Durham, DH1 3LE, UK}}

\author{Manimala Mitra} \email{manimala.mitra@durham.ac.uk} \iiserm \ippp
\author{Richard Ruiz}\email{richard.ruiz@durham.ac.uk}\ippp
\author{Darren~J.~Scott} \email{d.j.scott@durham.ac.uk}\ippp
\author{and Michael Spannowsky}\email{michael.spannowsky@durham.ac.uk} \ippp

\begin{abstract}
{We re-examine the discovery potential at hadron colliders of high-mass right-handed (RH) gauge bosons $W_R$ 
- an inherent ingredient of Left-Right Symmetric Models (LRSM).
We focus on the regime where the $W_R$ is very heavy compared to the heavy Majorana neutrino $N$,
and  investigate an alternative signature for $W_R \rightarrow N$ decays.
The produced neutrinos are highly boosted in this mass regime. 
Subsequently, their decays via off-shell $W_R$ bosons to jets, i.e., $N \rightarrow \ell^\pm j j$ are highly collimated, 
forming a single \textit{neutrino jet} $(j_N)$.
The final-state collider signature is then  $\ell^\pm j_N$, instead of the widely studied $\ell^\pm\ell^\pm jj$.
Present search strategies  are not sensitive to this hierarchical mass regime due to the  breakdown of the collider signature definition.
We  take into account QCD corrections beyond next-to-leading order (NLO) 
that are important for high-mass Drell-Yan processes
at the 13 TeV Large Hadron Collider (LHC).
For the first time, we evaluate $W_R$ production at NLO with threshold resummation at next-to-next-to-leading logarithm (NNLL) 
matched to the threshold-improved parton distributions. 
With these improvements, we find that a ${W_R}$ of mass $M_{W_R} = 3~(4)~[5]$ TeV and mass ratio of $(m_N/M_{W_R})<0.1$ 
can be discovered with a $5-6 \sigma$ statistical significance at 13 TeV after ${10~(100)~[2000]~\text{fb}^{-1}}$ of data. 
Extending the analysis to the hypothetical 100 TeV Very Large Hadron Collider (VLHC), 
$5\sigma$ can be obtained for $W_R$ masses up to $M_{W_R}=15~(30)$ with approximately $100~\text{fb}^{-1}~(10~\text{ab}^{-1})$. 
{
Conversely, with $0.9~(10)~[150]\invfb$ of 13 TeV data, $M_{W_R} < 3~(4)~[5]\TeV$ and $(m_N/M_{W_R})<0.1$ can be excluded at 95\% CL;
with $100\invfb~(2.5\invab)$ of 100 TeV data, $\MWR<22~(33)\TeV$ can be excluded.
}
}

\end{abstract}
\maketitle

\tableofcontents

\section{Introduction}
The observation of nonzero neutrino masses $m_\nu$ that have hierarchically smaller masses
than all other elementary fermions in the Standard Model of Particle Physics (SM),
and their non-trivial mixing provide unambiguous experimental evidence of physics beyond the SM (BSM).
The natural explanation for such tiny masses is the so-called Seesaw Mechanism, 
where eV neutrino masses are generated from the  $(B-L)$-violating operators at dimension-5~\cite{Weinberg:1979sa,Wilczek:1979hc}.
At tree level, these operators can be generated by extending minimally~\cite{Ma:1998dn} the SM field contents by
right-handed (RH) neutrinos $N_R$ (Type I)~\cite{Minkowski:1977sc,Mohapatra:1979ia,Yanagida:1979as,GellMann:1980vs,Schechter:1980gr,Shrock:1980ct},
scalar SU$(2)_L$ triplets $\Delta_L$ (Type II)~\cite{Magg:1980ut,Cheng:1980qt,Lazarides:1980nt,Mohapatra:1980yp},
or fermionic SU$(2)_L$ triplets $\Sigma$ (Type III)~\cite{Foot:1988aq}.
If kinematically accessible, these states 
can  be observed at the 13 TeV Large Hadron Collider (LHC) or a hypothetical 
100 TeV Very Large Hadron Collider (VLHC)~\cite{Arkani-Hamed:2015vfh,Golling:2016gvc}, 
thus giving conclusive 
evidence of the mass generation mechanism. For reviews of TeV-scale Seesaw models and their phenomenology, see Refs.~\cite{Chen:2011de}.

An appealing renormalizable framework in which both Types I and II Seesaws can be embedded 
 is the Left-Right Symmetry Model (LRSM)~\cite{Pati:1974yy,Mohapatra:1974gc,Senjanovic:1975rk,Duka:1999uc}. 
This is based  on the gauge group
\begin{equation}
 {\rm SU}(3)_c \otimes {\rm SU}(2)_L\otimes {\rm SU}(2)_R \otimes {\rm U}(1)_{B-L},
 \label{eq:lrsmGroup}
\end{equation}
and postulates the restoration of parity symmetry at high energies. 
In addition to the SM particle content,
the model consists of three generations of  $N_R$, one $\Delta_L$, 
and an SU$(2)_R$ triplet scalar $\Delta_R$, all with non-trivial charges under the $B-L$ symmetry. 
After $\Delta_R$ acquires a vev $\vR$,  much larger than the electroweak (EW) scale, $\vSM\approx246\GeV$,
the $SU(2)_R \times U(1)_{B-L}$ symmetry breaks down to the $U(1)_Y$ of the SM.
Subsequently, the neutrinos $N_R$ and gauge bosons $\WR$ and $\ZR$ acquire  masses $M_R$, $\MWR$ and $\MZR$, 
respectively, that are proportional to $v_R$.
While the masses of the gauge bosons depend on the weak gauge coupling $g_R=g$, 
the   masses of  $N_R$ are dependent on   the  Yukawa coupling $f_R$ of the $\Delta_R$ and lepton doublet interaction. 
The RH neutrino also interacts with the SM neutrino via its Yukawa  interaction, 
generating Dirac masses $M_D$ after EW symmetry breaking (EWSB). 
For the Majorana mass $M_R$ much heavier than the Dirac mass $M_D$, 
a Type I Seesaw is triggered~\cite{Minkowski:1977sc,Mohapatra:1979ia,Yanagida:1979as,GellMann:1980vs,Schechter:1980gr,Shrock:1980ct},
giving rise to Majorana masses for 
light neutrinos $\nu_m$ with $m_{\nu} \sim M^2_D/M_R$ and
heavy neutrinos $N$ with  $m_N \sim M_R$. 
As no symmetry relates the RH gauge and triplet Yukawa couplings, the $W_R$ and heavy neutrino may have widely separated masses, 
and offers a wide parameter space to test the LRSM.

\begin{figure}[!t]
\begin{center}
\includegraphics[width=.7\textwidth]{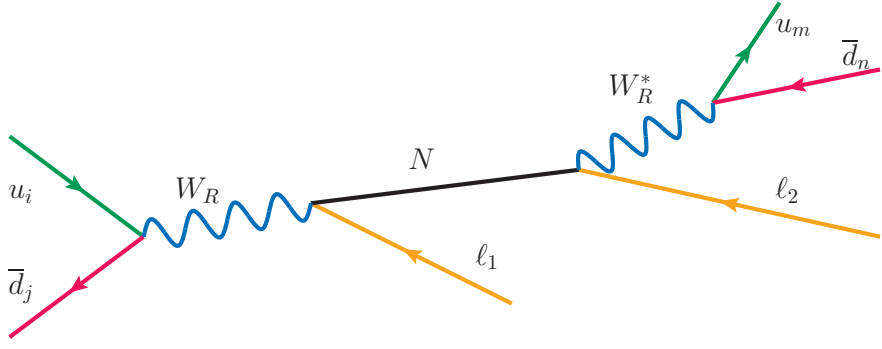}	
\end{center}
\caption{Born diagram of $W_R$ production in hadron collisions and decay via $N$ 
to leptons and quarks. Figures are drawn using JaxoDraw~\cite{Binosi:2003yf}.}
\label{fig:qqWR_Nl_WRll_Diagram}
\end{figure}

The  LRSM model can be tested either  indirectly, through
low energy experiments \cite{Chen:2005jx,Zhang:2007fn,Zhang:2007da,Chen:2008jg,Tello:2010am,Nemevsek:2011aa,Deppisch:2012nb,Chakrabortty:2012mh, 
Nemevsek:2012iq, Dev:2013vxa,Barry:2013xxa,Maiezza:2014ala,Bertolini:2014sua, Awasthi:2015ota,Awasthi:2016kbk}, or directly, through 
 searches at high energy colliders 
 \cite{Keung:1983uu,Ferrari:2000sp,Frank:2010cj,Das:2012ii,Nemevsek:2012iq,Han:2012vk,
 Chen:2013fna,Vasquez:2014mxa,Dev:2014iva,Dev:2015kca,Maiezza:2015lza,Gluza:2015goa,Ng:2015hba,Mondal:2015zba,Lindner:2016lxq,Maiezza:2016bzp,Chakrabortty:2016wkl,Golling:2016gvc,FileviezPerez:2016erl} 
 (and references therein). 
In this work, we focus on the direct detection of the $\WR$ and $N$.
For $\MWR>\mN$, the hallmark hadron collider test of the LRSM is the spectacular lepton number $(L)$ violating process~\cite{Keung:1983uu}
\begin{equation}
 q_1 ~\overline{q_2} ~\rightarrow ~\WR^{\pm} 
 ~\rightarrow ~\NR ~\ell^\pm_1 
 ~\rightarrow	~\ell^\pm_1 ~\ell^\pm_2 ~\WR^{\mp *}  
 ~\rightarrow	~\ell^\pm_1 ~\ell^\pm_2 ~q'_1 ~\overline{q'_2}.
 \label{eq:SSleps}
\end{equation}
The process, shown in Fig.~\ref{fig:qqWR_Nl_WRll_Diagram}, has been studied extensively.
Searches by the ATLAS~\cite{Aad:2015xaa} and CMS~\cite{Khachatryan:2014dka} collaborations
have excluded regions of the $(\MWR,\mN)$ parameter space for $\MWR$ $(\mN)$ up to several TeV (hundred GeV)~\cite{Patra:2015bga,Lindner:2016lpp}.
However, for hierarchical masses, i.e., $(\mN/\MWR)<0.1$, the present search strategy is no longer sensitive.
Complimentary dijet searches have similarly excluded $\MWR$ below ${2.5-3.5}\TeV$~\cite{Jezo:2014wra,ATLAS:2015nsi,Khachatryan:2015dcf}.

In light of such stringent bounds, we re-examine the discovery potential of high-mass $W_R$ at hadron colliders.
We focus on the situation where $N$ are hierarchically lighter than $\WR$, i.e., $(\mN/\MWR)<0.1$.
In the process $p ~p \rightarrow \WR \rightarrow N ~\ell$, this leads to boosted $N$ with transverse momentum $p_T^N \sim \MWR/2$. 
The decay products of $N$, which proceed dominantly through far off-shell $\WR$ to quarks, i.e.,
$ N ~\rightarrow~ \WR^{*} \ell ~\rightarrow~ q \overline{q'} \ell,$ 
are subsequently collimated with parton separations scaling as $\Delta R_{ij} \sim 2m_N / p_T^N \sim 4 \mN/\MWR$.
Hence, for $\mN/\MWR\lesssim0.1$, one has  $\Delta R_{ij} \lesssim 0.4$, 
which falls below the electron isolation threshold in standard high-$p_T$ lepton searches at the 13 TeV LHC~\cite{CMS:2015kjy}.
Indeed, the LHC sensitivity of Eq.~(\ref{eq:SSleps}) for such $(\MWR,\mN)$ is considerably weaker, particularly in the electron channel~\cite{Aad:2015xaa}.
This deficiency has been noted before, e.g., Refs.~\cite{Ferrari:2000sp,Han:2012vk,Kang:2015uoc,Lindner:2016lxq}, 
but never explored in substantial detail.

After hadronization, the decay products of  $N$ do not appear as individual, isolated objects, but instead as a single \textit{neutrino jet}~$\jN$. 
This is akin to the formation of top jets from boosted 
top quarks~\cite{Kaplan:2008ie,Plehn:2009rk,Plehn:2010st,Soper:2012pb,Schaetzel:2013vka}.
Thus, for $\mN\ll\MWR$, $\WR-N$ production and decay appear in $pp$ collisions as the distinctive 
\begin{equation}
 p ~p ~\rightarrow~ \WR^{\pm} ~\rightarrow~ \ell^\pm ~\jN.
 \label{eq:lepjN}
\end{equation}
Despite the inclusive channel's simplified topology, and hence larger SM backgrounds,
it inherits much of the strong discriminating power of Eq.~(\ref{eq:SSleps}), 
including a fully reconstructible final state and no missing $p_T$ (MET), other than the  hadronization and detector effects.
We consider a search strategy for $\WR-N$ production and decay when ${\MWR > 3\TeV}$ and $(\mN/\MWR) < 0.1$, 
while using a simple set of kinematical cuts on the effective two-body final state.
We explore the discovery potential of observing Eq.~(\ref{eq:lepjN}) for the c.m.~energies $\sqrt{s}=13$ and 100 TeV, relevant for the LHC and VLHC.

Furthermore, determining if the $\WR$ gauge coupling $g_R$ equals the SM weak coupling $g$, 
a postulate of Eq.~(\ref{eq:lrsmGroup}), requires precision knowledge of $\WR$ production rates.
However, for such large $W_R$ masses, QCD corrections beyond next-to-leading order (NLO) 
are important at 13 TeV because of soft gluon radiation off initial-state partons. 
In light of this, we also calculate, for the first time, 
$W_R$ production at NLO with threshold resummation at next-to-next-to-leading logarithm (NNLL) 
matched to threshold-improved parton distributions functions (PDFs)~\cite{Bonvini:2015ira,Beenakker:2015rna}.
Previous predictions~\cite{Sullivan:2002jt,Gavin:2012sy,Jezo:2014wra}
have considered  threshold resummation up to next-to-leading logarithm (NLL)~\cite{Jezo:2014wra} but never matched to resummed PDFs.
NLO+NNLL contributions improve the Born (NLO)-level predictions for $\MWR = 4-5\TeV$ by  $40-140~(4-72)\%$  at 13 TeV LHC.

With these improvements, we find that a ${W_R}$ of mass $M_{W_R} = 3~(4)~[5]$ TeV and $(m_N/M_{W_R})<0.1$ 
can be discovered with a $5-6 \sigma$ statistical significance at 13 TeV after ${10~(100)~[2000]~\text{fb}^{-1}}$. 
At 100 TeV with $0.1~(10)\invab$, the $5\sigma$ reach extends to $M_{W_R}={15~(30)}$ TeV.
Conversely, with $0.9~(10)~[150]\invfb$ of 13 TeV data, $M_{W_R} < 3~(4)~[5]\TeV$ and $(m_N/M_{W_R})<0.1$ can be excluded at 95\% CL;
with $100\invfb~(2.5\invab)$ of 100 TeV data, $\MWR<22~(33)\TeV$ can be excluded.

Our report is organized as follows:
In Sec.~\ref{sec:theory}, we briefly review the minimal LRSM (MLRSM) and current constraints.
In Sec.~\ref{sec:xsec}, we present predictions up to NLO+NNLL for $\WR-N$ production and decay rates at hadron colliders.
We explore the phenomenology of boosted heavy  neutrinos and present our signal-verses-background analysis in Sec.~\ref{sec:observability}.
In Sec.~\ref{sec:conclusion}, we summarize and conclude.
We relegate technical details of our resummation calculation to App.~\ref{app:threshold} and 
implementation of the LRSM model files by Ref.~\cite{Roitgrund:2014zka} to App.~\ref{app:modelFile}.

\section{Minimal Left-Right Symmetric Model}\label{sec:theory}
Here, we briefly review main aspects of the MLRSM relevant to our study.
For an expanded discussion, see, e.g., Refs.~\cite{Zhang:2007da}. 
In Secs.~\ref{sec:WRMass} and \ref{sec:Nmass}, we address the masses of $\WR$ and $N$.
In Sec.~\ref{sec:constraints}, experimental constraints are reviewed.
We reserve discussing the model's scalar potential and its implementation into publicly available simulation model files~\cite{Roitgrund:2014zka}
for App.~\ref{app:modelFile}. As we use the files of Ref.~\cite{Roitgrund:2014zka}, we adopt their notation.

The MLRSM~\cite{Pati:1974yy,Mohapatra:1974gc,Senjanovic:1975rk} is based on the extended gauge group
\begin{equation}
 {\rm SU}(3)_c \otimes {\rm SU}(2)_L\otimes {\rm SU}(2)_R \otimes {\rm U}(1)_{B-L}.
 \label{eq:lrsmGroupCopy}
\end{equation}
In addition to the SM fermion field content, there are three generations of RH neutrinos $N_R$.
Quark and lepton multiplets are assigned the following gauge group representations:
\be
& Q_{L,i} \ = \ \left(\begin{array}{c}u_L\\d_L \end{array}\right)_i : \: \left({ \bf 3}, {\bf 2}, {\bf 1}, \frac{1}{3}\right), \qquad \qquad  
Q_{R,i} \ = \ \left(\begin{array}{c}u_R\\d_R \end{array}\right)_i : \: \left({ \bf 3}, {\bf 1}, {\bf 2}, \frac{1}{3}\right), \nonumber \\
& \psi_{L,i} \ = \  \left(\begin{array}{c}\nu_L \\ e_L \end{array}\right)_i : \: \left({ \bf 1}, {\bf 2}, {\bf 1}, -1 \right), \qquad \qquad 
\psi_{R,i} \ = \ \left(\begin{array}{c} N_R \\ e_R \end{array}\right)_i : \: \left({ \bf 1}, {\bf 1}, {\bf 2}, -1 \right).
\ee
In the above,  $i=1,\dots,3,$ is the family index.
${(B-L)}$ charges are normalized such that the electric charge is given by $Q=I_{3L}+I_{3R}+(B-L)/2$, 
with $I_{3L (3R)}$ being the third isospin components of $SU(2)_{L (R)}$.
The scalar sector consists of the following multiplets:
\be
\Phi = \left(\begin{array}{cc}\phi^0_1 & \phi^+_2\\\phi^-_1 & \phi^0_2\end{array}\right) : ({\bf 1}, {\bf 2}, {\bf 2}, 0),
\ee
\be
\Delta_L=\left(\begin{array}{cc}\Delta^+_L/\sqrt{2} & \Delta^{++}_L\\\Delta^0_L & -\Delta^+_L/\sqrt{2}\end{array}\right) : ({\bf 1}, {\bf 3}, {\bf 1}, 2), 
\qquad 
\Delta_R=\left(\begin{array}{cc}\Delta^+_R/\sqrt{2} & \Delta^{++}_R\\\Delta^0_R & -\Delta^+_R/\sqrt{2}\end{array}\right) : ({\bf 1}, {\bf 1}, {\bf 3}, 2).  
\nonumber
\ee

At a scale much higher than the EW scale, $\Delta_R$ acquires a vev $v_R = \sqrt{2} \langle \Delta_R\rangle$.
This triggers spontaneous breaking of the $LR$- and $(B-L)$-symmetries,
and reduces Eq.~(\ref{eq:lrsmGroupCopy}) to the SM gauge group, i.e., ${\rm SU}(2)_R\times {\rm U}(1)_{B-L} \rightarrow {\rm U}(1)_Y$.
The bidoublet $\Phi$ is responsible for Dirac masses and EW symmetry breaking (EWSB) after it acquires the vevs 
$\langle\Phi\rangle={\rm diag}( k_1,  k_2)/\sqrt{2}$, where
\begin{equation}
 k_\pm^2 \equiv k_1^2 \pm k_2^2 \quad\text{and}\quad	k_+ = \vSM \approx 246\GeV.
 \label{eq:ewvevDef}
\end{equation}
In the absence of fine-tunning, $k_1,~k_2$ naturally scale as
\begin{equation}
 \frac{k_2}{k_1}\sim \frac{m_b}{m_t} \ll 1.
 \label{eq:bidoubletvevs}
\end{equation}
$\Delta_L$ can also acquire a vev $v_L = \sqrt{2}\langle \Delta_L \rangle$.
However, precision measurements of the $\rho/T$-parameter indicate $v_L$ is much smaller than the EW scale~\cite{Chen:2005jx,Chen:2008jg}.
For simplicity, we take $v_R$ and $k_{1,2}$ to be real, i.e., no CP violation, and $v_L = 0$.

\subsection{Charged Gauge Boson Masses}\label{sec:WRMass}
After LR and EWSB, the charged vector boson (squared) mass matrix in the gauge, i.e., $(W_L, W_R)$, basis is given by
\be
{\cal M}_W = \frac{g^2}{4}\left(\begin{array}{cc}
 k_1^2+ k_2^2+2v_L^2 & 2 k_1 k_2 \\
2 k_1 k_2 &  k_1^2+ k_2^2+2v_R^2
\end{array}\right).
\ee
The gauge states are related to the mass eigenstates, i.e., $(W_1, W_2)$ with $M_{W_2} > M_{W_1}$, by
\be
\left(\begin{array}{c}
W_1 \\
W_2 
\end{array}\right) = \left(\begin{array}{cc}
\cos \xi & \sin \xi \\
-\sin \xi & \cos \xi 
\end{array}\right) 
\left(\begin{array}{c}
W_L \\
W_R 
\end{array}\right),
\ee
where the $W_L-W_R$ mixing parameter $\xi$ is 
\be
\tan{2\xi} = \frac{2 k_1  k_2}{v_R^2-v_L^2}. 
\label{kappa}
\ee
Under the vev hierarchy 
\begin{equation}
\vR \gg k_+ \gtrsim k_1 \gtrsim k_- \gg k_2 \gg v_L\sim 0,
\label{eq:vevHierarchy}
\end{equation}
the vector boson masses simplify to 
\begin{equation}
 M_{W_1} \approx M_{W_L} = \frac{g}{2} k_+ \quad\text{and}\quad M_{W_2} \approx M_{W_R} = \frac{g}{\sqrt 2}v_R,
\end{equation}
implying that the $W_1~(W_2)$ mass state is closely aligned with the $W_L~(W_R)$ gauge state.
Hence, for the remainder of the text, we refer to  $W_1~(W_2)$ as $W_L~(W_R)$.

\subsection{Neutrino Masses}\label{sec:Nmass}
The leptonic Yukawa couplings for generations $i$ and $j$ are given by
\begin{eqnarray}
{\cal L}_Y = 
&~-~ h_{ij}\bar{\psi}_{L_i}\Phi \psi_{R_j}  
&~-~ \tilde{h}_{ij}\bar{\psi}_{L_i}\tilde{\Phi} \psi_{R_j}		
\nonumber\\
&~-~ f_{L_{ij}} \psi_{L_i}^{\sf T} C i\sigma_2 \Delta_L \psi_{L_j} 
&~-~ f_{R_{ij}} \psi_{R_i}^{\sf T} C i\sigma_2 \Delta_R \psi_{R_j}  ~+~ {\rm H.c.},
\label{eq:yuk}
\end{eqnarray}
where $C$ denotes the charge conjugation operator and $\tilde{\Phi}=\sigma_2\Phi^*\sigma_2$. 
After LR and EWSB, RH Majorana, LH Majorana, and Dirac neutrino mass matrices, respectively, of the form
\be 
M_R = \sqrt 2 v_R f_R, ~\quad 
M_L = \sqrt 2 v_L f_L, ~\quad
M_D = \frac{1}{\sqrt 2}\left( k_1 h +  k_2 \tilde{h} \right),
\label{eq:nuSeesawMasses}
\ee
are spontaneously generated. 
The $3\times3$ matrices in Eq.~(\ref{eq:nuSeesawMasses}) can be combined such that in the gauge basis, i.e., $(\nu_{L1},\dots,N_{R1}^c,\dots)$, 
the $6\times 6$ neutrino mass matrix is given by
\be
{\cal M}_\nu = \left(\begin{array}{ccc}
M_L & M_D \\
M_D^{\sf T} & M_R
\end{array}\right),
\label{eq:big}
\ee
and can be diagonalized via the unitary matrix ${\widetilde V}$: 
\be
\mathcal{M}_\nu^{diag} 
~=~ {\widetilde V}^{\sf T}\mathcal{ M}_\nu {\widetilde V} 
~=~  \ \left(\begin{array}{cc} {M}^{diag}_\nu & {\bf 0} \\ {\bf 0} & {M}^{diag}_N \end{array}\right).
\label{eq:mnu} 
\ee
${M}^{diag}_\nu = {\rm diag}(m_1,m_2,m_3)$ and ${M}^{diag}_N = {\rm diag}(m_{N_1},m_{N_2},m_{N_3})$ 
are the light neutrino and heavy neutrino masses, respectively. 
For the vev hierarchy of Eq.~(\ref{eq:vevHierarchy}), ${\widetilde V}$ is \cite{Korner:1992zk, Grimus:2000vj}
\be
{\widetilde V} \ = \ \left(\begin{array}{cc}
({\bf 1}+\zeta^*\zeta^{\sf T})^{-1/2} & \zeta^*({\bf 1}+\zeta^{\sf T}\zeta^*)^{-1/2} \\
-\zeta^{\sf T}({\bf 1}+\zeta^*\zeta^{\sf T})^{-1/2} & ({\bf 1}+\zeta^{\sf T}\zeta^*)^{-1/2}
\end{array}  \right)
\left(\begin{array}{cc}
U_L & {\bf 0} \\
{\bf 0} & Y_R
\end{array}\right) \ \equiv \ \left(\begin{array}{cc}
U & V \\
X & Y 
\end{array}\right) \;,
\label{eq:NuMixing}
\ee
where $\zeta^*= M_D M_R^{-1}$  and $U_L$, $Y_R$ are unitary matrices that diagonalize $\widetilde{M}_\nu$ and $\widetilde{M}_R$:
\begin{eqnarray}
 {M}^{diag}_\nu	=	U^{\sf T}_L 	\widetilde{M}_\nu	U_L
 \quad\text{and}\quad
 {M}^{diag}_N 	=	Y^{\sf T}_R 	\widetilde{M}_N 	Y_R
\end{eqnarray}
$\widetilde{M}_\nu$ and $\widetilde{M}_N$ are related to the mass matrices in Eq.~(\ref{eq:nuSeesawMasses}) by the Seesaw relations
~\cite{Minkowski:1977sc,Mohapatra:1979ia,Yanagida:1979as,GellMann:1980vs,Schechter:1980gr,
Shrock:1980ct,Magg:1980ut,Cheng:1980qt,Lazarides:1980nt,Mohapatra:1980yp}
\begin{eqnarray}
\widetilde{M}_\nu	\simeq M_L - M_D M_R^{-1} M_D^{\sf T} 
\quad\text{and}\quad 
\widetilde{M}_N 	\simeq M_R. 
\label{eq:mnuint}
\end{eqnarray}
In the notation of Refs.~\cite{Atre:2009rg,Han:2012vk}, 
after rotating the charged leptons from the flavor basis into the mass basis, which for simplicity we take to be a trivial rotation,
the $U_{\ell \nu_m}~(Y_{\ell N_{m'}})$ of Eq.~(\ref{eq:NuMixing}) denotes the large, $\mathcal{O}(1)$ mixing between the 
LH~(RH) lepton flavor state $\ell~(\ell=e,\mu,\tau)$ and light (heavy) neutrino mass eigenstate $\nu_m~(N_{m'})$.
Similarly, $V_{\ell N_{m'}}~(X_{\ell \nu_{m}})$ denotes the suppressed, $\mathcal{O}(m_\nu/m_N)$ mixing between the 
LH~(RH) lepton flavor state $\ell$ and heavy (light) neutrino mass eigenstate $N_{m'}~(\nu_m)$.

\subsection{Experimental Constraints} \label{sec:constraints}
Here, we review the most stringent constraints on the MLRSM.
\begin{enumerate}
 \item  \textit{Collider Bounds on $(\MWR,\mN)$ from $\ell^\pm\ell^\pm jj$ searches}:
 Searches by the ATLAS experiment for $pp\rightarrow e^\pm e^\pm j j ~(\mu^\pm\mu^\pm jj)$ mediated by $\WR$ and $N$  
 excludes at $\sqrt{s} = 8\TeV$~\cite{Aad:2015xaa}:
\begin{equation}
 \MWR \lesssim 1.5~(2.7)\TeV	\quad\text{at}\quad 95\%~\text{C.L}\quad\text{with}\quad\mathcal{L}=20.3\invfb.
\end{equation}
The sensitivity disparity is due a failing isolated electron-jet criterion when $\mN/\MWR\lesssim0.1$~\cite{Aad:2015xaa} and is the point of our study.
Limits from CMS are comparable~\cite{Khachatryan:2014dka}.

 \item  \textit{Collider Bounds on $\MWR$ from dijet searches}:
Searches by the ATLAS (CMS) experiment for a sequential SM $W'\rightarrow jj$, 
excludes at $\sqrt{s} = 13\TeV$\cite{ATLAS:2015nsi,Khachatryan:2015dcf}:
\begin{equation}
 M_{W'_{\rm SSM}} \lesssim 2.6~(2.6)\TeV	\quad\text{at}\quad 95\%~\text{C.L.}\quad\text{with}\quad\mathcal{L}=3.6~(2.4)\invfb.
\end{equation}

\item  \textit{Limits on $\WR$ and Higgs masses from neutral hadron transitions}:
Analyses of $\Delta F=2$ transitions in neutral $K$ and $B_{d,s}$ systems and neutron EDM assuming generalized charge (parity)
in the MLRSM exclude~\cite{Maiezza:2014ala,Bertolini:2014sua}:
\begin{eqnarray}
 \MWR 		 <& 2.9-20\TeV	&\quad\text{at}\quad 95\%~\text{C.L},\\
 m_{\rm FCNH} 	 <& 20\TeV	&\quad\text{at}\quad 95\%~\text{C.L},
 \label{eq:fcnhBound}
 \end{eqnarray}
 where the range over $\MWR$ is based on theoretical arguments and
 $m_{\rm FCNH}$ is the mass of the lightest Higgs mediating flavor changing neutral transitions.

\item \textit{Searches for $0\nu\beta\beta$}:
In MLRSM, the gauge boson $W_R$ together with $N_i$ can give a saturating contribution in $0\nu \beta \beta$. Non-observation of this
LNV process hence constrains the  masses of $W_R$ and $N_i$ as $\sum_i \frac{Y^2_{ei}}{M_i M^4_{W_R}} \leq  (0.082-0.076) \,\rm{TeV}^{-5}$, using 
the 90$\%$ C.L half-life limit from KamLAND-Zen  $T^{0\nu}_{1/2} \ge 1.07 \times 10^{26} \rm{yrs}$ \cite{KamLAND-Zen:2016pfg}. 
For a $\MWR$ of  3 TeV (5 TeV) this implies a lower limit on the $m_N \ge 150-162$ GeV (19.5 - 21 GeV) \cite{Dev:2013vxa}.

\end{enumerate}

\section{Properties of $W_R$ and $N$ at Hadron Colliders}\label{sec:xsec}
In this section, we present production and decay rates of $\WR$ and $N$ to leptons and jets, 
with $\mN\ll\MWR$, at the 13 TeV LHC and 100 TeV VLHC.

In the MLRSM, the $\WR$ interaction with quarks is given by the Lagrangian
\begin{eqnarray}
 \mathcal{L}_{\WR-q-q'} = \frac{-g}{\sqrt{2}}\sum_{i,j=u,d,\dots}\overline{u}_i V_{ij}^{\rm CKM'}~W_{R \mu}^+ \gamma^\mu P_R~ d_j + \text{H.c.},
\end{eqnarray}
where $u_i (d_j)$ is an up-(down-)type quark of flavor $i (j)$; 
$V_{ij}^{\rm CKM'}$  is the RH Cabbibo-Kobayashi-Masakawa (CKM) matrix; and
$P_{R(L)} = \frac{1}{2}(1\pm\gamma^5)$ denotes the RH(LH) chiral projection operator.
For either generalized charge conjugation or parity, one expects $\vert V_{ij}^{\rm CKM'}\vert \approx \vert V_{ij}^{\rm CKM}\vert$,
up to $\mathcal{O}(m_b/m_t)$ contributions for the latter case~\cite{Zhang:2007fn,Zhang:2007da,Maiezza:2010ic,Senjanovic:2014pva,Senjanovic:2015yea}.
Hence, we can assume, for simplicity, four massless quarks and take $V_{ij}^{\rm CKM'}$ to be diagonal with unit entries.

The $\WR$ coupling to six heavy $(N_{m'})$ and light $(\nu_{m})$  neutrinos is parametrized by~\cite{Atre:2009rg,Han:2012vk}
\begin{eqnarray}
 \mathcal{L}_{\WR-\ell-\nu/N} = \frac{-g}{\sqrt{2}}
 \sum_{\ell=e,\mu,\tau}
 \left[
 \sum_{m=1}^3	\overline{\nu^{c}_m} X_{\ell m} ~+~
 \sum_{m'=4}^6	\overline{N_{m'}} Y_{\ell m'}
 \right] ~W_{R \mu}^+ \gamma^\mu P_R~ \ell^-+\text{H.c.},
\end{eqnarray}
where mixing matrices $X_{\ell m}$ and $Y_{\ell m'}$  are defined in Sec.~\ref{sec:Nmass}.
We consider only the lightest heavy neutrino, denoted simply by $N$, and neglect heavier mass eigenstates.
For simplicity, we assume diagonal neutrino mixing with maximal coupling to electron-flavor leptons:
\begin{equation}
  \vert Y_{e N}\vert = 1, \quad \vert Y_{\mu N}\vert = \vert Y_{\tau N}\vert = \vert X_{\ell m}\vert = 0.
  \label{eq:NuMixingAssignments}
\end{equation}
{Choosing instead maximal coupling to muons, i.e., $\vert Y_{\mu N}\vert=1$,
or large $e-\mu$ mixing, i.e., $\vert Y_{e N}\vert\sim\vert Y_{\mu N}\vert$, 
has little impact on our analysis due to the long lifetime of the muon.
On the other hand, the $\tau \ell$ final state requires specialized cuts to account for $\tau$ decays to light neutrinos.
For more details, see Sec.~\ref{sec:pheno}.}

For numerical results, SM inputs are taken from the 2014 Particle Data Group~\cite{Agashe:2014kda}:
\begin{eqnarray}
 \alpha^{\rm \overline{MS}}(M_Z)		= 1/127.940,
 ~M_{Z}=91.1876\GeV, 	
 ~\sin^{2}_{\rm \overline{MS}}(\theta_{W};M_Z) 	= 0.23126. 
 \label{eq:smInputs}
\end{eqnarray}
PDFs and $\as(\mu_r)$ are extracted using the LHAPDF 6.1.6 libraries~\cite{Buckley:2014ana}.
The factorization $(\mu_f)$ and renormalization $(\mu_r)$ scales are set to $\mu_0 = \MWR$ everywhere.
For LO- and NLO-accurate calculations, we use the NNPDF 3.0 NLO $n_f=4$~(\texttt{lhaid=260400}) PDF set~\cite{Ball:2014uwa}.
For NLO+NNLL calculations, we use the threshold-improved NNPDF 3.0 NNLO+NNLL PDF set~\cite{Bonvini:2015ira}.
This choice follows from the unavailability of an NLO+NNLL PDF set and our desire to ascertain the effects of resummation at NNLL.
Formally, the induced uncertainty from our PDF-mismatching in the LO and NLO+NNLL calculations is 
$\mathcal{O}(\alpha_s)$ and $\mathcal{O}(\alpha_s^2)$, respectively, and beyond our claimed accuracy.
Numerically, this leads to LO cross sections that are $10\%$ smaller than those calculated with LO PDFs.

For additional computational details, see App.~\ref{app:threshold}.

\subsection{$W_R$ Production at NLO+NNLL}
At Fixed Order (FO) accuracy, we calculate the inclusive production cross section for 
\begin{equation}
p ~p ~\rightarrow ~\WR^\pm ~+~ X,
\label{eq:ppWRX}
\end{equation}
where $X$ is anything, via the usual application of the Collinear Factorization Theorem:
\begin{eqnarray}
 \sigmaFO(pp\rightarrow \WR+X) = \sum_{a,b=q,\overline{q'},g}\int_{\tau_0}^1 d\tau ~\mathcal{L}_{ab}(\tau,\mu_f)
  ~\hat{\sigma}^{\rm FO}(ab\rightarrow \WR),  \quad \tau_0 \equiv \frac{\MWR^2}{s}.
  \label{eq:foFactThm}
\end{eqnarray}
The luminosity $\mathcal{L}(\tau)$ of parton pair $ab$ in $pp$ collisions given by
\begin{eqnarray}
 \mathcal{L}_{ab}(\tau,\mu_f) &=& \frac{1}{1+\delta_{ab}}\int_{\tau}^1 \frac{d\xi_1}{\xi_1}
 \left[f_{a/p}(\xi_1,\mu_f)f_{b/p}(\xi_2,\mu_f) + f_{a/p}(\xi_2,\mu_f)f_{b/p}(\xi_1,\mu_f)\right], 
 \label{eq:lumiDef}\\
 \xi_2 &\equiv& \frac{\tau}{\xi_1}.
\end{eqnarray}
The PDFs $f_{a/p}(\xi_i,\mu_f)$ represent the likelihood of observing
parton $a$ in proton $p$ possessing a longitudinal momentum fraction $\xi_i = E_a / E_{p_i} = p^z_a / E_{p_i}	$, 
and (re)sum arbitrary collinear parton emissions up to a factorization scale $\mu_f$.
The partonic c.m.~energy $\sqrt{\hat{s}}$ is related to the hadronic (beam) c.m.~energy $\sqrt{s}$ by the hadronic threshold variable
\begin{equation}
 \tau = \xi_1 \xi_2 = \frac{\hat{s}}{s}, \quad \tau_0 \leq \tau < 1,
\end{equation}
and extends to the kinematic threshold $\tau_0$, below which Eq.~(\ref{eq:ppWRX}) is kinematically forbidden.

Partonic scattering rates $\hat{\sigma}$ are evaluated via helicity amplitudes,
and use the CUBA libraries~\cite{Hahn:2004fe} to handle Monte Carlo integration.
NLO in QCD corrections are obtained using the Phase Space Slicing
method~\cite{Fabricius:1981sx,Kramer:1986mc,Baer:1989jg,Harris:2001sx} and exploit factorization properties of Drell-Yan (DY) currents; 
see appendices of Refs.~\cite{Harris:2001sx,Ruiz:2015zca}.
LO and NLO results are checked against literature~\cite{Sullivan:2002jt,Jezo:2014wra,Roitgrund:2014zka} and  
MG5\_aMC@NLO~v2.3.3 (MG5)~\cite{Alwall:2014hca} assuming $\MWR=M_{W}$.

\begin{figure}[!t]
\begin{center}
\includegraphics[width=.97\textwidth]{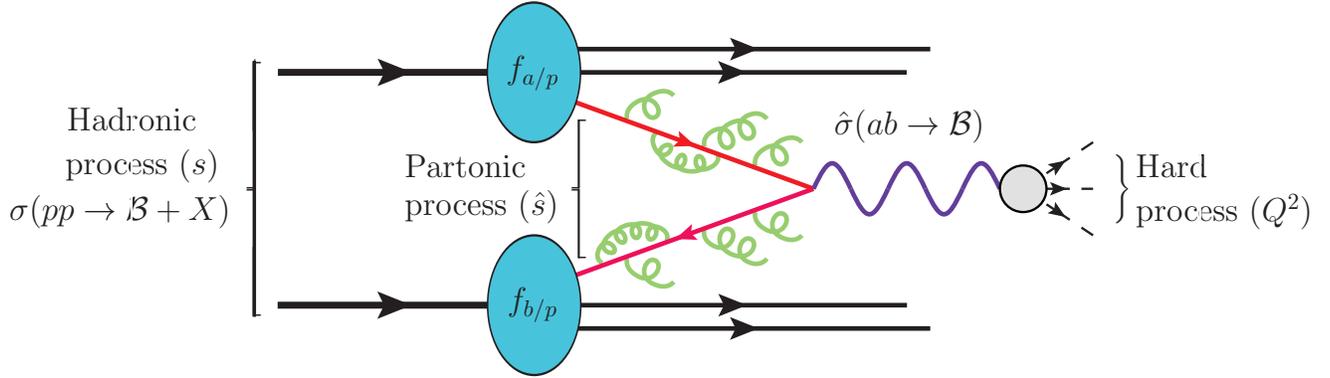}	
\end{center}
\caption{Schematic definitions of the hadronic $(s)$, partonic $(\hat{s})$ and hard scatting $(Q^2)$ components 
for inclusive production of a generic color-singlet boson $\mathcal{B}$ in $pp$ collisions.}
\label{fig:factorThm}
\end{figure}

Beyond FO, Eq.~(\ref{eq:foFactThm}) can be generalized~\cite{Sterman:1986aj,Catani:1989ne,Catani:1990rp} 
to include the arbitrary, initial-state emission of soft gluons, i.e., 
with energies much smaller than the hard scattering process scale $Q$.
The interpretation of $\hat{\sigma}$ also generalizes to include both the hard process,
\begin{equation}
 q ~\overline{q'} ~\rightarrow ~\WR^\pm \quad~\text{with}\quad Q = \MWR,
\end{equation}
and the factorized soft radiation off the $q,\overline{q'}$ initial states.
Schematically, the definitions of the hadronic, partonic, and hard components 
for the inclusive production of a generic color-singlet boson $\mathcal{B}$ are drawn in Fig.~\ref{fig:factorThm}.
Necessarily, the inequality $s > \hat{s} \geq Q^2$ holds.

Soft radiation becomes important when the hard scale approaches the partonic scale, i.e.,
when the partonic threshold variable $z$ approaches one:
\begin{equation}
 z \equiv \frac{Q^2}{\hat{s}} = \frac{\MWR^2}{\hat{s}} = \frac{\tau_0}{\tau}\rightarrow1.
\end{equation}
In this kinematic regime, 
which can be satisfied at $Q^2 \ll s$ as in  Higgs production via GF
or when $Q^2\sim s$ as in the present case of high-mass DY,
soft radiation give rise to numerically large logarithms that require resummation in order to restore perturbativity of Eq.~(\ref{eq:foFactThm}).

To carry out the resummation, we follow the procedure (and largely notations) of Refs.~\cite{Catani:1996yz,Catani:2003zt,Bonvini:2010tp},
and write a generalized form of Eq.~(\ref{eq:foFactThm}) in terms of $\tau,z,$ and $\tau_0$:
\begin{equation}
 \sigmaFO(pp\rightarrow \WR+X) = \sum_{a,b=q,\overline{q'},g}\int_{\tau_0}^1 d\tau ~\int_0^1 dz ~\delta\left(z - \frac{\tau_0}{\tau}\right) 
  ~\mathcal{L}_{ab}(\tau)  ~ \hat{\sigma}^{\rm FO}_{ab}(ab\rightarrow\WR).
 \label{eq:resFactThm}
\end{equation}
For inclusive $W_R$ production, $\hat{\sigma}^{\rm FO}$ can be expressed as
\begin{eqnarray}
  \hat{\sigma}^{\rm FO}_{ab} ~\equiv~ \hat{\sigma}^{\rm FO}(ab\rightarrow \WR) ~=~ \sigma_0 \times z \times \Delta_{ab}^{\rm FO}(z).
 \label{eq:foPartonicXSec}
\end{eqnarray}
The constant term $\sigma_0$ for  gauge coupling $g_R^2 = g^2 = 4\pi\alpha/\sin^2\theta_W$ is 
\begin{equation}
 \sigma_0 = \frac{g^2_R \pi ~ \vert V_{ab}^{\rm CKM'}\vert^2}{4N_c \MWR^2},
\end{equation}
and is related to the usual LO partonic formula by
\begin{equation}
 \hat{\sigma}^{\rm LO}(ab\rightarrow \WR) = \sigma_0 ~\times~ \MWR^2 ~\times~ \delta(\hat{s} -  \MWR^2) = 
 \sigma_0 ~\times~ z ~\times~ \delta(1 - z).
\end{equation}
Hence, one may identify up to $\mathcal{O}(\as)$, $\Delta^{\rm FO}_{q\overline{q'}}(z) \approx \delta(1-z) + \mathcal{O}(\as)$.

If working with pQCD, the threshold resummed cross section can be efficiently obtained after writing 
the hadronic cross section in so-called Mellin-space. 
For the function $h(x)$, the $N$th-moment of its Mellin transform and inverse Mellin transform with respect to $x$ are,
\begin{eqnarray}
 h_N \equiv \mathcal{M}[h(x);N] &=&  \int^1_0 dx ~x^{N-1}~h(x), 
 \label{eq:mellinDef}
 \\
 h(x) = \mathcal{M}^{-1}[h_N;x] &=& \frac{1}{2\pi i} \int_{c - i\infty}^{c + i\infty}dN ~x^{-N}~h_N, 
 \label{eq:invMellinDef}
\end{eqnarray}	
where $c\in\mathbb{R}$ is to the right of all singularities in $h_N$.
The Mellin transform of Eq.~(\ref{eq:resFactThm}) at LO with respect to $\tau_0$, gives
\begin{eqnarray}
  \sigmaLO_N &=&  \int^1_0 d\tau_0~\tau_0^{N-1} ~\times~ \sigmaLO(\tau_0) = 
  \sigma_0 ~\mathcal{L}_{q\overline{q'},(N+1)} ~\times~ \Delta_{q\overline{q'},(N+1)}^{\rm LO},
  \label{eq:foMellin}
\end{eqnarray}
revealing an explicit factorization into a product of the luminosity and soft coefficient, normalized by the Born weight $\sigma_0$. 
We drop the summation over $a,b=g$ as the $gq, ~g\overline{q'},$ and $gg$ initial states do not contribute to $\WR$ production at LO.

\begin{figure}[!t]
\begin{center}
\subfigure[]{\includegraphics[scale=1,width=.48\textwidth]{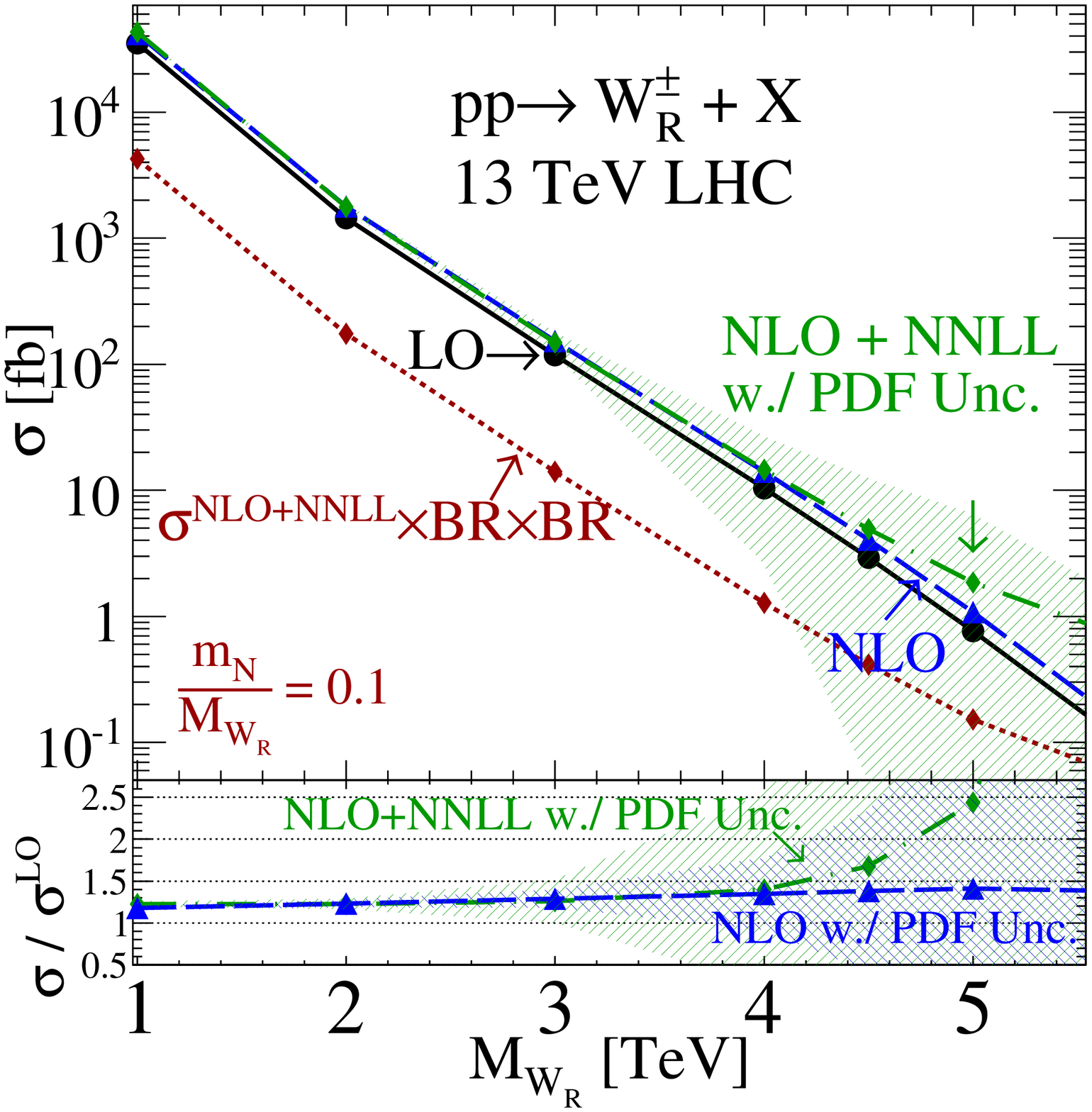}	\label{fig:xsec13TeV}}
\subfigure[]{\includegraphics[scale=1,width=.48\textwidth]{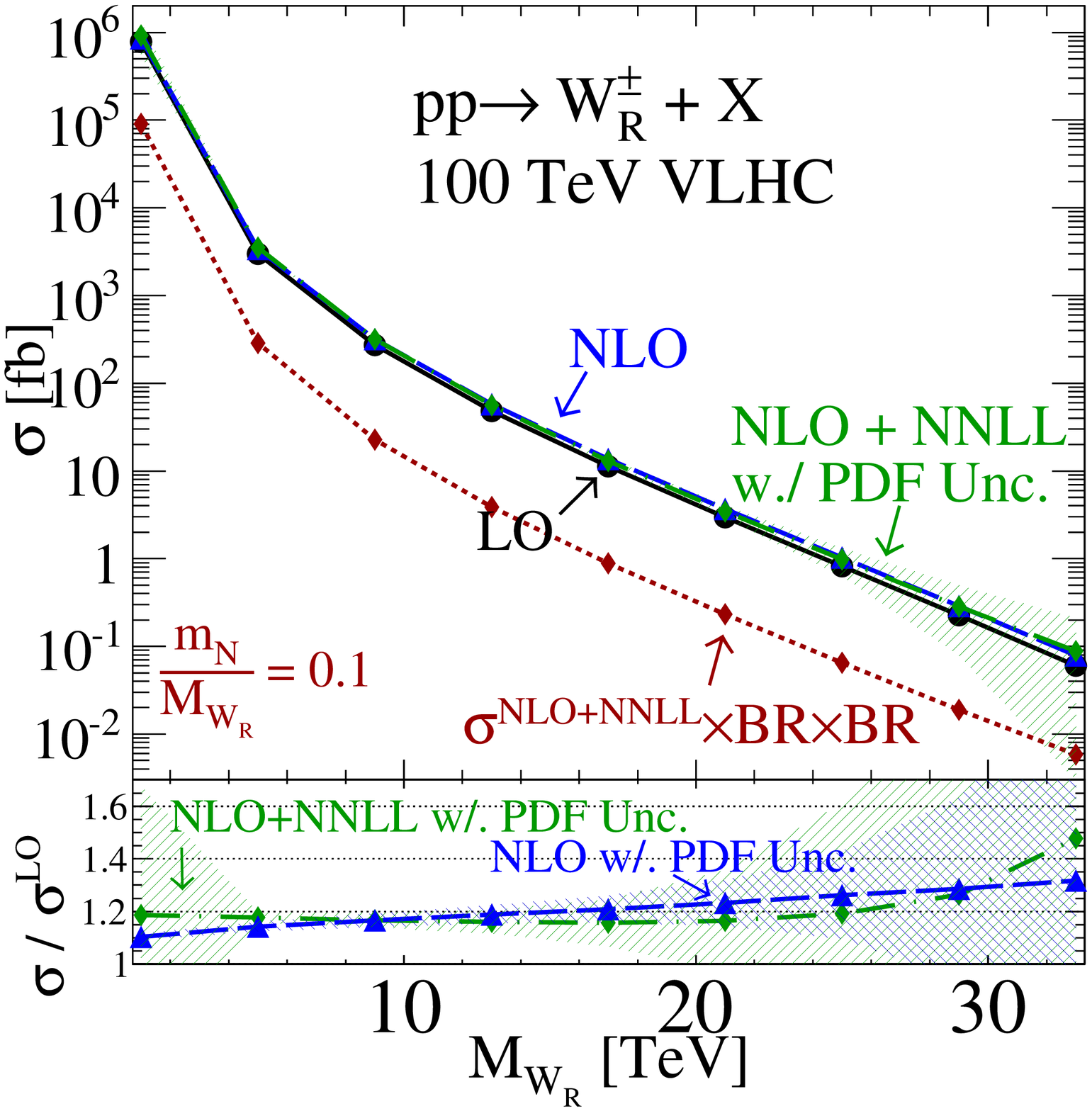}	\label{fig:xsec100TeV}}
\end{center}
\caption{
Upper panel: As a function of $\MWR$, $pp\rightarrow \WR$ production cross section for $\sqrt{s} =$ (a) 13 and (b) 100 TeV,
at LO (solid), NLO (dash), and NLO+NNLL (dash-dot) with $1\sigma$ PDF uncertainty (shaded), as well as 
$\sigma^{\rm NLO+NNLL}(pp\rightarrow \WR)
\times\text{BR}(\WR\rightarrow Ne)
\times\text{BR}(N\rightarrow eq\overline{q'})$ (dot).
Lower: NLO (dash) and NLO+NNLL (dash-dot) $K$-factors and PDF uncertainties.
} 
\label{fig:xsec}
\end{figure}

The advantage of working in Mellin-space is this explicit factorization.
Exploiting that in the soft limit gauge radiation amplitudes reduce to their color-connected Born amplitudes, 
resummation reduces to the simple procedure of replacing the LO soft coefficient $\Delta_{ab,N}^{\rm LO}$ 
with its resummed analogue $\Delta_{ab,N}^{\rm Res.}$~\cite{Sterman:1986aj,Catani:1989ne,Catani:1990rp}.
Thus, the threshold-resummed $pp\rightarrow \WR$ cross section in Mellin-space is 
\begin{eqnarray}
  \sigmaRes_N &=&   \sigma_0 ~ \mathcal{L}_{q\overline{q'},(N+1)} ~\times~ \Delta_{q\overline{q'},(N+1)}^{\rm Res.},
  \label{eq:resMellin} 
\end{eqnarray}
and in momentum space by  Mellin inverse of the above with respect to $\tau_0$:
\begin{eqnarray}
 \sigmaRes(pp\rightarrow \WR+X) 
 =
 \frac{\sigma_0}{2\pi i} \int_{c - i\infty}^{c + i\infty}dN 
 ~\tau_0^{-N} ~\times~ \mathcal{L}_{q\overline{q'},(N+1)} ~\times~ \Delta_{q\overline{q'},(N+1)}^{\rm Res.}
\end{eqnarray}
We approximate the luminosity function $\mathcal{L}(\tau)$ using the Chebyshev polynomial approximation~\cite{Bonvini:2010tp,Bonvini:2012sh},
which can be Mellin-transformed analytically,
and choose the integration path according to the Minimal Prescription (MP) procedure~\cite{Catani:1996yz}.
See App.~\ref{app:threshold} for more details.

Matching resummed and FO calculations beyond LO 
requires subtracting the soft contributions common to both calculations to avoid phase space double counting.
For a FO result at N$^k$LO, this can be done  by Taylor-expanding
$\sigmaRes$ up to $\mathcal{O}(\alpha_s^k)$, subtracting these terms from $\sigmaRes$,
and adding the N$^k$LO calculation to the residual resummed expression.
One may interpret this procedure as augmenting with approximate $\mathcal{O}(\alpha_s^k)$ terms in $\sigmaRes$, i.e., soft/non-hard, 
with the full $\mathcal{O}(\alpha_s^k)$ calculation, which describes accurately both soft and hard radiation.
Subsequently, $\WR$ production  matched at N$^k$LO+N$^j$LL is given by
\begin{eqnarray}
 \sigma^{\rm  N^{\it k}LO+N^{\it j}LL}(pp\rightarrow \WR +X) &=& \sigma^{\rm N^{\it k}LO} + \sigma^{\rm N^{\it j}LL} - 
 \sum_{l=0}^k \frac{\alpha_s^l}{l!}\left[\frac{d^l}{d\alpha_s^l}\sigma^{\rm N^{\it j}LL} \right]_{\alpha_s =0}.
\end{eqnarray}

\begin{table}[!t]
\small
\begin{center}
 \begin{tabular}{ c || c | c | c | c | c  }
\hline \hline
\multicolumn{6}{c}{$\sigma(pp\rightarrow \WR^\pm)$ [fb]}	\tabularnewline\hline
		 \multicolumn{6}{c}{13 TeV LHC}		\tabularnewline\hline
$\MWR$	& $\sigmaLO$	& $\sigmaNLO$	& $\kNLO$	& $\sigma^{\rm NLO+NNLL}$	& $K^{\rm NLO +}_{\rm NNLL}$	\tabularnewline\hline
1 TeV	& $3.52 \times 10^4 $		& $4.15^{+0.08~+0.08}_{-0.07~-0.08} \times 10^4 $		& $1.18$	
					& $ 4.33^{+(<0.5\%)~+0.09}_{-(<0.5\%)~-0.09} \times 10^4$	& $1.23 $		\tabularnewline\hline
3 TeV	& $1.18 \times 10^2 $		& $1.51^{+0.06~+0.13}_{-0.06~-0.13} \times 10^2 $		& $1.29$	
					 & $ 1.48^{+(<0.5\%)~+0.3}_{-(<0.5\%)~-0.3} \times 10^2$	& $ 1.25$		\tabularnewline\hline
5 TeV	& $0.765 $			& $1.08^{+0.07~+1.75}_{-0.07~-1.75} $				& $1.41$	
					& $1.86^{+(<0.5\%)~+4.55}_{-(<0.5\%)~-4.55} $			& $ 2.43$	\tabularnewline\hline
	 \multicolumn{6}{c}{100 TeV VLHC}	\tabularnewline\hline
$\MWR$	& $\sigmaLO$	& $\sigmaNLO$	& $\kNLO$	& $\sigma^{\rm NLO+NNLL}$	& $K^{\rm NLO+}_{\rm NNLL}$	\tabularnewline\hline
1 TeV	& $7.78 \times 10^5 $		& $8.60^{+0.06~+0.09}_{-(<0.5\%)~-0.09} \times 10^5 $		& $1.11$	
					& $9.25^{+0.23~+3.81}_{-0.19~-3.81}\times 10^5 $		& $1.19$		\tabularnewline\hline
5 TeV	& $2.98 \times 10^3 $		& $3.40^{+0.04~+0.05}_{-0.03~-0.05} \times 10^3 $		& $1.14$	
					& $3.50^{+0.02~+0.06}_{-(<0.5\%)~-0.06} \times 10^3 $		& $1.17$		\tabularnewline\hline
25 TeV	& $0.818 $			& $1.03^{+0.03~+0.14}_{-0.03~-0.14} $				& $1.26$	
					& $0.970^{+(<0.5\%)~+0.342}_{-(<0.5\%)~-0.342} $		& $1.19$		\tabularnewline\hline
33 TeV	& $5.98\times 10^{-2}$		& $7.86^{+0.31~+4.66}_{-0.34~-4.66} \times 10^{-2}$	& $1.31$
					& $8.81^{+(<0.5\%)~+12.2}_{-(<0.5\%)~-12.2} \times 10^{-2}$	& $1.47$	\tabularnewline\hline
\hline
\end{tabular}
\caption{$pp\rightarrow\WR$ production cross sections and $K$-factors at various accuracies for representative $\MWR$ and $\sqrt{s} = 13,~100\TeV$,
with absolute scale (first) and PDF (second) uncertainties. Exceptionally small uncertainties are noted by $(<0.5\%)$.}
\label{tb:wprimeXSec}
\end{center}
\end{table}

In Fig.~\ref{fig:xsec}, we show the total inclusive $pp\rightarrow\WR$ cross section at NLO+NNLL (dash-dot) with PDF uncertainty (shaded),
NLO (dash), and LO (solid) at (a) 13 and (b) 100 TeV.
The production rates at 13 (100) TeV span approximately:
\begin{equation}
 2\fb - 40\pb~(90\ab  - 930\pb) \quad\text{for}\quad \MWR = 1-5~(1-33)\TeV.
\end{equation}
In the lower panel are the NLO+NNLL and NLO $K$-factors, defined respectively as
\begin{equation}
 K^{\rm NLO+NNLL} \equiv \frac{\sigma^{\rm NLO+NNLL}}{\sigma^{\rm LO}}
 \quad\text{and}\quad
 K^{\rm NLO} \equiv \frac{\sigma^{\rm NLO}}{\sigma^{\rm LO}}.
 \label{eq:kFactors}
\end{equation}
The NLO+NNLL (dash-dot) and NLO (dash) $K$-factors with uncertainties span roughly:
\begin{eqnarray}
 K^{\rm NLO+NNLL} 	&:& 1.2 - 2.4 ~(1.2 - 1.5)\\	
 K^{\rm NLO} 		&:& 1.2 - 1.4 ~(1.1 - 1.3).
\end{eqnarray}
At 13 and 100 TeV, we observe that the effects of resummation become important with respect to the NLO rate at $\tau_0\approx0.3$.
At 13 TeV, the resummed corrections for $\tau_0>0.3$ are very large, 
increasing the Born (NLO) predictions by $40-140~(4-70)\%$ for $M_{W_R} = 4-5\TeV$.
The largeness of the 13 TeV $K$-factors for $\MWR \gtrsim 4\TeV$ does not indicate the breakdown of perturbation theory.
Rather, it demonstrates the importance of soft radiation as $\tau_0\rightarrow 1$, 
and is typical for processes near the boundaries of phase space~\cite{Catani:1996yz}.
For the DY process, this is particularly important for $\tau_0\gtrsim0.1$~\cite{Appell:1988ie}.
This is exemplified at 100 TeV by the reduced importance of resummation for comparable $\MWR$ (smaller $\tau_0$).
Despite the largeness of the PDF uncertainties at large $\MWR$, the NLO+NNLL central value remains within the NLO uncertainty,
as seen in the lower panel of Fig.~\ref{fig:xsec13TeV}. See Sec.~\ref{sec:uncertainties} for further discussions on uncertainties.
Away from threshold, the resummed calculation converges to the FO result, consistent with expectations~\cite{Bonvini:2015ira}.
For select $\MWR$, we summarize our NLO and NLO+NNLL results in Tb.~\ref{tb:wprimeXSec}.

\subsection{$W_R$ Decay}
\begin{table}[!t]
\begin{center}
 \begin{tabular}{ c | c | c | c | c | c }
\hline \hline
$(\MWR,\mN)$ [TeV,GeV]	& $(3,30)$		& $(3,150)$	& $(3,300)$	& $(4,400)$	& $(5,500)$	\tabularnewline\hline \hline 
$\Gamma_{\WR}$	[GeV]	& $84.4 $		& $84.3 $	& $84.2 $	& $112 $	& $141 $	\tabularnewline\hline 
$\Gamma_{N}$	[eV]	& $3.41\times10^{-3}$	& $10.7 $	& $355 $ 	& $513 $ 	& $687 $	\tabularnewline\hline 
 \hline 
\end{tabular}
\caption{Total $\WR$ and $N$ decay widths for representative $\MWR$ and $\mN$.}
\label{tb:widths}
\end{center}
\end{table}

As discussed in Sec.~\ref{sec:WRMass} and in Sec.~\ref{sec:Nmass},  $W_R-W_L$ and $N_i-\nu_i$ mixing are negligibly small and $m_{\rm FCNH} \gg \MWR$.
Subsequently, for $\mN < \MWR$, the only open $\WR$ decay modes are to quark and $\ell^\pm N$ pairs.
The corresponding partial widths are
\begin{eqnarray}
 \Gam{\WR \rightarrow q\overline{q'}} &=& N_c \vert V^{\rm{CKM'}}_{qq'}\vert^2 \frac{g^2 \MWR}{48 \pi},
 \\
 \Gam{\WR \rightarrow tb} 		&=& N_c \vert V^{\rm{CKM'}}_{tb}\vert^2 \frac{g^2 \MWR}{48 \pi}(1-r_t)^2(1+\frac{1}{2}r_t),
 \\
 \Gam{\WR \rightarrow \ell N} 	&=&     \vert Y_{\ell N}\vert^2 \frac{g^2 \MWR}{48 \pi}(1-r_N)^2(1+\frac{1}{2}r_N), \quad r_i = \frac{m_{i}^2}{\MWR^2}.
\end{eqnarray}
For our choice of quark and lepton mixing, the total $\WR$ width is
\begin{eqnarray}
 \Gamma_{\WR} &=& 2\Gamma(\WR \rightarrow q\overline{q'}) + \Gamma(\WR \rightarrow tb) + \Gamma(\WR \rightarrow e N) 
 \\
	      &=& \frac{g^2 \MWR}{48 \pi}\left[2N_c + N_c(1-r_t)^2(1+\frac{1}{2}r_t) + (1-r_N)^2(1+\frac{1}{2}r_N) \right].
\end{eqnarray}
We calculate the total $W_R$ and $N$, decay widths for representative masses in Tb.~\ref{tb:widths}.

The branching fraction of $A$ to final-state $X_i$ is defined as
\begin{equation}
 \text{BR}(A \rightarrow X_i) \equiv \cfrac{\Gam{A \rightarrow X_i}}{\sum_i \Gam{A \rightarrow X_i}}.
\end{equation}
In the large $\MWR$ limit, the $\WR$ branching fractions converge to the asymptotic values,
\begin{eqnarray}
\text{BR}(\WR \rightarrow q\overline{q'})	&\approx& 2\times\text{BR}(\WR \rightarrow tb) \approx \frac{2N_c}{3N_c + 1} = 60\%,
\\
\text{BR}(\WR \rightarrow Ne)			&\approx& \frac{1}{3N_c + 1} = 10\%.
\end{eqnarray}

\begin{figure}[!t]
\begin{center}
\subfigure[]{\includegraphics[width=.48\textwidth]{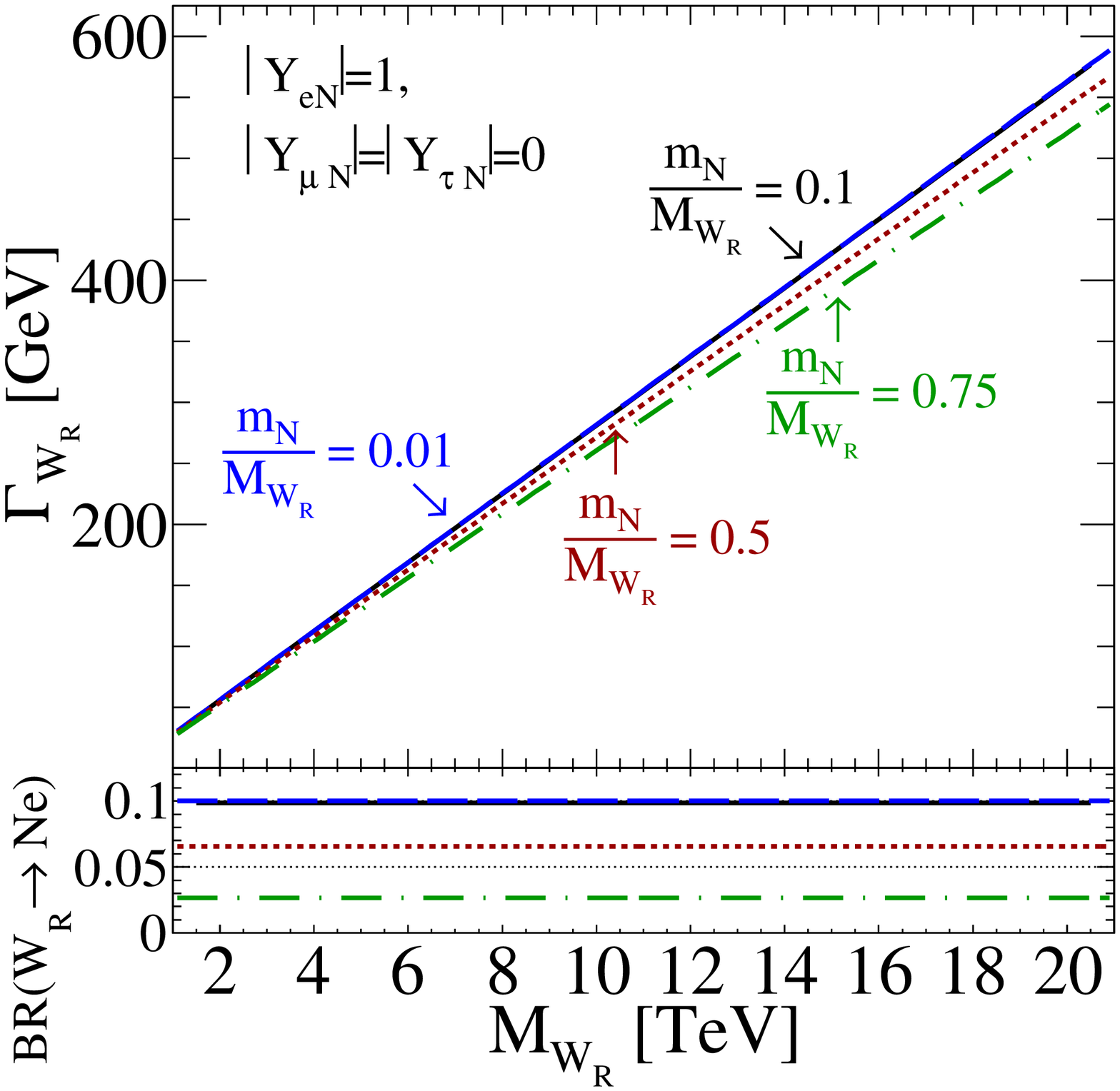}	\label{fig:dcWR}}
\subfigure[]{\includegraphics[width=.48\textwidth]{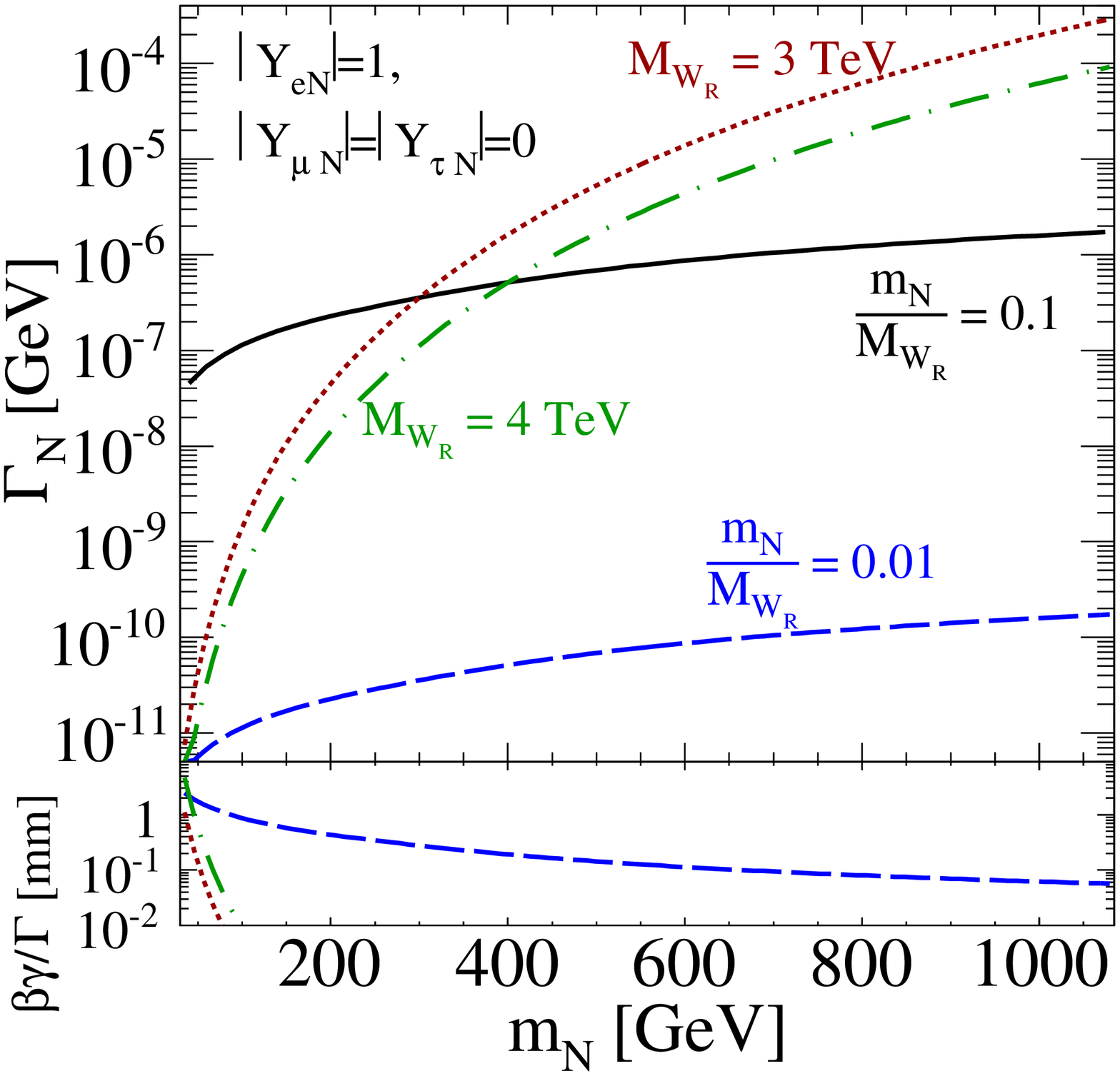}	\label{fig:dcNR}}
\end{center}
\caption{
Total decay widths for representative $\sqrt{r_N} = m_N / \MWR$ of (a) $\WR$ as a function of $\MWR$ and (b) $N$ as a function of $\mN$.
Lower: 
(a) $\WR \rightarrow Ne$ branching fraction.
(b) $N$ mean lifetime $d = \beta\gamma/\Gamma_N$ in $\WR$'s frame [mm].
} 
\label{fig:dcNRWR}
\end{figure}

In the upper (lower) panel of Fig.~\ref{fig:dcWR}, we show the total $\WR$ decay width (branching fraction) for $\MWR>1\TeV$ and fixed $m_N/\MWR$ ratios of 
$\sqrt{r_N} = 0.01$ (dash), $0.1$ (solid), $0.5$ (dot), and $0.75$ (dot-dash). 
Similar to the EW gauge bosons, the $\WR$ in this model has a narrow width for all values of $\MWR$, with $\Gamma_{\WR}/\MWR$ scaling as:
\begin{equation}
 \cfrac{\Gamma_{\WR}}{\MWR} \sim \frac{g^2}{48\pi}(3N_c + 1) \approx 2.8\%.
\end{equation}
This justifies the use of the Narrow Width Approximation (NWA). 
Furthermore, as $pp\to\WR$ is a DY process, its factorization properties imply that the
NLO and NLO+NNLL corrections to its on-shell production and decay to $N$ 
are equivalent to the production-only corrections, i.e.,
\begin{equation}
 \sigma^{\rm NLO(+NNLL)}(pp\rightarrow\WR\rightarrow e^\pm N) \approx \sigma^{\rm NLO(+NNLL)}(pp\rightarrow\WR) \times \text{BR}(\WR\rightarrow e^\pm N).
 \label{eq:sigmaBR}
\end{equation}
In the lower panel of Fig.~\ref{fig:dcWR}, we observe that the $\WR$ branching fractions remain virtually independent of $m_N$
and attain its maximum branching of BR$(\WR\rightarrow Ne^\pm)\approx 0.1$.
For 13 (100) TeV and $(\mN/\MWR) = 0.1$ the $pp\rightarrow \WR\rightarrow Ne$ cross section [Eq.~(\ref{eq:sigmaBR})] spans:
\begin{equation}
 180 \pb - 15\fb~(100 \pb - 350\fb) \quad\text{for}\quad \MWR = 3-5~(5-25)\TeV.
\end{equation}
For representative $(\MWR,\mN)$, we summarize our results in column 2 of Tb.~\ref{tb:wprimeXsec_decay}.

\subsection{$N$ Decays}
In our scenario, the heavy neutrino  dominantly decays  to the three-body final state
\begin{equation}
 N \rightarrow  ~e^\pm ~W^{\mp *}_R \rightarrow  ~e^\pm ~q ~\overline{q'}.
\end{equation}
Both $e^+$ and $e^-$ are allowed in the final state due to the Majorana nature of $N$. 
If kinematically accessible, the heavy neutrino can also decay to $t$ and $b$ quarks, with the final state $e^\pm t b$.
{
In principle, $N$ can also decay to SM EW bosons via mixing with SM neutrinos;
the rate is controlled by the tiny mixing parameter $\vert X_{\ell N}\vert^2 \sim 1 - \vert Y_{\ell N}\vert^2 \sim \mathcal{O}(m_\nu^2 /m_N^2)$.
Following Eq.~(\ref{eq:NuMixingAssignments}), such decays vanish at tree-level and, therefore, are not considered in the analysis.
}
For $\mN\ll\MWR$, the partial widths of $N$ are
\begin{eqnarray}
\Gam{N\rightarrow e^\pm ~q ~\overline{q'}} &=& 
2 N_c\cfrac{\vert Y_{\ell N}\vert^2 \vert V^{\rm{CKM'}}_{qq'}\vert^2 g^4 \mN^5}{3\cdot2^{11}\cdot\pi^3 \MWR^4}, \qquad y_t = \cfrac{m_t^2}{\mN^2}, 
\\
\Gam{N\rightarrow e^\pm ~t ~b} &=& 
2 N_c\cfrac{\vert Y_{\ell N}\vert^2 \vert V^{\rm{CKM'}}_{tb}\vert^2 g^4 \mN^5}{3\cdot2^{11}\cdot\pi^3 \MWR^4} 
\left(1-8y_t+8y_t^3-y_t^4 -12y_t^2\log y_t\right). 
\end{eqnarray}
The validity of this approximation for $\mN/\MWR\sim 0.1$ has been checked against MG5.
For our choice of mixing, the total $N$ width is
\begin{eqnarray}
 \Gamma_N &=& 2\Gam{N\rightarrow e^\pm ~q ~\overline{q'}} +  \Gam{N\rightarrow e^\pm ~t ~b} \\
	  &=& 2 N_c\cfrac{g^4 \mN^5}{3\cdot2^{11}\cdot\pi^3 \MWR^4}
 \left[3-8y_t+8y_t^3-y_t^4 -12y_t^2\log y_t\right].
 \label{eq:nWidth}
\end{eqnarray}
and implies that $\Gamma_N/\mN$ scales as
\begin{equation}
 \cfrac{\Gamma_N}{\mN} = \cfrac{g^4}{2^{10}~\pi^3}\left(\frac{\mN}{\MWR}\right)^4 \sim 5\cdot10^{-6} \times\left(\frac{\mN}{\MWR}\right)^4  \ll 1.
\end{equation}
Hence, application of the NWA in $N$ decays is justified but suggests $N$ may be long-lived.
Values of $\Gamma_N$ for representative $\MWR$ and $\mN$ used in this study are given in Tb.~\ref{tb:widths}.

In Fig.~\ref{fig:dcNR}, we plot $\Gamma_N$ as a function of $\mN$ for representative $\MWR$ and $(\mN/\MWR)$ ratios;
in the lower panel we show the mean flight distances
\begin{equation}
 d_0 = v \tau_0 = \beta \gamma \hbar c /\Gamma_N, \quad \beta\gamma = \frac{(1-r_N)}{2\sqrt{r_N}}.
\end{equation}
For $\mN = 30-1000\GeV$, we find
\begin{eqnarray}
 \cfrac{\mN}{\MWR}=0.1~\text{(solid)}	&:&	\confirm{\Gamma_N \sim 10^{-8} - 10^{-6}\GeV}, \\
 \cfrac{\mN}{\MWR}=0.01~\text{(dash)}	&:&	\confirm{\Gamma_N \sim 10^{-12} - 10^{-10}\GeV}.
\end{eqnarray}
The corresponding mean flight distances span
\begin{eqnarray}
 \cfrac{\mN}{\MWR}=0.1~\text{(solid)}	&:&	\confirm{d_0 \sim 10^{-7} - 10^{-5}~\text{mm},}\\
 \cfrac{\mN}{\MWR}=0.01~\text{(dash)}	&:&	\confirm{d_0 \sim 10^{-2} - 3~\text{mm}.}
\end{eqnarray}
This implies that for $N$ much lighter than $\MWR$, i.e., $\mN/\MWR < 0.01$, 
heavy neutrinos appear in detector experiments as displaced vertices, not prompt decays.
However, such a scenario is not reasonable within the spirit of the LRSM model. 
Supposing $\mN/\MWR < 0.01$ and using expressions for $\mN,\MWR$ in Sec.~\ref{sec:Nmass} and Sec.~\ref{sec:WRMass},
the Yukawa couplings of the heavy neutrino $N$ to the triplet Higgs are restricted to $f_R < 3\times10^{-3}$.
This  is comparable to generation I and II quark SM Yukawa couplings.
However, taking $\mN\sim\mathcal{O}(10)\GeV$, a (vanilla) Type I Seesaw then requires for light neutrino masses $m_{\nu_{m}}\sim 0.1$ eV 
a Dirac neutrino mass of $m_D \sim 30\KeV$, or a Yukawa coupling $\mathcal{O}(15-20)\times$ smaller than the SM electron Yukawa.
Though not forbidden, this is contrary to the Seesaw spirit of explaining light neutrino masses without excessively small couplings.

\begin{table}[!t]
\centering
\begin{tabular}{c||c|c}
\hline \hline
\multicolumn{3}{c}{$13$ TeV LHC [fb]}                                                                                                                                                                                        \\ \hline 
($M_{W_R}$,$m_N$) [TeV,GeV] & $\sigma^{\rm NLO+NNLL}$ $\times$ BR($W_R \rightarrow Ne$) &  $\times$ BR($N \rightarrow e^\pm q \overline{q'}$)  
\\ \hline
$(3,30)$                       & $14.8$                                                                & $14.8$                                                                                                           \\ \hline
$(3,150)$                      & $14.8$                                                                & $14.8$                                                                                                           \\ \hline
$(3,300)$                      & $14.6$                                                                & $14.1$                                                                                                           \\ \hline
$(4,400)$                      & $1.44$                                                                & $1.28$                                                                                                           \\ \hline
$(5,500)$                      & $0.184$                                                               & $0.152$                                                                                                          \\ \hline
\multicolumn{3}{c}{$100$ TeV VLHC [fb]}                                                                                                                                                                                      \\ \hline
($M_{W_R}$,$m_N$) [TeV,GeV] & $\sigma^{\rm NLO+NNLL}$ $\times$ BR($W_R \rightarrow Ne$) & $\times$ BR($N \rightarrow e^\pm q \overline{q'}$) \\ \hline
$(5,500)$                      & $345$                                                                 & $286$                                                                                                            \\ \hline
$(25,2500)$                    & $95.7 \times 10^{-3}$                                                 & $64.6 \times 10^{-3}$                                                                                            \\ \hline \hline
\end{tabular}
\caption{Cross section times branching ratio predictions for $pp\rightarrow \WR^\pm \rightarrow Ne^\pm$,
with subsequent decay of $N$ to leptons and quarks, for select $(\MWR,\mN)$.
}
\label{tb:wprimeXsec_decay}
\end{table}

From Eq.~(\ref{eq:nWidth}) the $N$ branching fractions are  independent of $\MWR$ and are given by
\begin{eqnarray}
\text{BR}(N\rightarrow e^\pm ~q ~\overline{q'}) &=&
\begin{cases}
  ~\qquad ~ \qquad ~\quad ~1, 				& m_N\leq m_t,\\
\cfrac{2}{3-8y_t+8y_t^3-y_t^4 -12y_t^2\log y_t},	& m_N>m_t,
\end{cases}
\\
\text{BR}(N\rightarrow e^\pm ~t ~b) &=& \cfrac{1-8y_t+8y_t^3-y_t^4 -12y_t^2\log y_t}{3-8y_t+8y_t^3-y_t^4 -12y_t^2\log y_t}, \quad m_N>m_t,
\end{eqnarray}
For $\MWR \gg \mN \gg m_t$, one finds asymptotically
\begin{equation}
 \text{BR}(N\rightarrow e^\pm ~q ~\overline{q'}) \approx 2\times \text{BR}(N\rightarrow e^\pm ~t ~b) \approx \frac{2}{3}.
\end{equation}
Consequently, the 13 and 100 TeV cross sections for the process
\begin{equation}
p ~p ~\rightarrow ~\WR ~\rightarrow N ~e ~\rightarrow ~e ~e ~q ~\overline{q'}
\end{equation}
in the NWA approximation can be given in terms of Eq.~(\ref{eq:sigmaBR}):
\begin{eqnarray}
 \sigma^{\rm NLO(+NNLL)}(pp\rightarrow \WR^\pm \rightarrow N e^\pm \rightarrow e^\pm e^\pm q \overline{q'} )
 &\approx& ~\sigma^{\rm NLO(+NNLL)}(pp\rightarrow \WR^\pm) 
 \nonumber\\
 &\times& ~\text{BR}(\WR \rightarrow N e)
 \nonumber\\
 &\times& ~\text{BR}(N \rightarrow e^\pm q\overline{q'})
 \label{eq:totppllqqXSec}
\end{eqnarray}
The total production rate for Eq.~(\ref{eq:totppllqqXSec}) for representative $(\MWR,\mN)$ 
are summarized in column 3 of Tb.~\ref{tb:wprimeXsec_decay} and for $m_N/\MWR = 0.1$ plotted in Fig.~\ref{fig:xsec} (dot).
We find that the total 13 (100) TeV rate spans approximately
\begin{equation}
 10^{-1} - 4\times10^4 ~(10^{-3} - 10^{5})\fb \quad\text{for}\quad \MWR = 1-5~(35)\TeV.
\end{equation}

\subsection{PDF and Scale Uncertainties}\label{sec:uncertainties}
\begin{figure}[!t]
\begin{center}
\subfigure[]{\includegraphics[width=.48\textwidth]{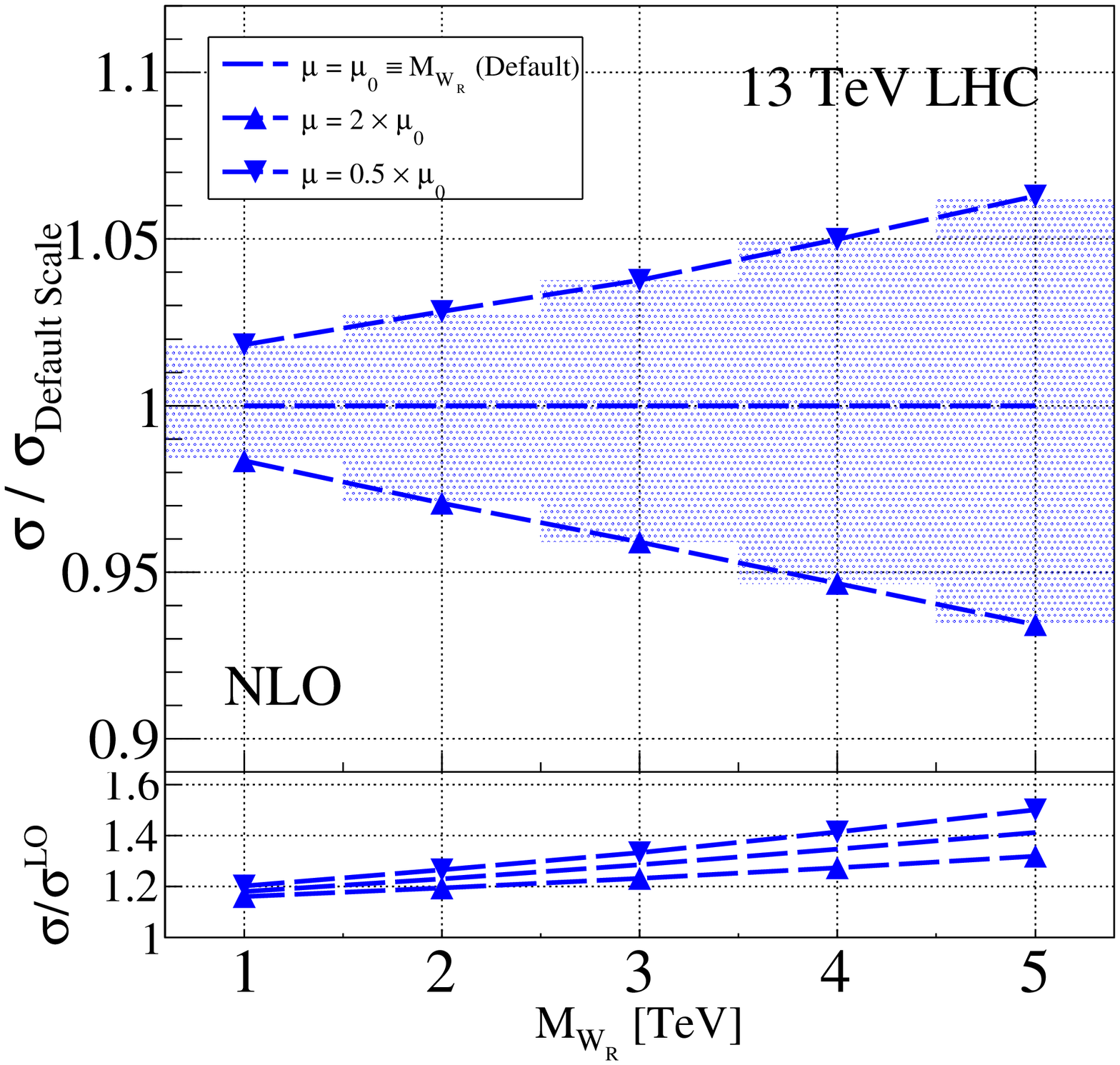}	\label{fig:ppWR_LHC13}}
\subfigure[]{\includegraphics[width=.48\textwidth]{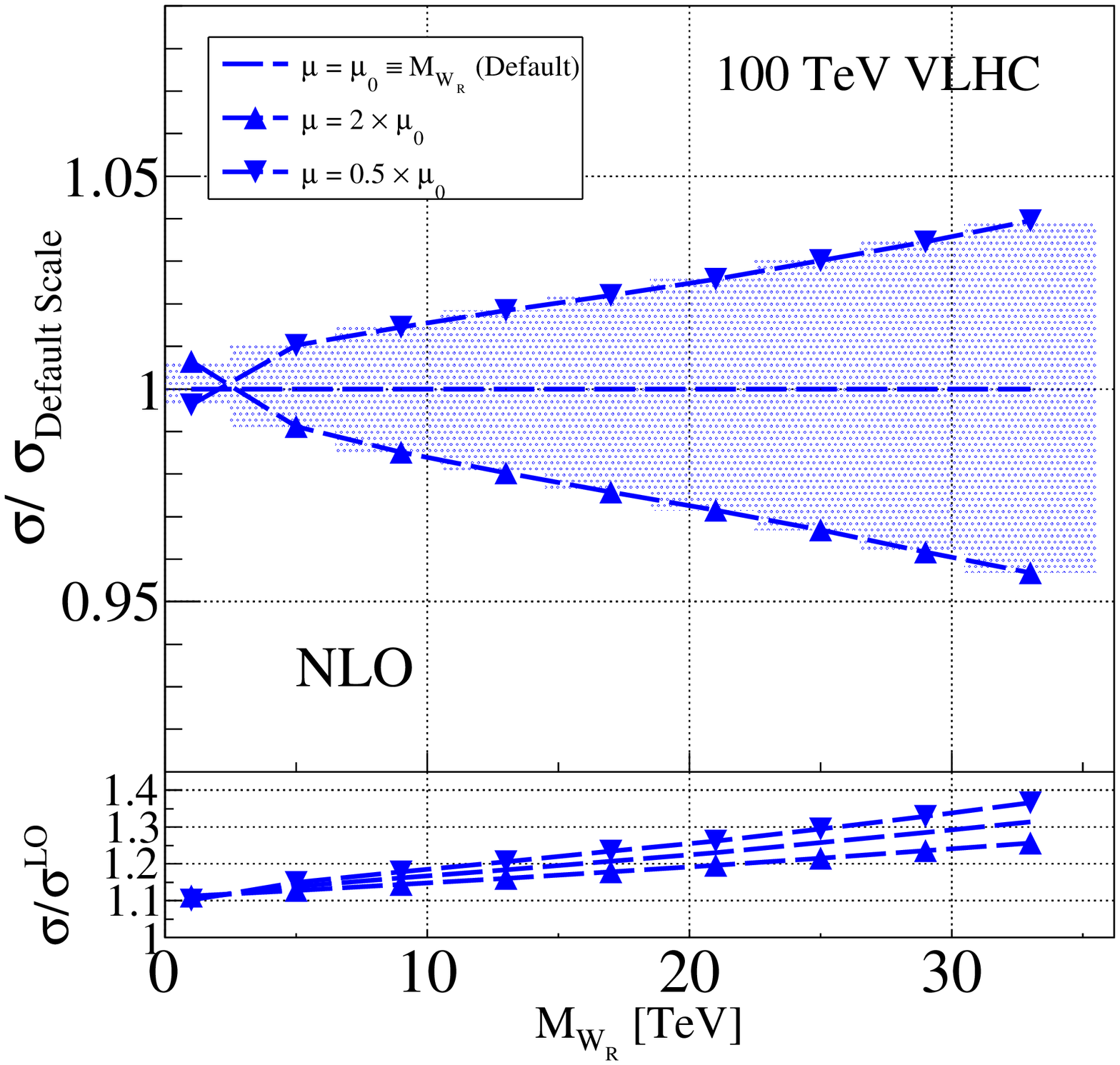}	\label{fig:ppWR_LHC100}}
\\
\subfigure[]{\includegraphics[width=.48\textwidth]{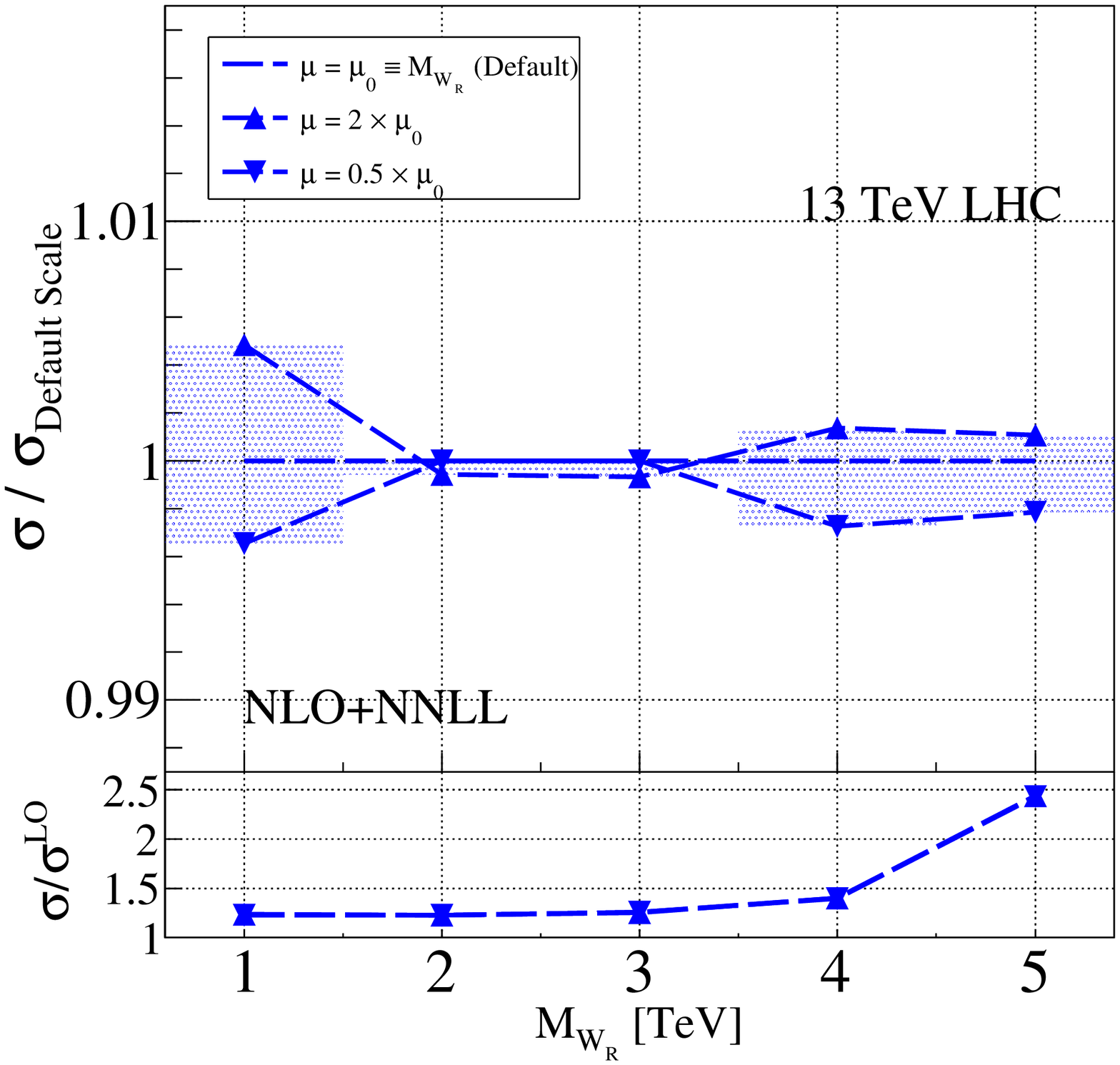}	\label{fig:ppWRres_LHC13}}
\subfigure[]{\includegraphics[width=.48\textwidth]{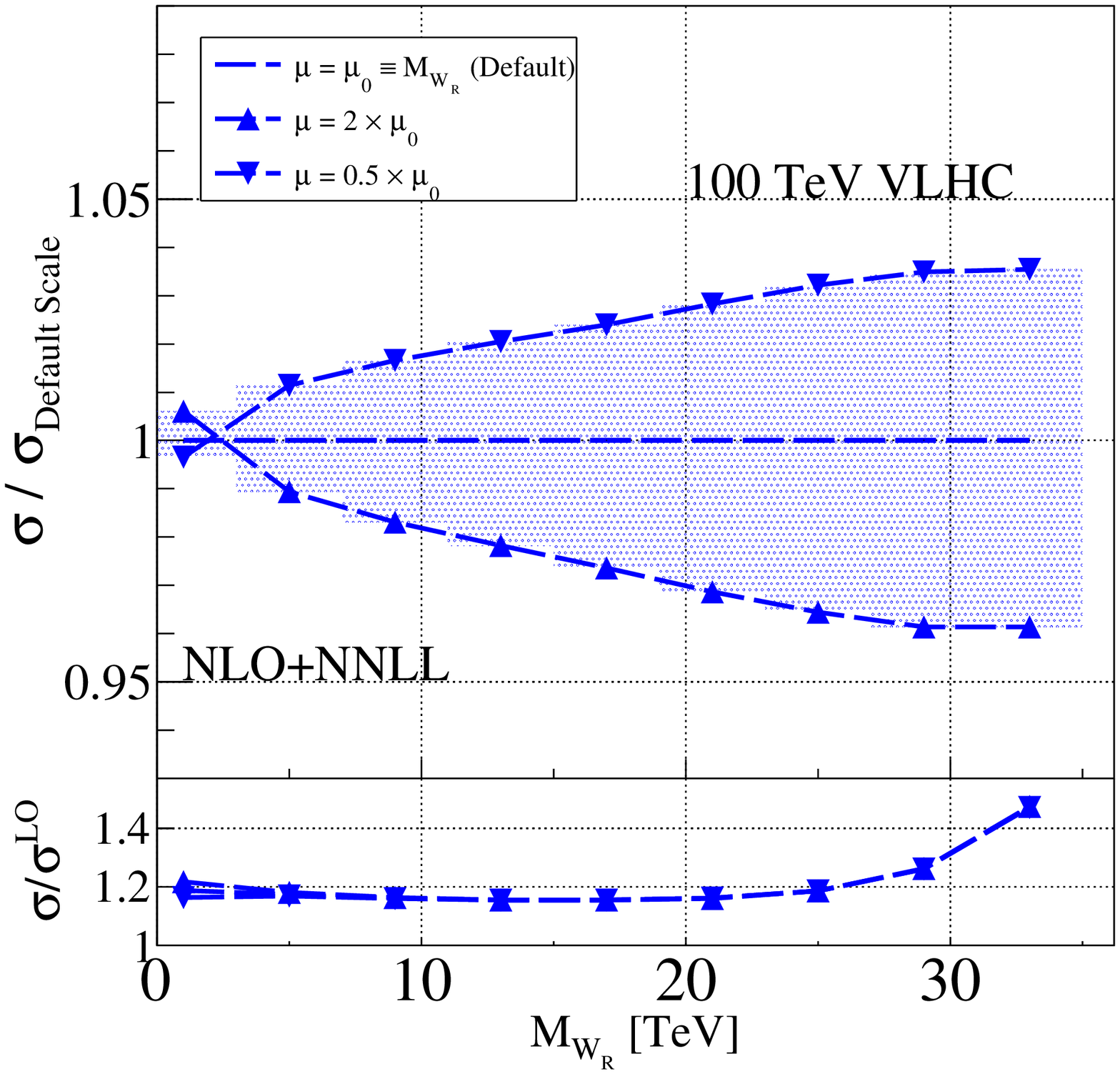}	\label{fig:ppWRres_LHC100}}
\end{center}
\caption{Scale dependence of the total $\WR$ cross section and associated $K$-factors at 
NLO for (a) $13$ and (b) 100 TeV, and NLO+NNLL for (c) $13$ and (d) 100 TeV.} 
\label{fig:ppWR_prod}
\end{figure}

To estimate the impact of higher order terms in the QCD perturbative series that are not calculated in the $\WR$ production cross section, 
we vary the factorization and renormalization scales about the default choice of $\mu_0=M_{W_R}$ up and down by a factor of two.
We present results normalized to the cross section at the default scale. 
In the lower panel of each plot is the $K$-factor as defined in Eq.~(\ref{eq:kFactors}). 

In Fig.~\ref{fig:ppWR_prod} we show the effect of scale variation on the NLO cross section at (a) $13$ and (b) 100 TeV for a range of $W_R$ masses. 
At NLO, it can be seen at both $13$ and $100$ TeV that increasing (decreasing) the default scale lowers (raises) the total cross section, 
except for very low $W_R$ masses at $100$ TeV, a feature common to high-mass DY processes~\cite{Ruiz:2015zca}. 
In addition, the $K$-factor also steadily increases with mass indicating the growing importance of higher order corrections in such scenarios. 
In both the $13$ and $100$ TeV cases, the scale variation results in a $2-5\%$ uncertainty to the total cross section.

The effect of scale variations on the NLO+NNLL result is presented in (c) $13$ and (d) 100 TeV for the same $\MWR$.
The effect of the resummation on the scale variation is manifest in the reduction of the associated uncertainty. 
For the $13$ TeV case, uncertainty is reduced to the sub per-cent level, while at $100$ TeV the impact is comparable (but smaller) than the NLO dependence. 
This is because resummed contributions are less important away from threshold.
Indeed, the observed reduction in scale uncertainty is consistent with what one expects from including higher order terms in the perturbative series.

We calculate the symmetric  PDF uncertainties from the NNPDF member sets following the recommended procedure of Ref.~\cite{Buckley:2014ana}.
The 68\% $(1\sigma)$ uncertainty bands are represented by the shaded regions in Fig.~\ref{fig:xsec}.
In the upper panel, only the NLO+NNLL uncertainty are shown; in the lower panel, both the NLO and NLO+NNLL uncertainties are shown.
At 13 TeV, for $\MWR = 4~(4.5)\TeV$, the NLO+NNLL uncertainty is approximately $\pm80~(240)\%$.
At 100 TeV, the uncertainties breach $100\%$ for $\MWR$ between $20$ and $30\TeV$.

The larger uncertainties in the threshold calculation compared to the NLO result is due in part to the less data used to constrain 
the threshold-improved PDFs~\cite{Bonvini:2015ira,Beenakker:2015rna}.
This follows from the limited threshold calculations available for processes that the enter into global fit PDFs,
and demonstrates their need for accurate LHC predictions. 

For representative $\MWR$, scale and PDF uncertainties are given in Tb.~\ref{tb:wprimeXSec}.

\begin{figure}[!t]
\begin{center}
\subfigure[]{\includegraphics[scale=1,width=.48\textwidth]{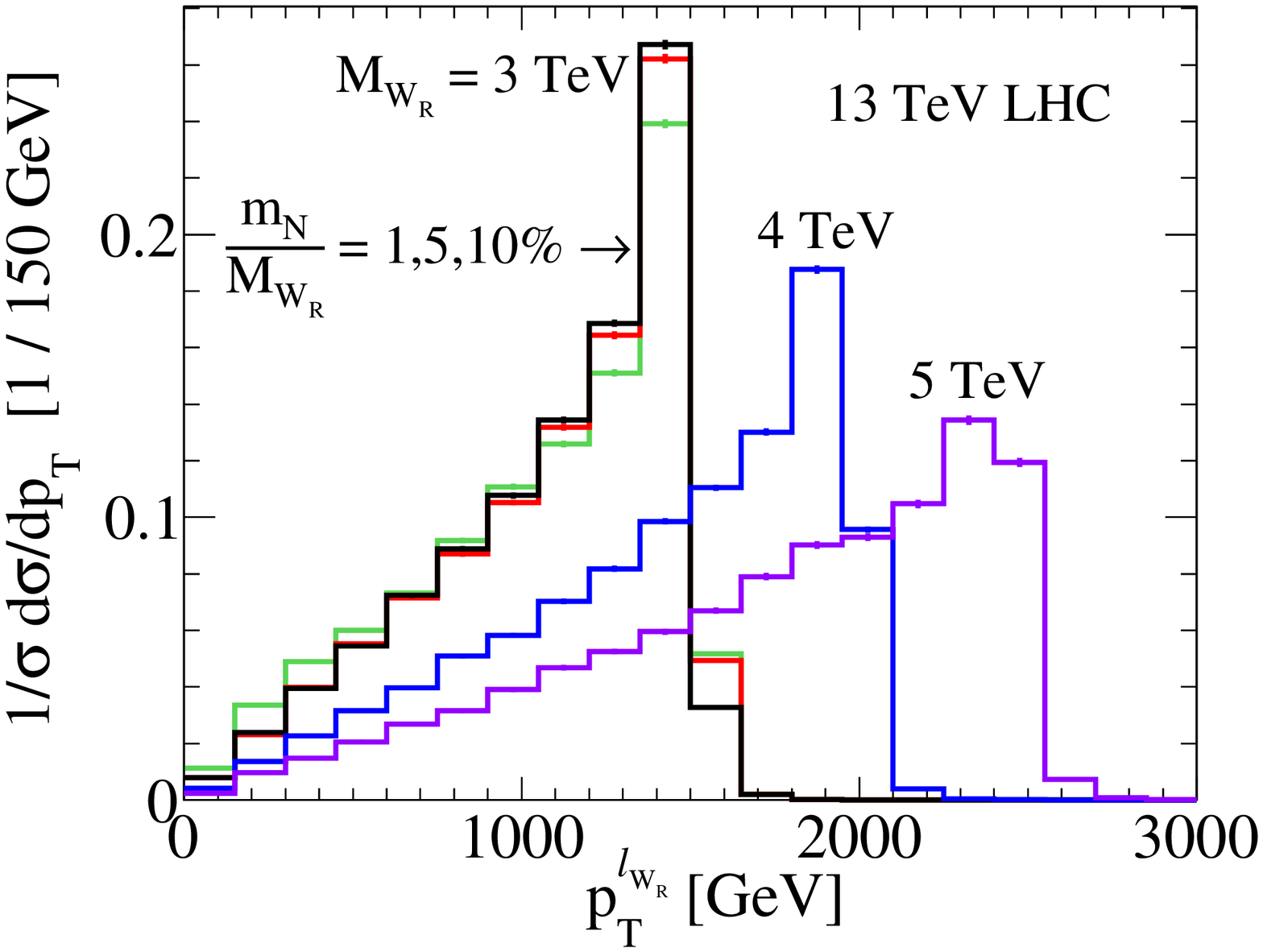}	\label{fig:lWR_pT}}
\subfigure[]{\includegraphics[scale=1,width=.48\textwidth]{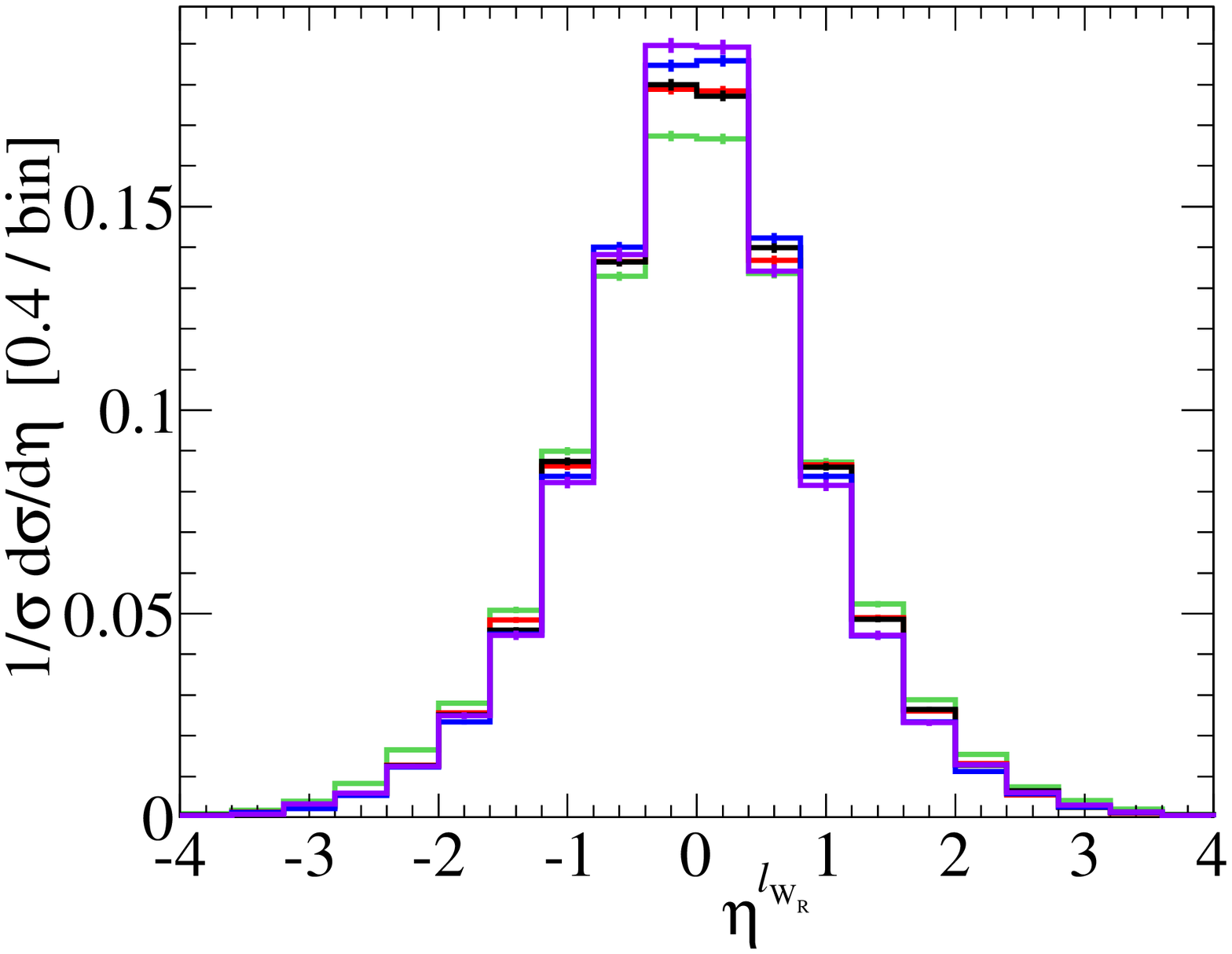}	\label{fig:lWR_eta}}
\end{center}
\caption{Normalized (a) transverse momentum $(p_T)$ and (b) pseudorapidity $(\eta)$ distributions of the charged lepton 
from $pp\rightarrow\WR\rightarrow N\ell$ for representative $\MWR$ and $\mN$ at 13 TeV.} 
\label{fig:lWR_kinematics}
\end{figure}

\section{Observability of Boosted $N$ at Hadron Colliders}\label{sec:observability}
In this section we study the observability at hadron colliders of $\WR$ and $N$ in the LRSM for $\mN/\MWR \lesssim 0.1$.
We start with production- and decay-level kinematics of $N$ at LO. 
After constructing several observables with strong background-discriminating power, 
we perform a full parton shower (PS)/detector-level signal-to-background analysis.

For signal event generation, we modify the Manifest LRSM FeynRules (FR) model file v1.1.6\_mix by Ref.~\cite{Roitgrund:2014zka}
(see App.~\ref{app:UFO}) and use FR v2.3.10~\cite{Alloul:2013bka,Christensen:2008py} to generate Universal File Object (UFO) inputs~\cite{Degrande:2011ua}.
LO events are simulated using MG5~\cite{Alwall:2014hca}.
Rates are scaled by the NLO+NNLL $K$-factors as defined in Eq.~(\ref{eq:kFactors}).
Application of $K$-factors is justified in the threshold regime as the dominant contribution, 
i.e., soft-radiation, largely leave kinematics unchanged.
Events are showered using PYTHIA 8.212~\cite{Sjostrand:2014zea} and jets are clustered with FastJet v3.20~\cite{Cacciari:2005hq,Cacciari:2011ma} 
using the Cambridge/Aachen (C/A) algorithm~\cite{Dokshitzer:1997in,Wobisch:1998wt} with a separation parameter of $R=1.0$.
SM background processes are simulated at LO+PS accuracy using the MG5,
and scaled by an appropriate NLO $K$-factor calculated via the MG5\_aMC@NLO framework.
Due to extreme phase space cuts, event generation at NLO+PS accuracy is impractical.

\subsection{Kinematic Properties of Boosted $N$}\label{sec:pheno}
To investigate the kinematics of boosted $N$ from $\WR$ decays, we simulate at 13 TeV 
\begin{equation}
 q_1 ~\overline{q_2} ~\rightarrow ~\WR ~\rightarrow ~e_1 ~N ~\rightarrow ~e_1 ~e_2 ~q'_1 ~\overline{q'_2},
 \label{eq:processllqq}
\end{equation}
where the two electrons possess any electric charge combination, for the representative $(\MWR,\mN)$ listed in Tb.~\ref{tb:widths}.
We focus on final-state electrons, which is the most problematic channel for ATLAS and CMS~\cite{Aad:2015xaa,Khachatryan:2014dka},
but our study is also applicable to the $e\mu$ and $\mu\mu$ final states.
{The largest change in those channels follows from the better muon identification compared to the electron~\cite{CMS:2015kjy};
this in fact extends the validity of standard dilepton searches.
}
Inclusion of the $N\rightarrow \ell t b$ final state is similarly straightforward.
To model detector response while keeping generator-level particle identification at LO, 
we smear final-state partons as done in~\cite{Alva:2014gxa},
which adopts the expected ATLAS detector performance parametrization~\cite{Aad:2009wy}.
Eq.~(\ref{eq:processllqq}) is free of kinematic poles and no generator-level cuts are applied.

In Fig.~\ref{fig:lWR_kinematics} we show the normalized differential distributions with respect to the
(a) transverse momentum $(p_T)$ and (b) pseudorapidity $(\eta)$ of the charged lepton in the $\WR^\pm\rightarrow Ne^\pm$ decay, denoted by $\ell_{\WR}$.
In the $p_T^{\ell_{\WR}}$ distribution, the Jacobian peak near $p_T\sim\MWR/2$ is unambiguous and is largely independent of such small $\mN$.
The $\eta^{\ell_{\WR}}$ distribution reveals that $\ell_{\WR}$ are very central,
with most electrons contained within $\vert\eta\vert<\confirm{1.0}$ and negligibly few with $\vert\eta\vert\geq\confirm{2.0}$.
Multi-TeV bounds on $\MWR$ (see Sec.~\ref{sec:theory}) nearly guarantee that $p_T^{\ell_{\WR}}$ is very large and ~$\vert\eta^\ell\vert$ small.
Consequently, Eq.~(\ref{eq:processllqq}) efficiently passes inclusive high-$p_T$ single-electron triggers, 
such as those used in Ref.~\cite{CMS:2015kjy}.

\begin{figure}[!t]
\begin{center}
\subfigure[]{\includegraphics[scale=1,width=.48\textwidth]{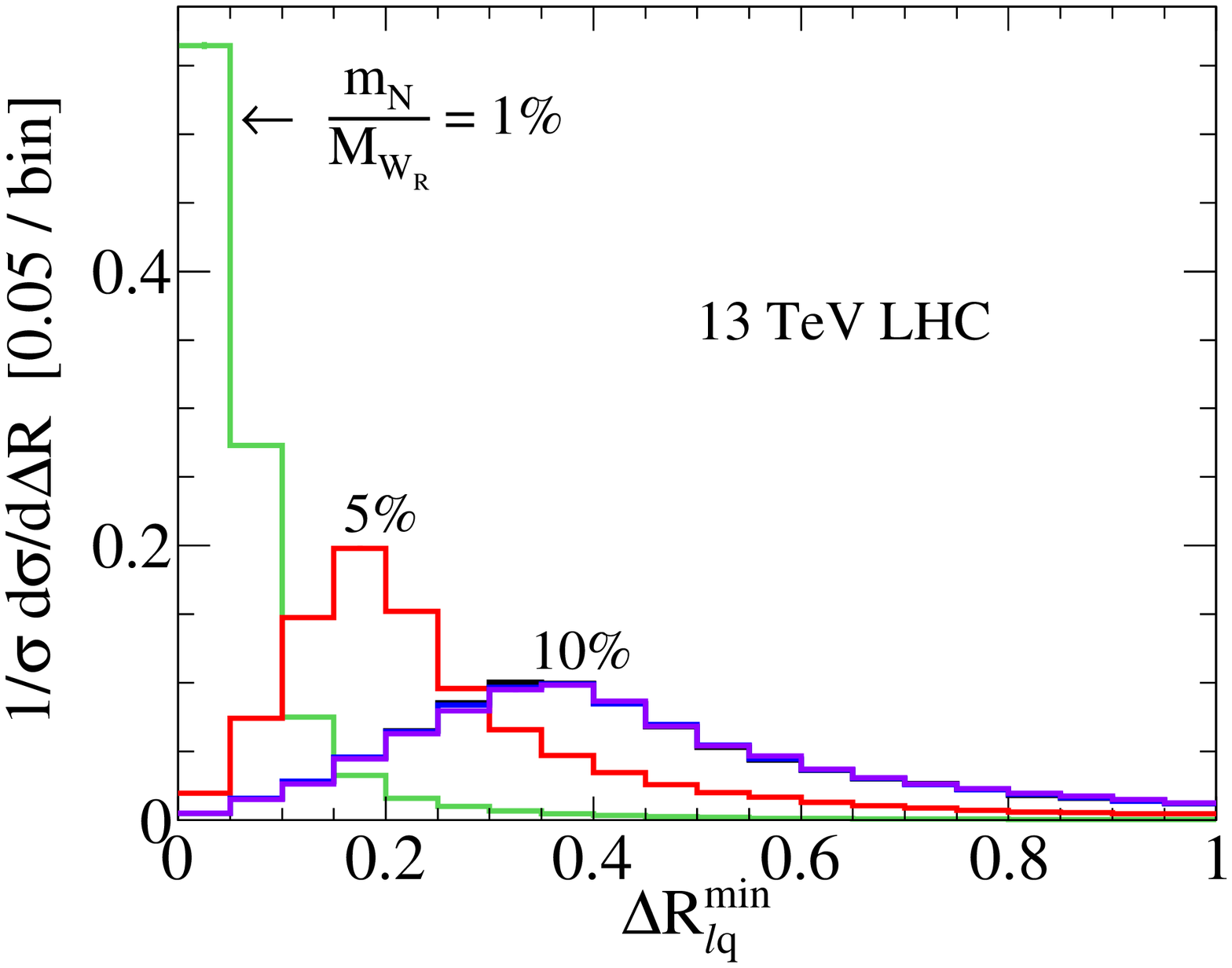}	\label{fig:n_dRlj}}
\subfigure[]{\includegraphics[scale=1,width=.48\textwidth]{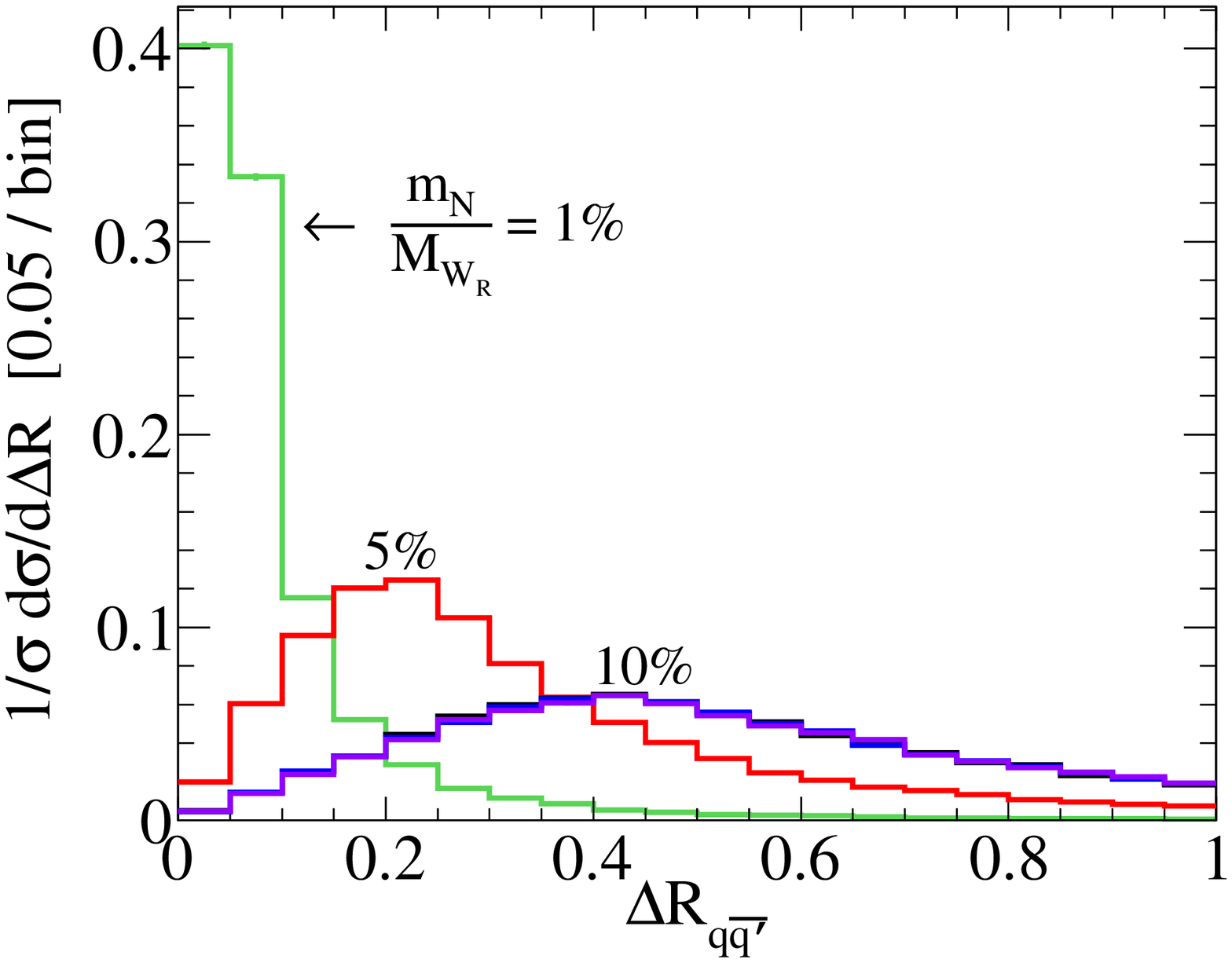}	\label{fig:n_dRjj}}
\end{center}
\caption{
Normalized distributions with respect to (a) $\Delta R_{\ell q}^{\min}$ and (b) $\Delta R_{q\overline{q'}}$ of $N$'s decay products
for the same configuration as Fig.~\ref{fig:lWR_kinematics}.}
\label{fig:n_kinematics}
\end{figure}

As $pp\rightarrow Ne^\pm$ is a $2\rightarrow2$ system, 
the heavy neutrino's $p_T$ and rapidity $(y)$ distributions are identical to Fig.~\ref{fig:lWR_kinematics}, up to mass corrections.
Hence, the decay products of the $N$ with high-$p_T$  are largely collimated due to its relative lightness.
For the $N\rightarrow \ell_N q\overline{q'}$ final state in Eq.~(\ref{eq:processllqq}),
we show in Fig.~\ref{fig:n_kinematics} the normalized 
separation\footnote{The separation between particle pair $(a,b)$ is defined as $\Delta R_{ab} \equiv \sqrt{(y_a - y_b)^2 + (\phi_a - \phi_b)^2 }$
for rapidity $y$ (or pseudorapidity $\eta$) and azimuthal angle $\phi$.}
distributions between 
(a) the charged lepton $\ell_N$ and its closest quark~ $(\Delta R_{q\ell_N}^{\min})$, as well as 
(b) the two quarks themselves ~$(\Delta R_{q \overline{q'}})$.
In both cases, the separation peaks at $\Delta R \sim 0.2~(0.4)$ for $\sqrt{r_N} =\mN/\MWR=\confirm{0.05~(0.1)}$,
and follows from the scaling relationship
\begin{equation}
 \Delta R_{qX} ~\sim~ 2p_{\perp}^{X}/p_T^N ~\sim~ 4\mN/\MWR,
\end{equation}
where $p_{\perp}^{X}$ is the perpendicular momentum of $X=\ell_N,\overline{q'}$ relative to its parent $N$.
Hence, for much of the phase space, these electrons fail particle identification criteria at 13 TeV~\cite{CMS:2015kjy}:
\begin{equation}
 p_T^\ell > 35\GeV, ~\quad~ \Delta R_{\ell X} > 0.3, ~\quad~ \vert\eta^\ell\vert < 2.4,
 \label{eq:eleCand}
\end{equation}
and leads to the breakdown of current ATLAS and CMS $\WR-N$ search strategies~\cite{Aad:2015xaa}.
Smaller $r_N=\mN^2/\MWR^2$, hadronization, and the presence of a $tb$ pairs exacerbate this issue.

\begin{figure}[!t]
\begin{center}
\subfigure[]{\includegraphics[scale=1,width=.48\textwidth]{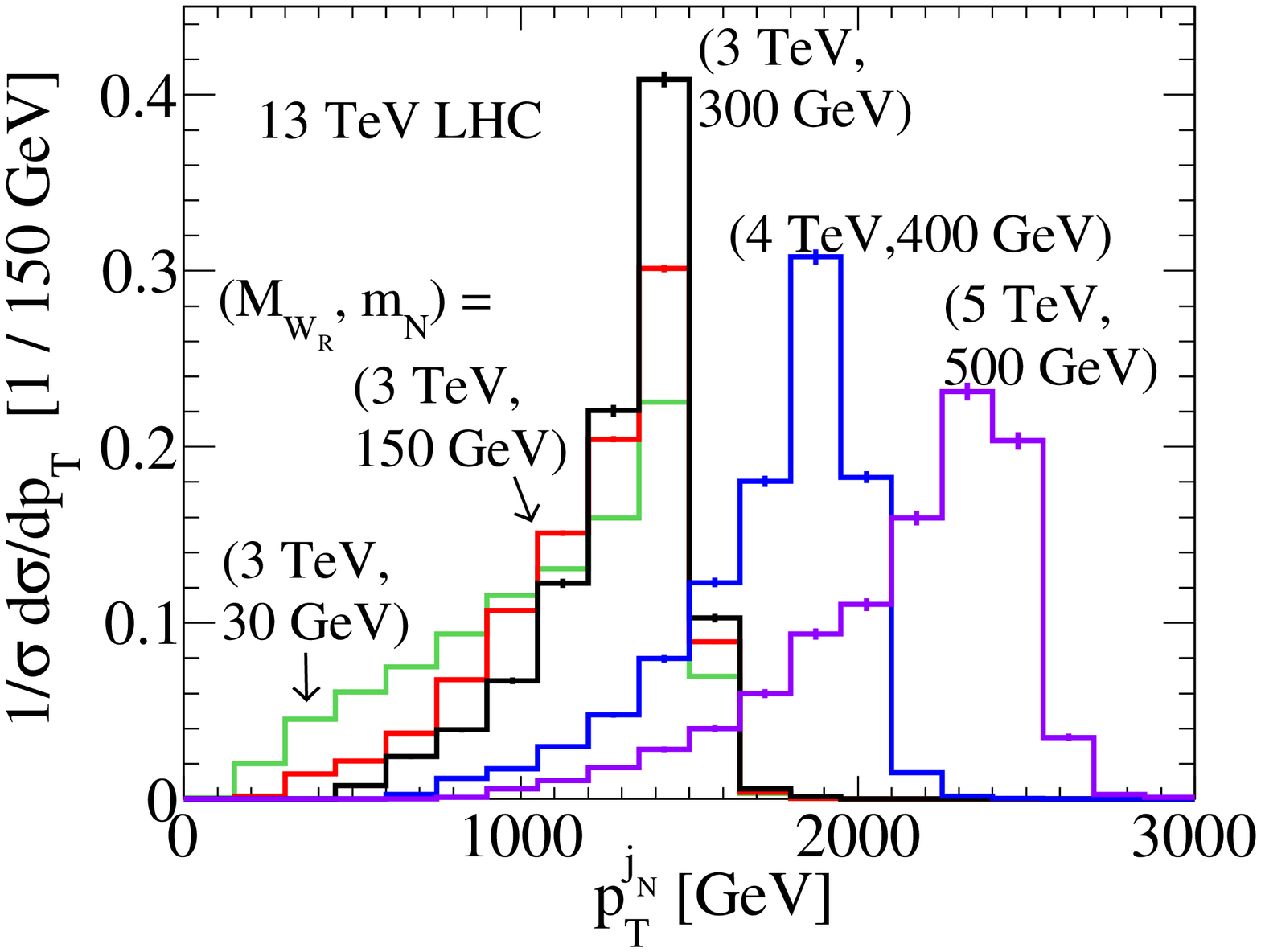}	\label{fig:jF_pT}}
\subfigure[]{\includegraphics[scale=1,width=.48\textwidth]{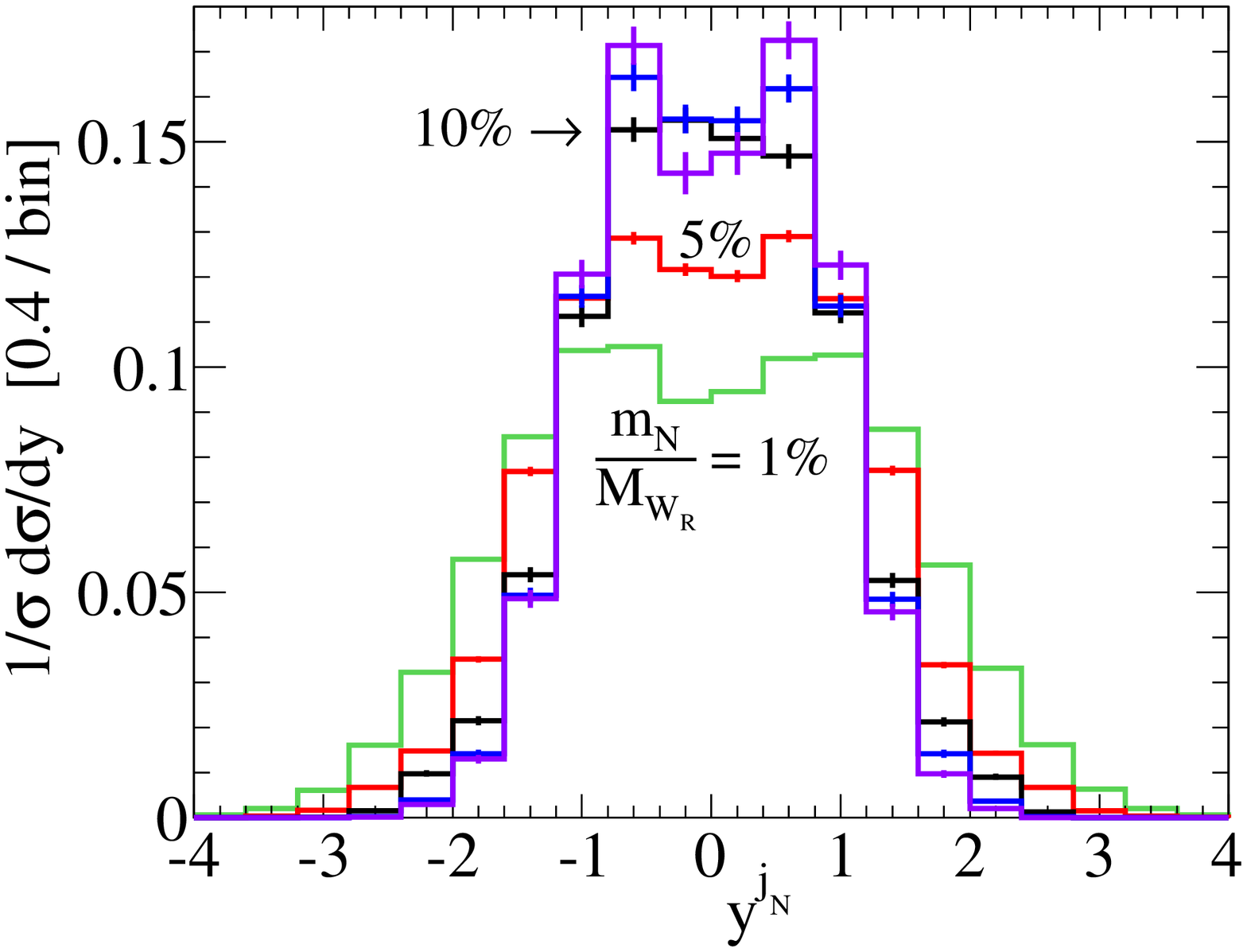}	\label{fig:jF_y}}
\\
\subfigure[]{\includegraphics[scale=1,width=.48\textwidth]{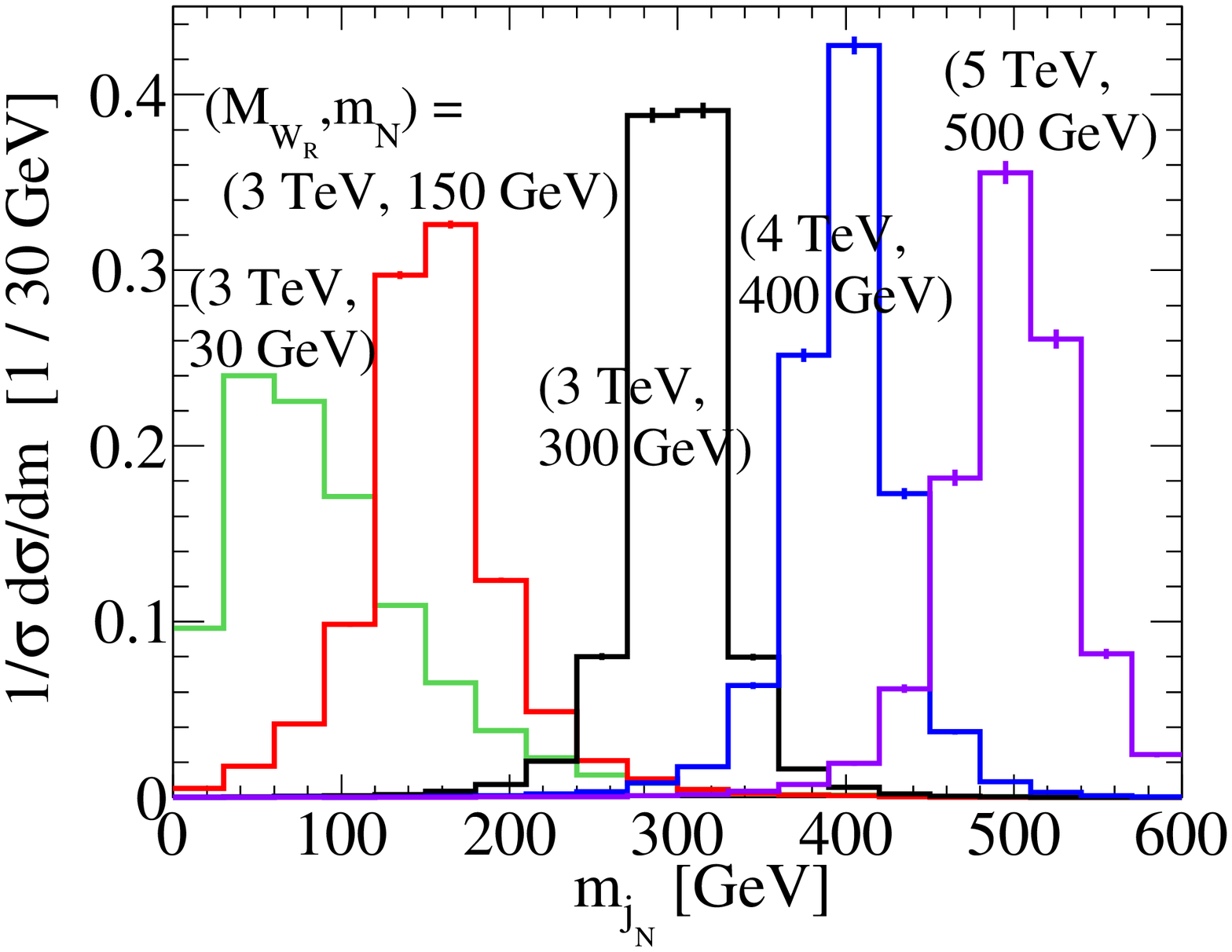}	\label{fig:jF_m}}
\subfigure[]{\includegraphics[scale=1,width=.48\textwidth]{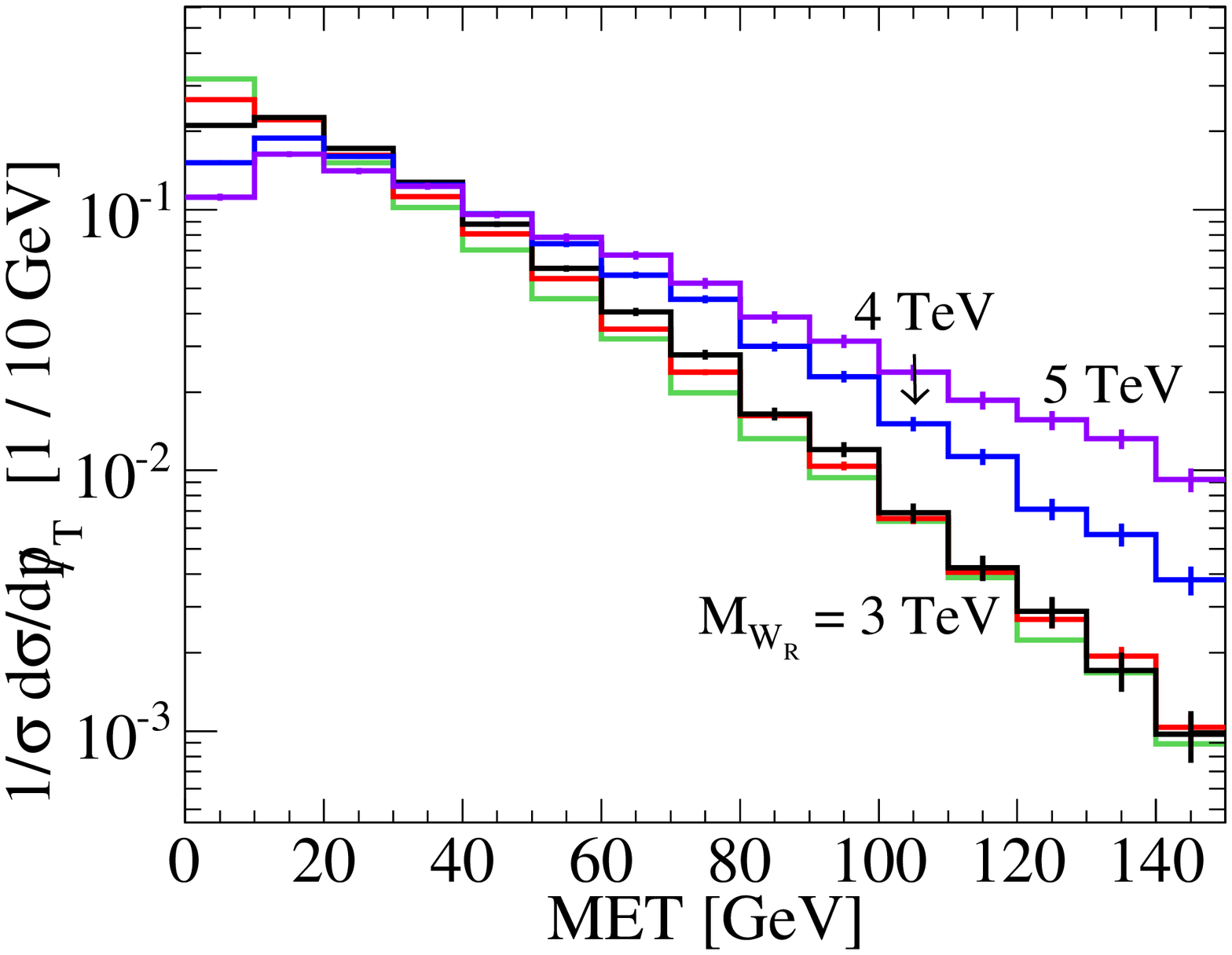}	\label{fig:lrsm_MET_multiMWR}}
\end{center}
\caption{Normalized distributions with respect to (a) $p_T^{\jN}$, (b) $y^{\jN}$, (c) $m_{\jN}$, and (d) MET
for same configuration as Fig.~\ref{fig:lWR_kinematics} but requiring exactly one electron and $\jN$ candidate.}
\label{fig:jN_kinematics}
\end{figure}

For such signal regions, we consider an alternate search strategy:
model the $N$ decay products as a single object, which we call a \textit{neutrino jet} $(\jN)$, and investigate instead the $2\rightarrow2$ process:
\begin{equation}
 p ~p ~\rightarrow ~\WR ~\rightarrow ~e^\pm ~\jN.
 \label{eq:processlj}
\end{equation}
The simplified signal topology alleviates the failing identification criteria
and retains the high signal-to-noise properties of the same-sign dilepton channel.
To build a qualitative picture of the new signal definition, 
we preliminarily define $\jN$ at the present FO parton-level via C/A clustering with $\Delta R=1$.
We cluster all final-state partons except any electron candidate satisfying Eq.~(\ref{eq:eleCand}).
$j_N$ is identified as the highest $p_T$ C/A jet.

In Fig.~\ref{fig:jN_kinematics}, we show the normalized distributions for $\jN$ with respect to
(a) $p_T$, (b) $y$, (c) invariant mass $(m_{\jN})$, and (d) missing transverse momentum (MET) 
for events with exactly one electron and $\jN$ candidate.
As anticipated, we observe strong similarities to the $\ell_{\WR}$ distributions and unambiguous 
Breit-Wigner resonances at the appropriate values in the $m_{\jN}$ distribution.
This indicates that $\jN$ is a good description of $N$ and 
that the signal definition of Eq.~(\ref{eq:processlj}) can be interpreted as Eq.~(\ref{eq:processllqq}) when $(\mN/\MWR) < 0.1$.

A cost of this new signal definition is the loss of the unambiguous \textit{smoking-gun} collider signature of two same-sign leptons 
and jets~\cite{Keung:1983uu}, which is intrinsically background-free up to detector effects as it violates $L$ by two units.
However, inherited from the original definition is the fact that, up to detector and hadronization effects, 
the process has no MET as no light neutrinos exist in the final state.
Requiring again exactly one electron candidate, we show in (d) the normalized MET distribution.
Due to smearing, we find moderate MET out to $10s$ of GeV and largely independent of $m_N$.
We observe that the peak MET shifts to larger values for larger $\MWR$ and is due
to the increased likelihood of more energy being mis-reconstructed for more energetic objects~\cite{Aad:2009wy}.
Present ATLAS detector capabilities~\cite{Khachatryan:2016olu} permit MET cuts as tight as
\begin{equation}
 \text{MET} < 35\GeV.
 \label{eq:metCut}
\end{equation}
In a realistic scenario (see Sec.~\ref{sec:analysis}), a more conservative cut is required due to pile up, etc.

\begin{figure}[!t]
\begin{center}
\subfigure[]{\includegraphics[scale=1,width=.48\textwidth]{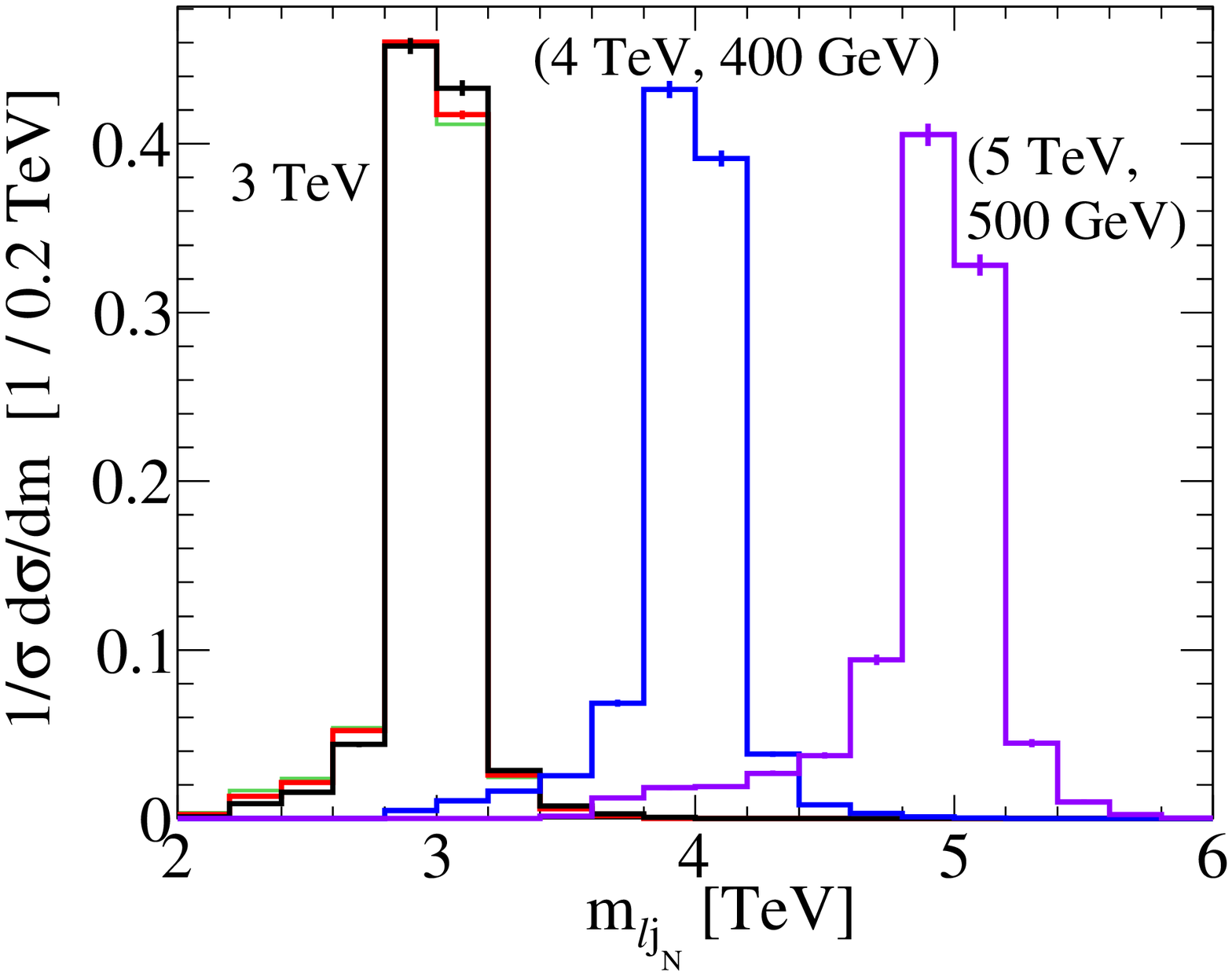}	\label{fig:lrsm_mljN_multiMWR}}
\subfigure[]{\includegraphics[scale=1,width=.48\textwidth]{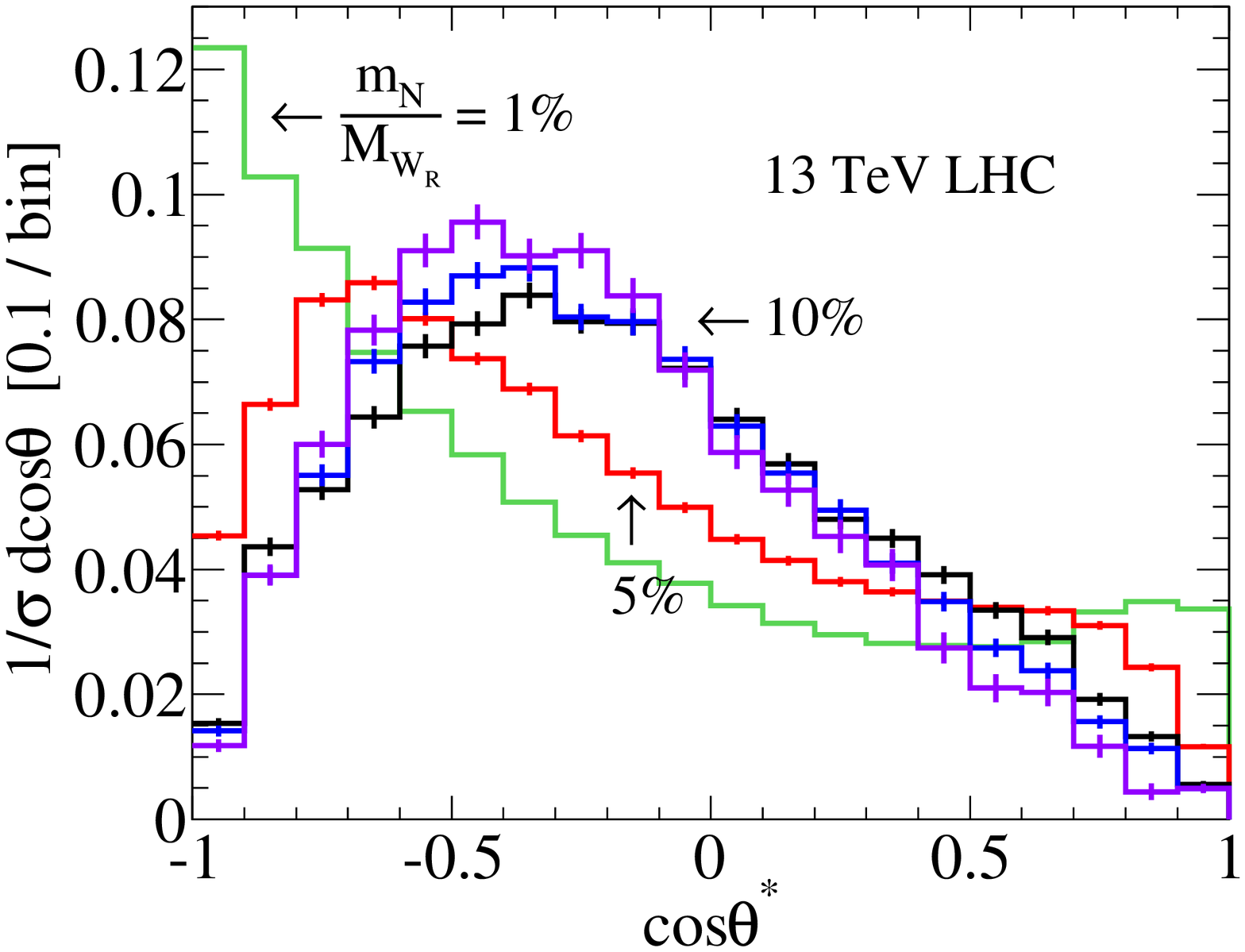}   \label{fig:lrsm_cosThStar_multiMWR}}
\end{center}
\caption{(a) Invariant mass $m_{\ell_{1}\jN}$ and (b) polar distribution of $\ell_1$ in the $(\ell_{1}-\jN)$ system's rest frame 
for the same configuration as Fig.~\ref{fig:jN_kinematics}.} 
\label{fig:reco_kinematics}
\end{figure}

In Fig.~\ref{fig:reco_kinematics}(a) the $\WR$ resonances built from the  $\ell_{1}-\jN$ invariant mass are clearly seen
for our representative masses, up to broadening due to mis-reconstruction of $N$ and detector smearing.
In (b), we show the polarization of $\ell_1$ in the $\ell_{1}-\jN$ system's rest frame.
We observe clearly the RH chiral structure of the $N\ell\WR$ vertex for $\sqrt{r_N} = m_N/\MWR = 0.01$.
At larger $r_N$, however, this becomes obfuscated due to the importance of opposite helicity states, 
which scale like $r_N$, and lead to spin-decorrelation. 

Altogether, this  demonstrates the viability of the new search procedure.

Aside from the application of micro-jets and substructure techniques,
it may be possible to verify the Majorana nature of heavier $N$ via its decays to top quarks.
For Dirac $N$, the off-shell $\WR^{*}$ to which it decays can only
carry the same electric charge as the charged lepton produced from the decay of the primary, on-shell $\WR$, i.e., $\ell_{\WR}$.
Decays of the $\WR^*$ to a top quark that subsequently decays leptonically can lead to 
final-state muons with the \textit{same} sign electric charge as $\ell_{\WR}$.
That is, for a fixed primary $\WR$ electric charge, one has
\begin{eqnarray}
 q ~\overline{q'} \rightarrow \WR^\pm \rightarrow &\ell_{\WR}^\pm ~N,
\quad\text{with}\quad
N\rightarrow \ell_N^\mp ~(t\to W_L^\pm b)  ~b \rightarrow \ell_N^\mp  ~b ~\overline{b} ~\mu^\pm ~\nu_\mu.
\end{eqnarray}
Hence, $\jN$ containing top quarks can be identified by their larger complexity, namely the presence of two $b$-subjets.
As the outgoing muon momentum scales like $p_T^\mu \sim \gamma_t m_t (1 + M_W^2/m_t^2)/4 \sim \gamma_t 50\GeV$, 
where $\gamma_t \sim m_t / p_T^N \sim m_t/\MWR$ is the top quark's Lorentz boost to the lab frame, it should be identifiable.
For a Majorana $N$, the off-shell $\WR^{*}$ can carry either electric charge.
Thus, observation of such muons with \textit{opposite} electric charge of the easily identifiable $\ell_{\WR}$ is evidence of $L$-violating transitions.
Further discussion of this topic is beyond the scope of this study.

We briefly note that the use of neutrino jets is also widely applicable to other situations:
In the MLRSM, high-mass $Z_R$ and $H_{FCNH}$ decays to boosted $NN$ pairs could give rise to two back-to-back $\jN$.
{If $N$ couples non-negligibly to EW bosons, then $j_N$ may also feature substructure topologies.}
In other models, such as the Inverse Seesaw, rare decays of $W/Z/h$ bosons to GeV-scale pseudo-Dirac neutrinos, as well as other processes, 
could also result in $\jN$.

\subsubsection{Estimation of Leading Standard Model Backgrounds}
\begin{table}[!t]
\begin{center}
 \begin{tabular}{ c | c | c | c | c | c | c | c}
\hline \hline
Cut $\backslash \quad \sigma^{\rm LO}$ [ab] & 
$Wj$ & $WZ$ & $t\overline{t}$ & $t\overline{t}j$ & $tbj$  & $eej$ & $WWj$ \tabularnewline\hline
\hline
$p_T^{j,b}>30\GeV, \vert\eta^{j,b}\vert<4.5$	& \multirow{2}{*}{ } & \multirow{2}{*}{ } & \multirow{2}{*}{ } 
				& \multirow{2}{*}{ } & \multirow{2}{*}{ } & \multirow{2}{*}{ } & \multirow{2}{*}{ } \tabularnewline
+$\Delta R_{jb}>0.4, \Delta R_{\ell X}>0.3$	& 2.17	& 11.0	& 63.8	& 44.0	& 4.18 		& 344 	&  327		\tabularnewline
No~Decay		& $\times10^9$	& $\times10^6$	&  $\times10^6$	&  $\times10^6$	& $\times10^6$	& $\times10^6$	& $\times10^3$ \tabularnewline\hline
+Decay+	& \multirow{3}{*}{ } & \multirow{3}{*}{ } & \multirow{3}{*}{ } & \multirow{3}{*}{ } 
				& \multirow{3}{*}{ } & \multirow{3}{*}{ } & \multirow{3}{*}{ } \tabularnewline
$p_T^{\ell~\max}>1\TeV$+	& 218	& 2.61	& 0.201	&  0.660	& 0.062 & 184	& 0.637	\tabularnewline
$\not\!\! E_T<50\GeV$	& 	& 	&  		&  	& 	&	&  	\tabularnewline\hline
+Smearing+$\vert\eta^\ell\vert<2.0$ & \multirow{2}{*}{ } & \multirow{2}{*}{ } & \multirow{2}{*}{ } 
				& \multirow{2}{*}{ } & \multirow{2}{*}{ } & \multirow{2}{*}{ } & \multirow{2}{*}{ } \tabularnewline				
+ $p_T^\ell>35\GeV$ +		& 218	& $-$	& $-$ 		& $-$  	& $-$ 	& 57	& $-$	\tabularnewline
+ 2nd $e^\pm$ veto 		& 	& 	&  		&  	& 	&	&  \tabularnewline\hline
$\not\!\! E_T<35\GeV$ 	& 85	& $-$ 	& $-$ 		& $-$ 	& $-$ 	& 25	& $-$  \tabularnewline\hline
$K^{\rm NLO} = 1.3$ 	& 111	& $-$ 	& $-$ 		& $-$ 	& $-$ 	& 33	& $-$  \tabularnewline\hline
\hline
\end{tabular}
\caption{Cross sections [ab] of SM background for $pp\rightarrow e^\pm \jN$ after decays and successive cuts.}
\label{tb:smBkgXSec}
\end{center}
\end{table}

Before simulating our full detector-level analysis, we are in position to estimate the leading SM backgrounds.
The simple lepton+jet topology of Eq.~(\ref{eq:processlj}) suffers from large SM backgrounds.
We sort the leading channels into three categories: 
(a) weak bosons, 
(b) top quarks, and 
(c) fake rates from electron misidentification:
\begin{eqnarray}
  \text{Weak~boson}	&~\quad:~\quad& W^\pm j~(\rightarrow e^\pm jX), ~\quad W^\pm Z (\rightarrow e^\pm jX) 
  \\
  \text{Top~quark}	&~\quad:~\quad& t\overline{t}+nj~\text{(semi-leptonic)}, ~\quad tbj (\rightarrow \ell^\pm +nb + mjX)
  \\
  \text{Fake~rates}	&~\quad:~\quad& e^+e^-j, ~\quad W^+W^-j (\rightarrow e^+ e^- jX)
\end{eqnarray}
Fake events correspond to regions of phase space where one electron candidate is identified according to Eq.~(\ref{eq:eleCand})
but additional electrons fail to pass the criteria. 

At the generator level and assuming the following (nominal) regulating cuts
\begin{equation}
 p_T^{j,b} > 30\GeV, \quad \Delta R_{jb} > 0.4, \quad \Delta R_{\ell X} > 0.3, \quad \vert \eta^{j,b}\vert < 4.5,
\end{equation}
the DY+$1j$ channels at LO, i.e., $Wj$ and $eej$, are found to dominate with cross sections reaching $\sigma^{\rm SM} \sim 0.3-2$ nb;
see row 1 of Tb.~\ref{tb:smBkgXSec}. 
The signal/noise ratio roughly translates to $S/N \sim 10^{-6} - 10^{-5}$. 
Background rates are dramatically reduced after decaying the top quark and EW bosons, 
and requiring that the $p_T$ of the leading charged lepton and process MET satisfy at the generator level
\begin{equation}
\label{eq:met}
 p_T^{\ell_1~\text{Generator-level}} > 1\TeV \quad\text{and}\quad \text{MET}^{\rm Generator-level} < 50\GeV.
\end{equation}
The $Wj$ and $eej$ channels remain dominant but now only reach $\sigma^{\rm SM} \sim 200$ ab; see row 2 of Tb.~\ref{tb:smBkgXSec}.
The top background is particularly neutralized owing to the cascade nature of their decays, 
which require TeV-scale charged leptons to be accompanied by TeV-scale light neutrinos from a multi-TeV top quark parent.
Subsequently, the top quark and diboson backgrounds can be neglected.

Requiring exactly one charged lepton to satisfy the electron identification of Eq.~(\ref{eq:eleCand}) 
and rejecting events with additional electrons leaves the $Wj$ rate largely unchanged but reduces the neutral current DY background to the 
$\sigma^{\rm SM}\sim 60$ ab level; see row 3 of Tb.~\ref{tb:smBkgXSec}. 
Imposing the MET requirement of Eq.~(\ref{eq:metCut}) after smearing indicates that the remaining SM background sums to a total of 
$\sigma^{\rm SM}\sim 110$ ab; see row 4 of Tb.~\ref{tb:smBkgXSec}.
{
We calculate an NLO $K$-factor of $K^{\rm NLO} = 1.30$ for the $Wj$ channel;
the same  $K$-factor is applicable to the $eej$ channel due to color symmetry.
This increases the total SM background to $\sigma^{\rm SM}\sim 140-150$ ab; see row 5 of Tb.~\ref{tb:smBkgXSec}.
After incorporating a loose $m_{\ell j_{\rm Fat}}$ cut around $\MWR$ and appropriate signal $K$-factor,}
the signal/noise ratio exceeds $S/N \gtrsim 10 - 100$.

\subsection{Detector-Level Signal Analysis and Neutrino Jet Definition}\label{sec:analysis}
 \begin{table}[!t]
\begin{center}
 \begin{tabular}{ c || c | c | c | c | c }
\hline \hline
\multicolumn{6}{c}{$\sigma(pp\rightarrow \WR^\pm \to \ell^\pm N \to \ell^\pm j_N)$ [fb]}	\tabularnewline\hline
					  \multicolumn{6}{c}{13 TeV LHC}			\tabularnewline\hline
	&	\multicolumn{5}{c}{$(\MWR,m_N)$ [TeV, GeV]}					\tabularnewline\hline
Cut	& $(3,30)$	& $(3,150)$	& $(3,300)$	& $(4,400)$	& $(5,500)$		\tabularnewline\hline
Fiducial+Kinematics  			& \multirow{3}{*}{6.87} & \multirow{3}{*}{6.76}  
					& \multirow{3}{*}{6.39} & \multirow{3}{*}{0.69} 
					& \multirow{3}{*}{0.06}				\tabularnewline
+Detector+$K$-Factor    	& 		& 		& 		& 		& 		\tabularnewline
~[Eq.~(\ref{cut:leadLepJet})]	& 		& 		& 		& 		& 		\tabularnewline\hline
MET [Eq.~(\ref{cut:MET})]	& 4.30~(63\%)	& 4.22~(62\%)	& 4.02~(63\%)	& 0.40~(58\%)	& 0.03~(50\%)	\tabularnewline\hline
$m_{\ell j_{\rm Fat}}$ 
~[Eq.~(\ref{cut:mass})]		& 3.64~(85\%)	& 3.59~(85\%)	& 3.41~(85\%)	& 0.30~(75\%)	& 0.02~(67\%)	\tabularnewline\hline\hline
$\mathcal{A} = \sigma^{\rm Cuts}/\sigma^{\rm Fid.+Kin.}$
				& 53\%	& 53\%	& 53\%	& 43\%	& 33\%		\tabularnewline\hline
$\frac{S}{\sqrt{S+B}}$ [$\mathcal{L}=10\invfb$]		& 5.9	& 5.9	& 5.7	& 1.7	& 0.4		\tabularnewline\hline
$\frac{S}{\sqrt{S+B}}$ [$100\invfb$]	& 19	& 19	& 18	& 5.4	& 5.7 [2$\invab$]	\tabularnewline\hline\hline
					  \multicolumn{6}{c}{100 TeV VLHC}			\tabularnewline\hline
	&	\multicolumn{5}{c}{$(\MWR,m_N)$ [TeV, GeV]}					\tabularnewline\hline
Cut	& $(3,30)$	& $(3,150)$	& $(3,300)$	& $(4,400)$	& $(5,500)$		\tabularnewline\hline
Fiducial+Kinematics  		& \multirow{3}{*}{1020} & \multirow{3}{*}{1010} 
				& \multirow{3}{*}{957} & \multirow{3}{*}{408}
				& \multirow{3}{*}{183}				\tabularnewline
+Detector+$K$-Factor    	& 		& 		& 		& 		& 		\tabularnewline
~[Eq.~(\ref{cut:leadLepJet})]   & 		&	 	& 		& 		& 		\tabularnewline\hline
MET [Eq.~(\ref{cut:MET})]	& 597~(58\%)	& 591~(58\%)	& 540~(56\%)	& 223~(55\%)	& 93.0~(51\%)	\tabularnewline\hline
$m_{\ell j_{\rm Fat}}$ 
~[Eq.~(\ref{cut:mass})]		& 483~(81\%)	& 476~(81\%)	& 433~(80\%)	& 164~(73\%)	& 61.2~(66\%)	\tabularnewline\hline\hline
$\mathcal{A} = \sigma^{\rm Cuts}/\sigma^{\rm Fid.+Kin.}$
				& 47\%	& 47\%	& 45\%	& 40\%	& 33\%			\tabularnewline\hline
$\frac{S}{\sqrt{S+B}}$ [$10\invfb$]	& 68	& 67	& 64	& 40	& 24		\tabularnewline\hline\hline
\end{tabular}
\caption{$pp\rightarrow e^\pm j_{\rm Fat}$ rates [fb] after successive cuts and QCD normalization, as well as 
acceptance rate and statistical significance after all cuts for representative $(\MWR,m_N)$ at $\sqrt{s} = 13,~100\TeV$.
}
\label{tb:acceptance}
\end{center}
\end{table}

Using a custom detector simulation, we model the effects of detector resolution and efficiency 
based closely on the ATLAS Krak\'ow-parameterization \cite{ATL-PHYS-PUB-2013-004}. 
The parametrization provides a conservative estimate of the ATLAS detector performance for the phase-II high-luminosity LHC. 
We model pile-up (with $\mu=80$) and $\Sigma E_T$-dependent resolutions for jets and MET.
We define an electron to be isolated if the hadronic energy deposit within a cone of size $R = 0.3$ 
is smaller than $10\%$ of the lepton candidate's $p_T$.
For benchmark points we use the $(\MWR,m_N)$ listed in Tb.~\ref{tb:widths}, i.e., $m_N/m_{W_R} \lesssim 0.1$ at $\sqrt{s}= 13$ and 100 TeV.
We summarize our analysis in Tb.~\ref{tb:acceptance}.

As described in Sec.~\ref{sec:pheno}, the angular separation between the charged lepton and the $\WR^*$ decay products 
in the chain $N \to \ell^\pm W^\mp \to \ell^\pm q \overline{q'}$, depends on the $W_R-N$ mass hierarchy. 
A significant amount of radiation from the $\WR^*$ decay enters the isolation cone of $\ell$ and can negatively affect the lepton's identification. 
While so-called mini-isolation requirements~\cite{Rehermann:2010vq} can be applied to recover the unidentified leptons, 
we adopt a more conservative approach and include the lepton's momentum as part of a fat jet $(j_{\rm Fat})$, 
recombined with the C/A algorithm and a cone size of $R=1.0$, i.e., $\jN$.
Hence, we focus on the inclusive process 
\begin{equation}
p ~p ~\to ~W_R~ \to ~e^{\pm} ~N ~\to ~e^{\pm} ~j_{\rm{Fat}}.
\label{eq:signalDef}
\end{equation}

We require the electron and $j_{\rm Fat}$ to further satisfy
\begin{equation}
 p_T^{\ell} > 1\TeV, \quad 
 p_T^{j_{\rm Fat}} > 1\TeV, \quad 
 \vert \eta^\ell \vert < 2.5, \quad
 \vert y^\ell \vert < 2.5.
\label{cut:leadLepJet}
\end{equation}
After kinematic and fiducial cuts, we see in row 5~(14) of Tb.~\ref{tb:acceptance}
that the 13 (100) TeV rate for our representative $(\MWR,m_N)$ spans $\confirm{60\ab - 7\fb~(0.2-1\pb)}$.
Including the detector response shifts the signal MET distribution to larger values than estimated in Sec.~\ref{sec:pheno}.
In order to not lose a majority of the events, we loosen the MET cut of Eq.~(\ref{eq:metCut}) to
\begin{equation}
 \mathrm{MET} < 100\GeV.
 \label{cut:MET}
\end{equation}
In rows 6 and 15 of Tb.~\ref{tb:acceptance}, we find that about $\confirm{50-60}\%$ of events survive the MET requirement, 
with heavier (lighter) $\WR$ having a lower (higher) survival likelihood. 
This behavior is due to the increase in momentum mis-measurement at larger $p_T$ scales, which necessarily occurs with heavier $\MWR$,
and is visible in the MET distribution of Fig.~\ref{fig:lrsm_MET_multiMWR}.
Similarly, higher collider energies lead to additional secondary radiation and larger MET.

\begin{figure}[!t]
\begin{center}
 \subfigure[]{\includegraphics[scale=1,width=.48\textwidth]{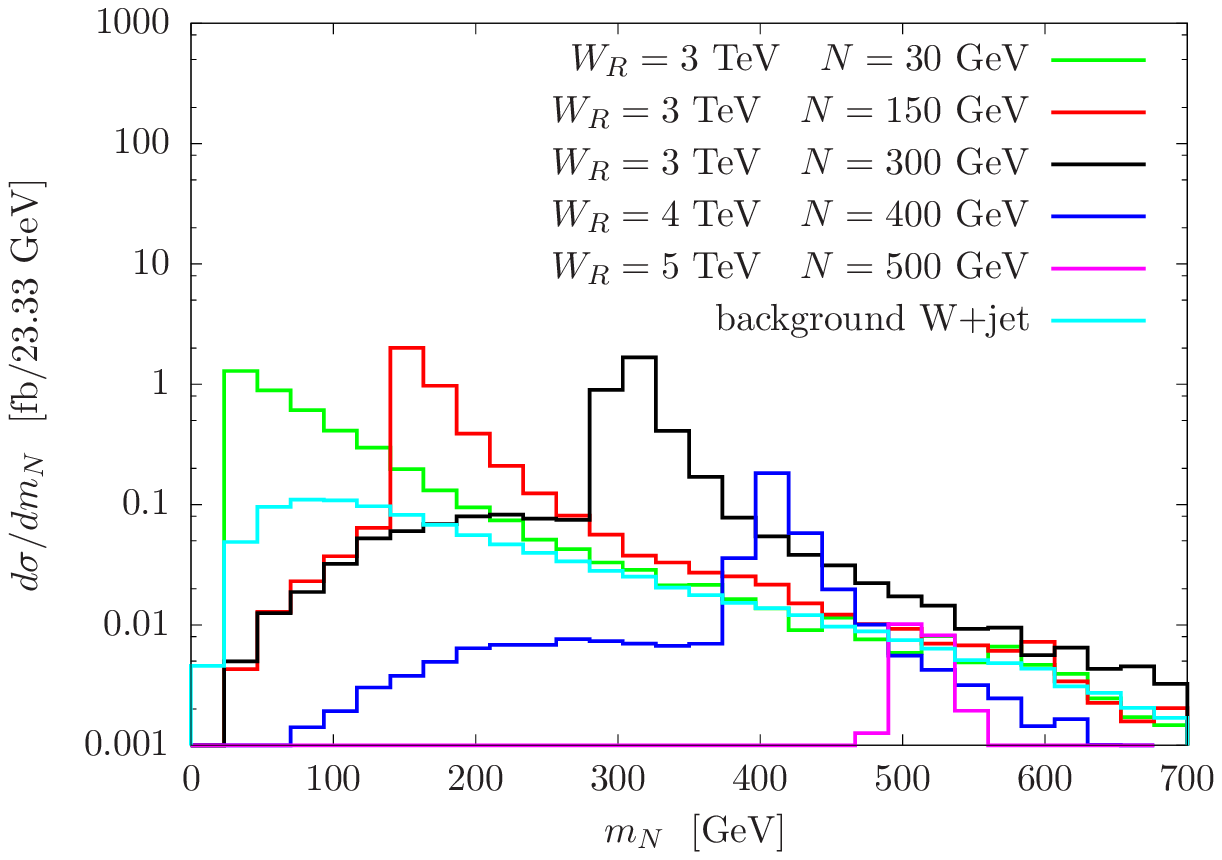}	\label{fig:mNRlog}}
 \subfigure[]{\includegraphics[scale=1,width=.48\textwidth]{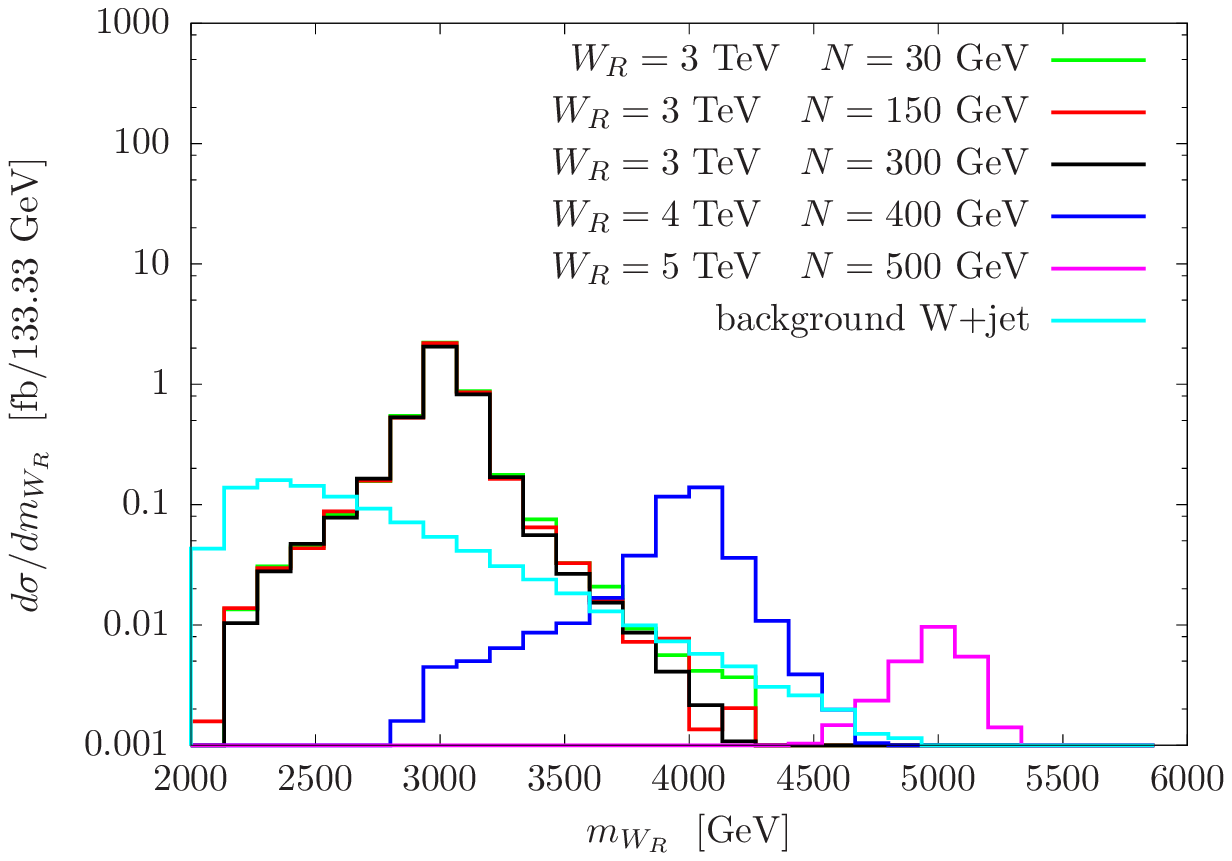}	\label{fig:mWRlog}}
 \end{center}
 \caption{Mass of the reconstructed (a) neutrino $N$ and (b) $W_R$ including detector effects at LO+PS, as detailed in Sec.~\ref{sec:analysis}.} 
 \label{fig:recWN}
\end{figure}

In Fig.~\ref{fig:recWN}, we show the invariant mass distributions at LO+PS for the reconstructed heavy neutrino $N =j_{\mathrm{Fat}}$ 
and $W_R = (\ell^\pm + j_{\mathrm{Fat}})$ systems. The signal is overlaid with the dominant SM $W+1j$ background, also at LO+PS.
At this more realistic level, we find that $j_{\mathrm{Fat}}$ indeed still recovers the desired distributions,
indicating that neutrino jets are indeed good descriptors of boosted heavy neutrinos and further validates our approach.

To further reduce the SM background, we apply the following cut around $m_{\ell j_{\rm Fat}}$:
\begin{equation}
 \vert m_{\ell j_{\rm Fat}} - \MWR \vert < 200\GeV.
 \label{cut:mass}
\end{equation}
The largeness of the mass window is motivated by the size of $W_R$'s total width $\Gamma_{\WR}$.
In row 7~(16) of Tb.~\ref{tb:acceptance}, we see that roughly $\confirm{60-85\%~(65-80\%)}$ of events at 13 (100) TeV rate pass this cut,
again with heavier (lighter) $\WR$ having a lower (higher) survival likelihood.
The behavior here can be understood by comparing the 200 GeV mass widow  to $\Gamma_{\WR}$ in Tb.~\ref{tb:widths}.
For heavier (lighter) $\WR$, we see that the mass window is about $\confirm{1.4~(2.4)}\times\Gamma_{\WR}$, and hence encapsulating fewer (more) $\WR$.
{As in the parton-level analysis, we find that the residual SM background is negligible.}

For 13 (100 TeV), we calculate in row 8~(17) the acceptance rate, defined as the ratio of rows 7 and 5 (16 and 14):
\begin{equation}
 \mathcal{A} \equiv \frac{\sigma^{\rm All~Cuts}}{\sigma^{\rm Fiducial+Kinematics+Detector~Response}}.
\end{equation}
We find that approximately $\confirm{33-50\%}$ events pass our selection criteria. 

\begin{figure}[!t]
\begin{center}
\subfigure[]{\includegraphics[scale=1,width=.48\textwidth]{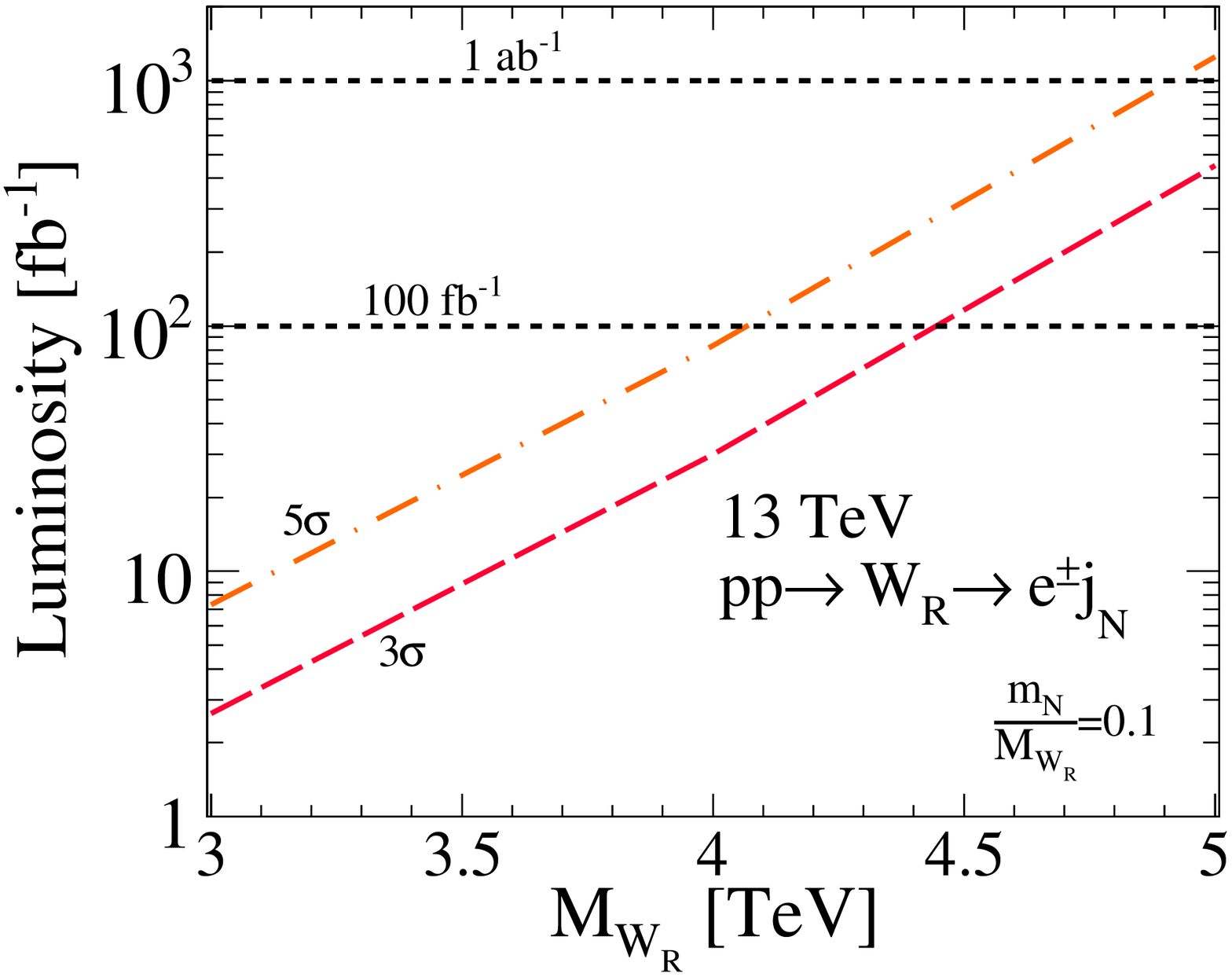}	\label{fig:disc13}}
\subfigure[]{\includegraphics[scale=1,width=.48\textwidth]{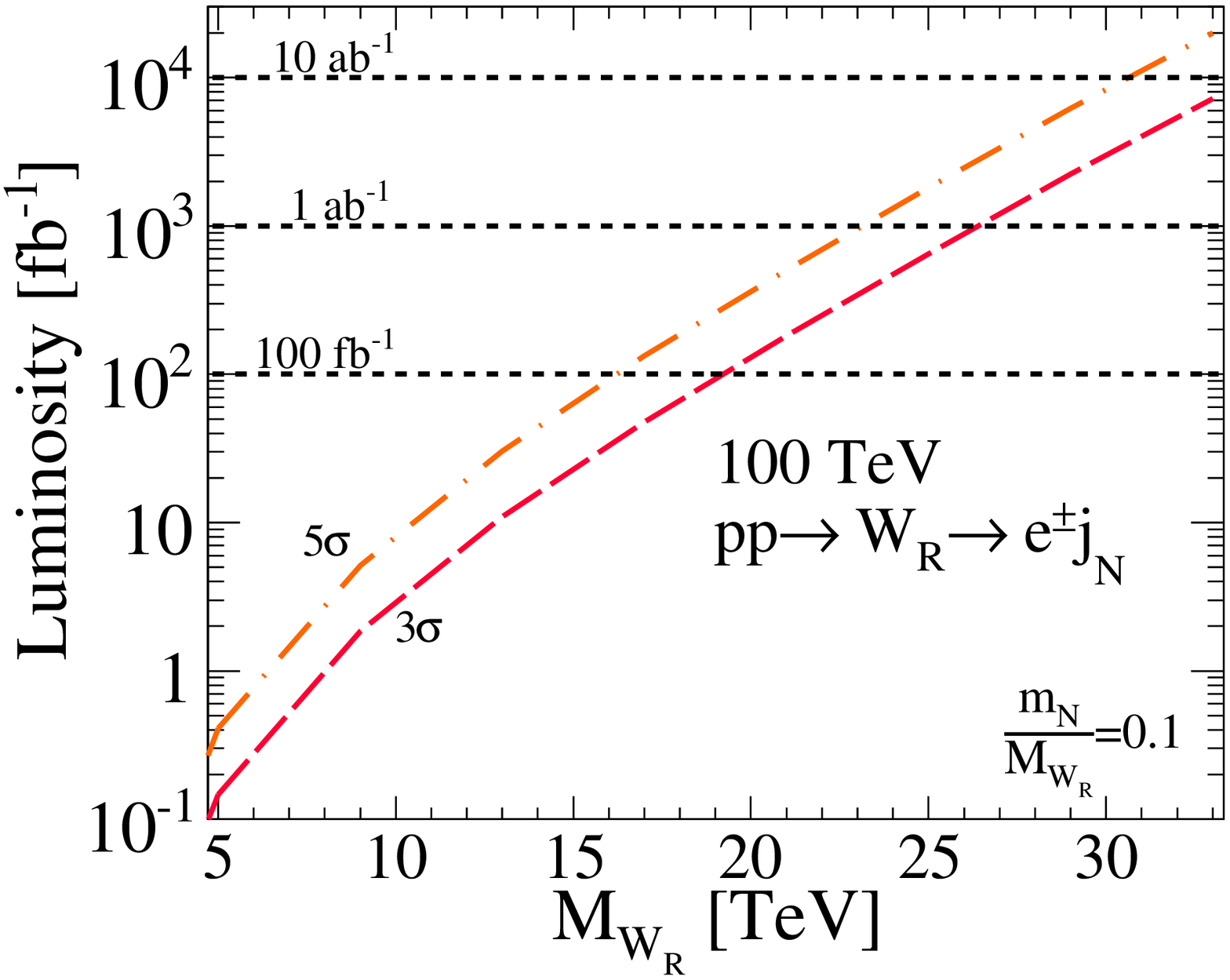}	\label{fig:disc100}}
\\
\subfigure[]{\includegraphics[scale=1,width=.48\textwidth]{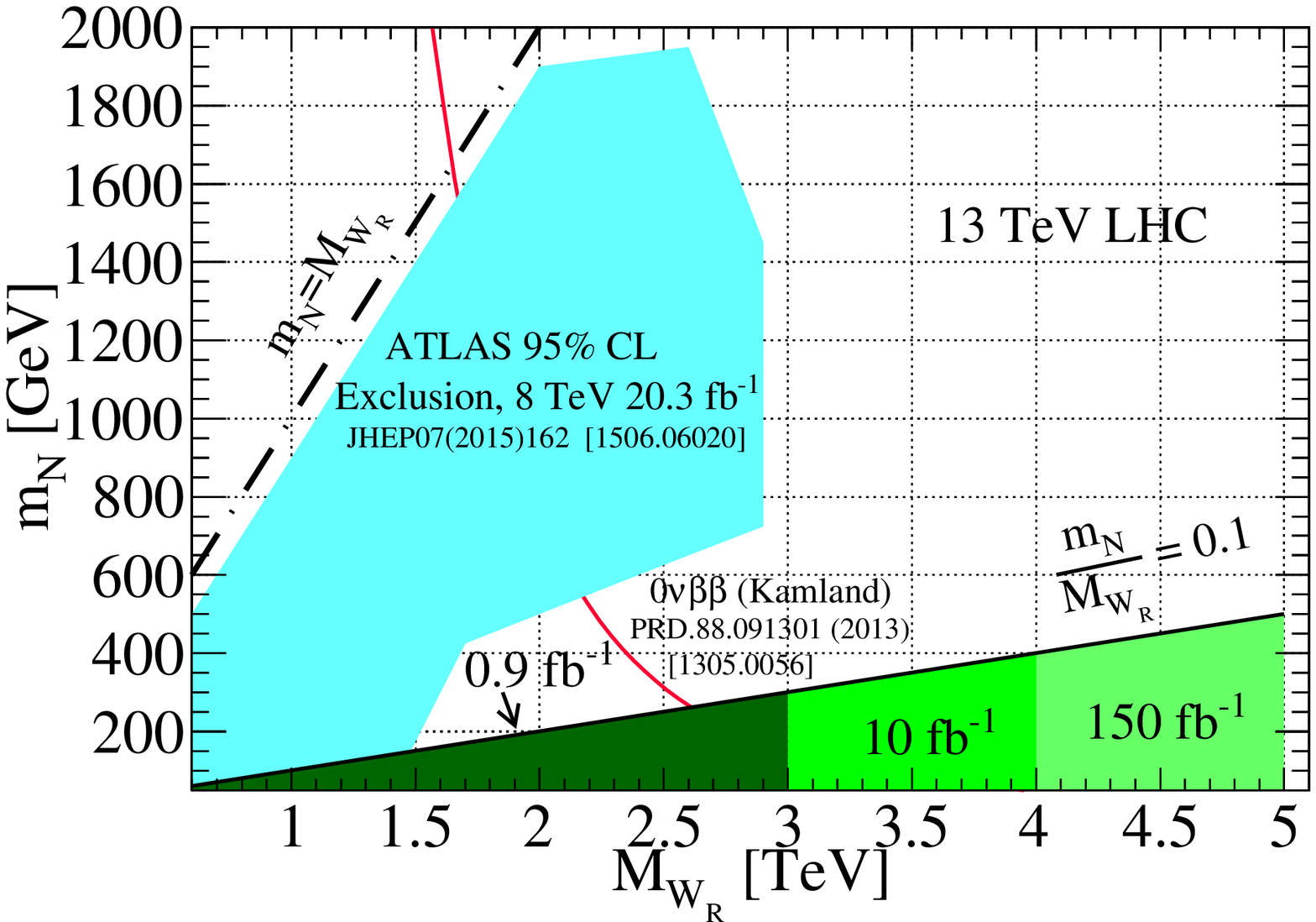}	\label{fig:exclusion13}}
\subfigure[]{\includegraphics[scale=1,width=.48\textwidth]{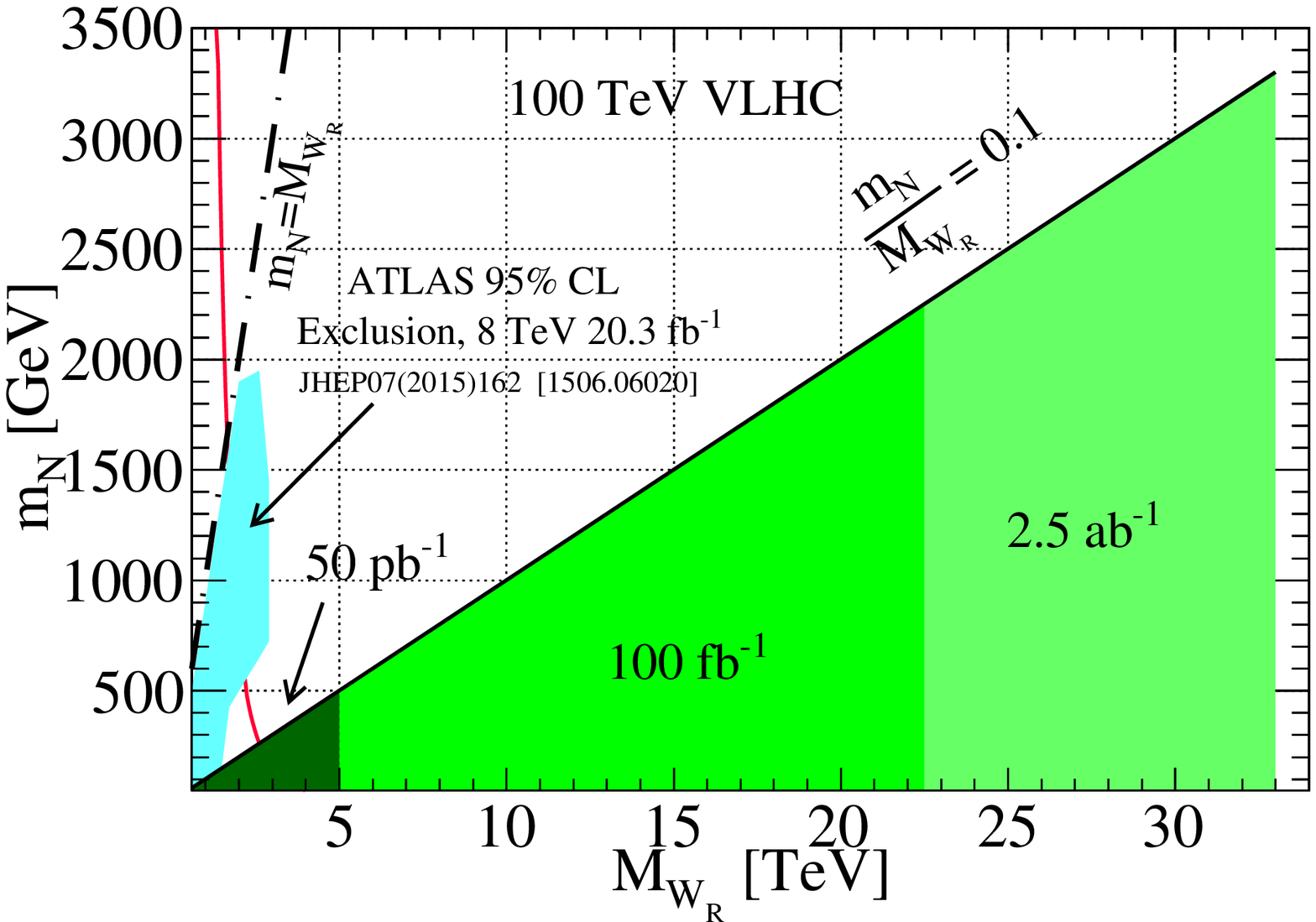}	\label{fig:exclusion100}}
\end{center}
\caption{Discovery (a,b) and 95\% CL exclusion (c,d) potential of $pp\rightarrow\WR\rightarrow e^\pm j_N$ searches  at (a,c) $\sqrt{s} =$ 13 and (b,d) 100 TeV. 
Also shown in (c,d), 
ATLAS experiment's 8 TeV 95\% CL~\cite{Aad:2015xaa}
and KamLAND-Zen 90\% CL~\cite{Dev:2013vxa,KamLAND-Zen:2016pfg} exclusion limits.}
\label{fig:discov}
\end{figure}

Using the Gaussian estimator,
\begin{equation}
 \sigma = \frac{S}{\sqrt{S+B}} \approx \sqrt{S}, \quad\text{for}\quad S(B) = \mathcal{L}\times \sigma^{\rm All~Cuts}_{\rm Signal~(SM~background)},
\end{equation}
we can determine the statistical significance of the signal process $(S)$ over 
the SM backgrounds $(B)$ after an integrated luminosity of $\mathcal{L}$.
At 13 TeV, we find a $>5\sigma$ statistical observation (discovery) for $\MWR = 3~(4)~[5]\TeV$, independent of $m_N$,
after $\mathcal{L} = 10~(100)~[2000]\invfb$. 
At 100 TeV and $\mathcal{L} = 10\invfb$, all benchmark points are in excess of 20$\sigma$.
This is summarized in rows 9, 10, and 18 of Tb.~\ref{tb:acceptance} and Figs.~\ref{fig:discov}(a) and (b).
We extrapolate the discovery potential for higher $\MWR$ by keeping fixed the efficiency,
$\varepsilon \equiv \sigma^{\rm Fid.+Kin.} / \sigma^{\rm Total},$
and acceptance for $(\MWR, \mN)=(5\TeV,500\GeV)$, in which case $(\varepsilon \approx 0.64,~\mathcal{A}\approx0.33)$.
As seen in Fig.~\ref{fig:disc100}, 
a $5\sigma$ discovery can be obtained for $W_R$ masses up to $M_{W_R}=15~(30)$ with approximately $100~\text{fb}^{-1}~(10~\text{ab}^{-1})$.

{
Finally, we discuss briefly the 13 and 100 TeV potential to exclude previously unconstrained regions of the $(\MWR,\mN)$ parameter space.
We use Poisson counting to deduce the required luminosity $\mathcal{L}_{95}$ for a $95\%$ Confidence Level (CL) exclusion:
For a SM background of $B\approx0$ events, we solve for the largest number of signal events $S$
such that the expected probability to observe $B$ events is at most $5\%(=1-\CL)$, 
i.e., find $S$ such that we satisfy:
\begin{equation}
 \Pr\left(n_{\rm observed}=B \vert n_{\rm expected} = S+B\right) = 
 \cfrac{(S+B)^B}{B!}~e^{-(S+B)}
 \leq 1-\CL = 0.05.
\end{equation}
For $B\approx0$, this yields $S=3$.
Given an efficiency $\epsilon$ and acceptance $\mathcal{A}$, $\mathcal{L}_{95}$ can then be determined by the relationship
\begin{equation}
\mathcal{L}_{95}=\frac{S}{\epsilon \cdot \mathcal{A} \cdot \sigma^{\rm NLO+NNLL}\times {\rm BR} \times {\rm BR}}.
\end{equation}
We then state that a $(\mN/\MWR)$ mass-hypothesis is excluded at 95\% CL if
\begin{equation}
N =  \mathcal{L}_{95} \times\sigma^{\rm NLO+NNLL}\times  {\rm BR} \times {\rm BR} \cdot \epsilon  \cdot \mathcal{A} \geq S=3 
\end{equation}
For $(\mN/\MWR)\leq0.1$, we show in Fig.~\ref{fig:exclusion13} that $\MWR < 3~(4)~[5]\TeV$ can be excluded at 95\% CL
with $\mathcal{L} = 0.9~(10)~[150]\invfb$ of 13 TeV data.
Also plotted are the
ATLAS experiment's 8 TeV 95\% CL~\cite{Aad:2015xaa}
and KamLAND-Zen 90\% CL~\cite{Dev:2013vxa,KamLAND-Zen:2016pfg} complimentary exclusion limits.
We find that regions of the $(\MWR,\mN)$ parameter space unconstrained by ATLAS and CMS are indeed covered by the present, complimentary analysis.
The open region between this analysis and ATLAS is an artifact of our choice to limit our study to $(\mN/\MWR)\leq0.1$; 
application of the neutrino jet analysis to larger mass ratios will close the region.
We note that the ability to exclude $\MWR<3\TeV$ at $\sqrt{s}=13\TeV$ with approximately $1/20$ of the 8 TeV data 
is consistent with the luminosity increase for DY-type processes~\cite{Martin:2009iq}.
In Fig.~\ref{fig:exclusion100}, we show the analogous 100 TeV exclusion potential:
with $\mathcal{L}=100\invfb~(2.5\invab)$, we find that $\MWR<22~(33)\TeV$ and $(\mN/\MWR)\leq0.1$ can be excluded at 95\% CL.
}

\section{Summary and Conclusion}\label{sec:conclusion}
The origin of tiny, nonzero neutrino masses remains an open question in particle physics.
In this study, we re-examine the discovery potential of a $\WR$ gauge boson decaying to a heavy Majorana neutrino $N$ in the MLRSM.
We focus on the case when $N$ is hierarchically lighter than $\WR$, i.e., $m_N/\MWR \lesssim 0.1$.
In this limit, $\WR\to N$ decays produce highly boosted $N$ that then decay to collimated final states.
Subsequently, the canonical collider definition 
\begin{equation}
 p ~p ~\rightarrow ~\WR~ ~\to ~e^\pm ~N(\to e^\pm j j) 
\end{equation}
breaks down due to failing isolation criteria of the final-state charged leptons.
For such a regime, we consider an alternative collider definition,
\begin{equation}
 p ~p ~\rightarrow ~\WR ~\to ~e^\pm ~N \rightarrow ~e^\pm ~\jN,
\end{equation}
where $\jN$ is a color-singlet \textit{neutrino jet} and consists of the collimated $N$ decay products.
Furthermore, we consider resummed QCD corrections that are important for high-mass DY processes.
We calculate, for the first time, inclusive $pp\rightarrow\WR$ production at NLO+NNLL matched to threshold-improved PDFs.
This captures dominant contributions beyond NLO, and are arguably the most precise predictions available for high-mass $\WR$ at 13 and 100 TeV. 
We summarize our findings:
\begin{enumerate}
 \item We introduce the concept of neutrino jets, which has widespread applicability to other processes and models; see Sec.~\ref{sec:pheno}.
 With our new collider signal definition, a $5-6 \sigma$ discovery is achievable at 13 TeV with $\confirm{10~(100)~[2000]~\text{fb}^{-1}}$
 for $M_{W_R} = 3~(4)~[5]$ TeV and $(m_N/M_{W_R})<0.1$. At 100 TeV, a $5\sigma$ discovery can be obtained for $W_R$ masses up to $M_{W_R}=15~(30)$ TeV
 with approximately $100~\text{fb}^{-1}~(10~\text{ab}^{-1})$. 
 {Conversely, with $0.9~(10)~[150]\invfb$ of 13 TeV data, $\MWR < 3~(4)~[5]\TeV$ can be excluded at 95\% CL;
 with $100\invfb~(2.5\invab)$ of 100 TeV data, $\MWR<22~(33)\TeV$ can be excluded.}
 See Sec.~\ref{sec:analysis}.

  \item At 13 TeV, the NLO+NNLL contributions increase the Born (NLO)-level predictions by $40-140~(4-70)\%$ for $M_{W_R} = 4-5\TeV$,
  well beyond the NLO scale uncertainty.
  At 100 TeV, threshold effects become important for $\MWR \gtrsim 30\TeV$, where resummation increases the Born (NLO) prediction by $\gtrsim40~(10)\%$.
  Away from threshold, we find that the resummed result converges to the NLO rate.
  See Sec.~\ref{sec:xsec}.
 
 \item The residual scale dependence at NLO+NNLL for $M_{W_R} = 1-5~(1-30)\TeV$ at 13 TeV is maximally sub-percent, and $\pm4\%$ at 100 TeV.
  The PDF uncertainty at NLO+NNLL exceeds 100\% in the threshold regions.
  Away from threshold, the PDF uncertainty is comparable to the NLO PDF uncertainty.
 See Sec.~\ref{sec:uncertainties}

\end{enumerate}

\vspace*{0.3cm}
\noindent {\it Acknowledgments:}{
B.~Pecjak, J.~Pavon, J.~Rojo, L.~Rottoli, C.~Tamarit, G.~Watt, and C.~W\'eiland 
are thanked for useful discussions.
This work was supported by the U.K. Science and Technology Facilities Council.
MM~acknowledges the support from Royal Society International Exchange Programme and thanks IPPP, Durham University, UK for the hospitality. 
The work of MM is partially supported by the DST INSPIRE grant INSPIRE-15-0074. 
The work of RR is partially supported by the European Union's Horizon 2020 research and innovation programme under 
the Marie Sklodowska-Curie grant agreements Nos. 690575 (InvisiblesPlus RISE), and 674896 (Elusives ITN).
}

\appendix

\section{Threshold Resummation for Inclusive $pp\rightarrow W'$ Production}\label{app:threshold}
Here, we review the details of threshold resummation  
for inclusive production of $W'$ bosons with arbitrary chiral couplings in $pp$ collisions.
Often labeled as soft-gluon or large-$x$ resummation~\cite{Sterman:1986aj,Catani:1989ne,Catani:1990rp}, 
the calculation should not be confused with small-$k_T$ (or recoil or Collins-Soper-Sterman) resummation~\cite{Collins:1981uk,Collins:1981va,Collins:1984kg}
nor joint recoil-threshold resummation~\cite{Li:1998is,Laenen:2000de,Laenen:2000ij}.
Many concise texts on the topic exist, 
e.g.,~Refs.~\cite{Catani:1996yz,Forte:2002ni,Catani:2003zt,Bozzi:2007qr,Bonvini:2010tp,Bonvini:2012sh,Luisoni:2015xha}.
We largely follow the notation and spirit of~Refs.~\cite{Catani:1996yz,Bonvini:2010tp,Catani:2003zt} 
and implement the numerical procedures of~Refs.~\cite{Catani:1996yz,Bonvini:2010tp},
but make explicit mass and coupling factors that are often omitted for simplicity.

\subsection{Threshold Resummation for $W'$ Bosons with Arbitrary Chiral Couplings}
The emission of soft gluons from initial-state partons participating in hard, hadronic collisions 
can spawn large, with respect to the expansion parameter $\alpha_s$, logarithmic enhancements in scattering cross sections. 
In certain kinematic configurations, these logarithms can become large enough to render a perturbative expansion unreliable, 
requiring that the divergent series be summed to all orders. 
In particular, soft logarithms near the partonic threshold take the form
\begin{equation}
 \as\log\left(\frac{\hat{s} - Q^2}{\hat{s}}\right) = \as\log(1 - z), \quad z \equiv \frac{Q^2}{\hat{s}},
\end{equation}
where $Q\sim \sqrt{\hat{s}} \gg \Lambda_{\rm QCD}$ is the scale of the hard scattering process,
$\sqrt{\hat{s}}$ is the partonic scattering scale, and the dimensionless variable $z$ quantifies the nearness of the partonic scale to the hard scale.
A schematic distinction of the hard, partonic, and hadronic scattering (beam) scale $\sqrt{s}$ is illustrated in Fig.~\ref{fig:factorThm}.
The purpose of threshold resummation is to perform a summation of such terms when $z\rightarrow1$
while accounting for the hierarchy of scales via renormalization group evolution (RGE).
We now briefly summarize the procedure directly in perturbative QCD (pQCD).

For a  generic color-singlet boson $\mathcal{B}$ (scalar or vector) produced in hadron collisions,
the total inclusive cross section is given by the usual Collinear Factorization Theorem
\begin{eqnarray}
 \sigmaFO(h_1 ~h_2 ~\rightarrow ~\mathcal{B} + X) = \sum_{a,b=q,\overline{q'},g}\int_{\tau_0}^1 d\tau ~\mathcal{L}_{ab}(\tau,\mu_f)
  \times \hat{\sigma}^{\rm FO}(ab\rightarrow \mathcal{B}),  \quad \tau_0 \equiv \frac{\MB^2}{s}, 
  \label{eq:appFOFactThm}
\end{eqnarray}
where the luminosity $\mathcal{L}$ of parton pair $ab,$ with $a,b\in\{q,\overline{q'},g\}$, at the LHC ($h_1 = h_2 = p$)  
is given in terms of the PDFs $f_{a/p}$ and $f_{b/p}$ jointly evolved to a factorization scale $\mu_f$:
\begin{eqnarray}
 \mathcal{L}_{ab}(\tau,\mu_f) &=& \frac{1}{1+\delta_{ab}}\int_{\tau}^1 \frac{d\xi_1}{\xi_1}
 \left[f_{a/p}(\xi_1,\mu_f)f_{b/p}(\xi_2,\mu_f) + f_{a/p}(\xi_2,\mu_f)f_{b/p}(\xi_1,\mu_f)\right], 
 \label{eq:lumiDef}\\
 \xi_2 &\equiv& \frac{\tau}{\xi_1},
\end{eqnarray}
and $\hat{\sigma}^{\rm FO}$ is the FO partonic cross section for the process
\begin{equation}
 a ~b ~\rightarrow ~\mathcal{B} \quad\text{with}\quad Q = \MB.
 \label{eq:appPartonScatt}
\end{equation}
Following the notation and methodologies of Refs.~\cite{Catani:1996yz,Catani:2003zt}, 
we account for the arbitrary emission of soft radiation by using a generalization of Eq.~(\ref{eq:appFOFactThm}):
\begin{eqnarray}
 \sigmaFO(pp \rightarrow \mathcal{B} + X)
  = \sum_{a,b=q,\overline{q'},g} \int^1_{\tau_0} d\tau ~\int^1_0 dz  ~\delta\left(z - \frac{\tau_0}{\tau}\right)
 \times    \mathcal{L}_{ab}(\tau)  \times \hat{\sigma}^{\rm FO}(ab\rightarrow \mathcal{B}).
 \label{eq:appGenFactThem}
\end{eqnarray}
Written this way, we identify the FO partonic cross section in the soft radiation limit as
\begin{equation}
\label{eq:partonic_Xsec}
\hat{\sigma}^{\rm FO}(ab\rightarrow \mathcal{B}) ~\equiv~ \hat{\sigma}_{ab}^{\rm FO} ~=~ \sigma_0 ~\times~ z ~\times~ \Delta_{ab}^{\rm FO}(z),
\quad
 \Delta_{ab}^{\rm FO}(z) =  \sum_{k=0}^\infty \left(\frac{\as}{\pi}\right)^k ~\Delta_{ab}^{(k)}(z).
\end{equation}
The soft threshold coefficient $\Delta_{ab}(z)$, which encapsulates the factorizable soft emissions, 
is often denoted as $C_{ab}$ and $G_{ab}$ in literature.
For a $W'$ gauge boson with arbitrary chiral couplings $g_L$ and $g_R$ to quarks and mass $\MB = M_{W'}$, the constant term is
\begin{equation}
 \sigma_0 = \vert V_{ab}^{\rm CKM'}\vert^2\frac{(g_L^2 + g_R^2) \pi}{4N_c M_{W'}^2}.
\end{equation}
We suppress the indices on $\sigma_0$ as the trivial generalization introduces an unnecessary notational complication.
The expression is related to the usual LO partonic expression by
\begin{equation}
 \hat{\sigma}^{LO}(q ~\overline{q'} ~\rightarrow~ W') 
 ~=~ \sigma_0 ~\times~ M_{W'}^2 ~\times~ \delta(\hat{s} -  M_{W'}^2)
 ~=~ \sigma_0 ~\times~ z ~\times~ \delta(1 - z).
\end{equation}
At LO, one may identify $\Delta$ with the above $\delta$-function, which is determined by kinematics alone.
This is because  the LO $2\rightarrow1$ process occurs identically at threshold.
Beyond LO, the structure of soft logarithms in the perturbative expansion of $\Delta(z)$ remains essentially kinematic in origin~\cite{Forte:2002ni}.
In terms of explicit scale dependence, one can also write Eq.~(\ref{eq:partonic_Xsec}) as
\begin{equation}
 \hat{\sigma}_{ab}^{\rm FO} ~=~ z ~\times~ \Delta_{ab}^{\rm FO}(\hat{s},Q^2)  ~\times~ \sigma_0(Q^2=M_{W'}^2).
\end{equation}
This suggestive form indeed implies, in the language of RGE, 
that $\Delta(\hat{s},Q^2)$ is an evolution operator that runs the hard process 
at $Q^2=M_{W'}^2$ up to the partonic scale $\hat{s} = \tau s$~\cite{Forte:2002ni}.

If working with pQCD, the threshold resummed cross section can be efficiently obtained after writing 
the hadronic cross section in so-called Mellin-(or $N$- or moment-)space. 
This is because such convolutions become products in Mellin space.
Applying the Mellin transform, as defined in Eq.~(\ref{eq:mellinDef}), to Eq.~(\ref{eq:appGenFactThem}) with respect to $\tau_0$ yields for LO $W'$ production
\begin{eqnarray}
\sigma^{\rm LO}_N(pp\rightarrow W') &=&  
\sum_{a,b=q,\overline{q'},g} 
\int_0^1 d \tau_0 ~\tau_0^{N-1} ~\int_{\tau_0}^1 d \tau ~\int_0^1 dz 
~\delta\left(z - \frac{\tau_0}{\tau}\right)
~\mathcal{L}_{ab}(\tau) ~\sigma_0 ~ z ~ \Delta_{ab}^{\rm LO}(z) 
\nonumber\\
&=& \sigma_0 \sum_{a,b=q,\overline{q'},g}  \mathcal{L}_{ab,(N+1)}  ~\Delta_{ab,(N+1)}^{\rm LO}
= \sigma_0 ~\mathcal{L}_{q\overline{q'},(N+1)} ~\Delta_{q\overline{q'},(N+1)}^{\rm LO}.
\label{eq:appFactThmMellin}
\end{eqnarray}
In the last step we have used the fact that vector boson production is due strictly to $q\overline{q}$ annihilation in the soft limit.
This follows from currents of massless fermions being proportional to external fermion energies, i.e., 
$J^\mu_{q_f q_i} \propto \sqrt{E_{q_f}E_{q_i}}$,
and therefore vanish in the soft radiation limit for initial-states $qg,\overline{q}g$ and $gg$.

Non-trivially, obtaining the resummed cross section in $N$-space 
is a simple matter of replacing the LO coefficient $\Delta_N^{\rm LO}$ in Eq.~(\ref{eq:appFactThmMellin}) 
by its resummed analogue~\cite{Sterman:1986aj,Catani:1989ne,Catani:1990rp}.
That is, 
\begin{equation}
\sigma^{\rm Res.}_N(pp\rightarrow W') = \sigma_0 ~\mathcal{L}_{q\overline{q'},(N+1)} ~\Delta_{q\overline{q'},(N+1)}^{\rm Res.}.
\end{equation}
Among other considerations, the resummation is usually performed in the large-$N$ limit 
as the $N \rightarrow \infty$ limit corresponds to the $z \rightarrow 1$ (threshold) limit for partonic cross sections. 
In this limit, additional gluon emission is constrained to be soft, and is therefore exactly where one finds a perturbative expansion rendered unreliable by large logarithms. 
Specifically, the divergent contributions at leading power in $(1-z)$ are plus distributions of the form
\begin{equation}
\Delta^{(j)}(z) \sim  \alpha_s^j(Q^2)\bigg[ \frac{\log^{m}(1-z)}{1-z}\bigg]_+ \; , \quad m \leq 2j-1.
\end{equation}
In Mellin space and in the large-$N$ limit, such distributions are transformed to a series of the form
\begin{equation}
\Delta^{(j)}_N ~\sim~  \alpha_s^j(Q^2) \sum_{r=0}^{2j}b_r \log^r N \; ,
\end{equation}
where $b_r$ is some $N$ independent coefficient.
To all orders in $\as$, resummation captures a number of these divergent logarithms, 
producing a finite result that can supplement FO calculations. 
For the $k$th term in the expansion, 
resummation corresponds at leading log (LL) accuracy to gathering all logarithms with power of $r=2k$;
at next-to-leading log (NLL), all logs such that $2k \geq r \geq 2k-2$; and generically at N$^j$LL, $2k \geq r \geq 2k - 2j$.
{Furthermore, this implies that in re-expanding N$^j$LL in $\as$,
one can identify the inclusive N$^{(j-1)}$LO calculation in the limit where all radiation is soft.
This necessitates a matching scheme when combining resummed and FO results beyond LO.}

In the notation of Ref.~\cite{Bonvini:2010tp}, the resummed coefficient $\DeltaNRes$ for color-singlet $q\overline{q'}$ pairs is
\begin{equation}
   \Delta_{q\overline{q'},N}^{\rm Res.}  = g_0(\as) ~\exp{\mathcal{S}(\lambda,\asBar)},  
   \quad\text{with}\quad \lambda = \overline{\alpha} \log \frac{1}{N} 
   \quad\text{and}\quad \overline{\alpha} = a \; \alpha_s(Q^2) \; \beta_0 \;,
\end{equation}
where $a=2~(1)$ for DY (DIS) accounts for the number of contributing initial-state hadrons,
and the \emph{Sudakov factor} $\mathcal{S}$ is given as an expansion in $\overline{\alpha}$, while treating $\overline{\alpha} \ln N \sim \mathcal{O}(1)$:
\begin{eqnarray}
   \mathcal{S}(\lambda,\asBar) &=& \sum_{m=0}^\infty ~\asBar^{m-1}~g_{m+1} = 
\underset{\rm NNLL}{\underbrace{\underset{\rm NLL}{\underbrace{\underset{\rm LL}{\underbrace{\frac{1}{\asBar}g_1(\lambda)}}
+ g_2(\lambda)}}
+ \asBar g_3(\lambda)}} + \mathcal{O}(\asBar^2)
\equiv
\sum_{k=0}^{\infty} \as^k \mathcal{S}_k.
\label{eq:appSudakov}
\end{eqnarray}
We note that it is possible to consistently re-expand $\mathcal{S}$ in terms of $\as$ and coefficients $\mathcal{S}_k$.
The normalization function $g_0$ is similarly perturbative and is given by
\begin{eqnarray}
 g_0(\as) = \sum_{n=0}^\infty ~g_{0n}~\as^n ~=~ 
\underset{\rm NNLL}{\underbrace{\underset{\rm NLL}{\underbrace{\underset{\rm LL}{\underbrace{g_{00}}} + \as g_{01}}} 
+ \as^2 g_{02}}}
+\mathcal{O}(\as^3)
\end{eqnarray}
Expressions and normalizations for $g_m$, $g_{0n}$, and the QCD $\beta$-function coefficient $\beta_0$ are detailed in~\cite{Bonvini:2010tp}. 
Acquiring a resummation of order\footnote{I.e., N$^j$LL in the ``$*$'' convention or N$^j$LL$'$~in the ``$~'~$'' convention, 
which are precisely defined in Ref.~\cite{Bonvini:2012sh}.}
N$^j$LL is achieved by including the matching functions in $g_0(\alpha_s)$ up to $\mathcal{O}(\as^j)$, i.e., all $g_{0n}$ up to $n=j$, 
and $g_m$ functions for $m$ up to $m=j+1$.
In this work, we resum soft radiation up to next-to-next-to-leading logarithmic (NNLL) accuracy.
Thus our resummed soft function is
\begin{equation}
\Delta_{(q\bar{q}')N}^{\rm NNLL} = (g_{00}+g_{01}\as+g_{02}\as^2) \exp{\left[ \frac{1}{\overline{\alpha}} g_1(\lambda,\asBar)
+g_2(\lambda,\asBar)+\overline{\alpha} \, g_3(\lambda,\asBar)  \right]},
\end{equation}
and our resumed cross section in Mellin-space at NNLL
\begin{eqnarray}
 \sigma^{\rm NNLL}_N(pp\rightarrow W') &=&  \sigma_0 ~\mathcal{L}_{q\overline{q'},(N+1)}~ \Delta_{(q\bar{q}')(N+1)}^{\rm NNLL}.
 \label{eq:appNNLLMellin}
\end{eqnarray}

\subsection{Inverse Mellin Transformation via Minimal Prescription Procedure}
Taking the inverse Mellin transformation of Eq.~(\ref{eq:appNNLLMellin}), as defined in Eq.(\ref{eq:invMellinDef}), gives
the resummed production cross section in momentum space:
\begin{eqnarray}
  \sigmaRes(pp\rightarrow W' + X) = \frac{\sigma_0}{2\pi i}\int_{c-i\infty}^{c+i\infty}dN ~\tau_0^{-(N-1)} \times \mathcal{L}_N \times \DeltaNRes.
  \label{eq:appResInvMellin}
\end{eqnarray}
Formally, the integration path, with $c\in\mathbb{R}$, is to the right of all singularities.
In practice, this is impossible due to the QCD Landau pole at $N = N_L \equiv  \exp[1/2\as \beta_0]$.
The situation can be remedied by adhering to the Minimal Prescription (MP) procedure~\cite{Catani:1996yz}, 
which entails choosing $c = C_{\rm MP}$ such that
\begin{equation}
 2 < C_{\rm MP} < N_L,
\end{equation}
to avoid the pomeron (Landau) pole as small (large) $N$, and deforming the path toward $\text{Re}[N]<0$. 
\begin{figure}[!t]
\begin{center}
\subfigure[]{\includegraphics[scale=1,width=.45\textwidth]{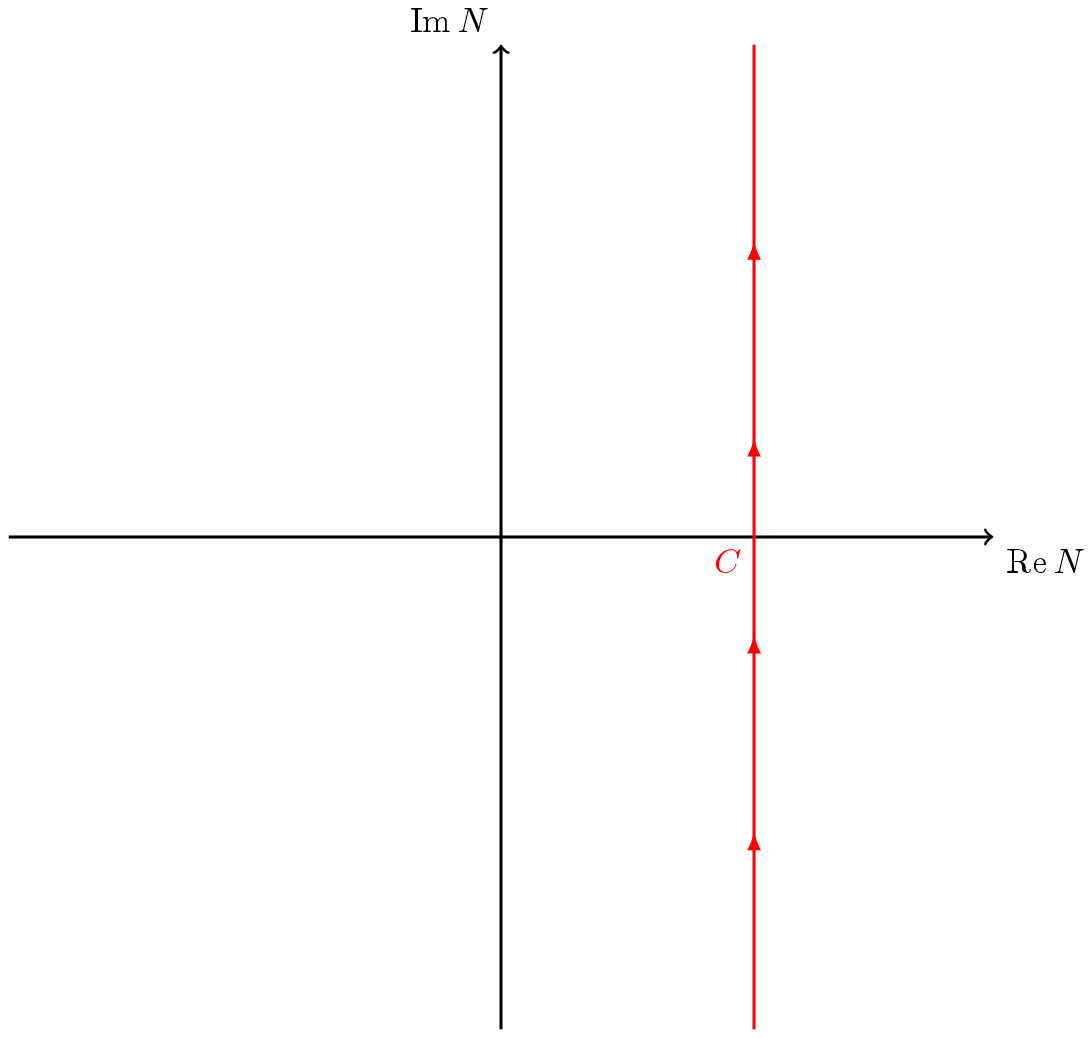}	\label{fig:melInv}}
\subfigure[]{\includegraphics[scale=1,width=.45\textwidth]{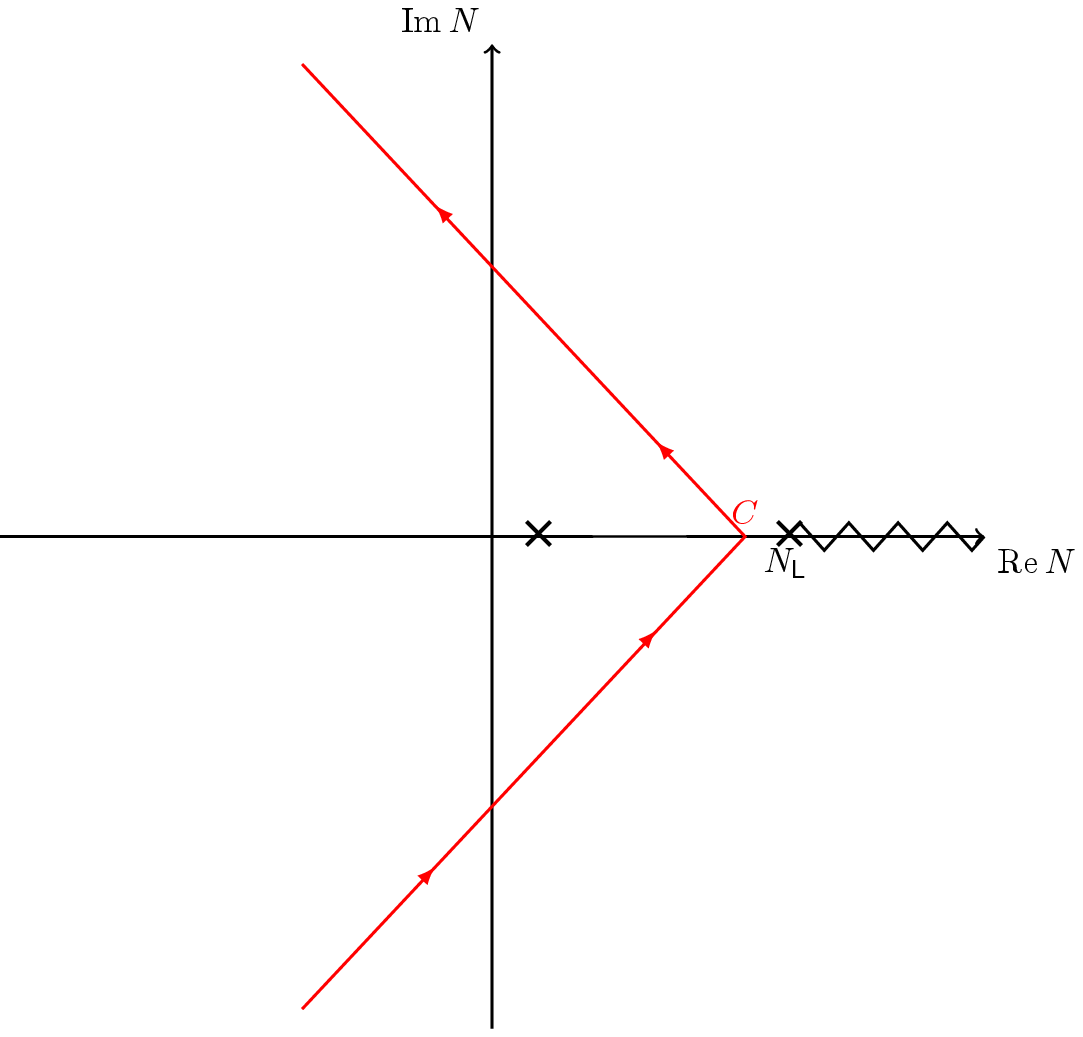}	\label{fig:appIntPath}}
\end{center}
\caption{Contours taken to perform the inverse Mellin transform. In (a) the choice of contour as defined in Eq.~(\ref{eq:invMellinDef}). 
The contour used to implement the Minimal Prescription for the resummed prediction is shown in (b).}
\label{fig:mellinConts}
\end{figure}
The path deformation is illustrated in Fig.~\ref{fig:appIntPath}. 
Subsequently, Eq.~(\ref{eq:appResInvMellin}) becomes
\begin{eqnarray}
  \sigmaRes(pp\rightarrow W' + X)  &\overset{\rm MP}{=}&
 \frac{\sigma_0}{2\pi i}\int_{C_{\rm MP}}dN ~\tau_0^{-(N-1)} \times \mathcal{L}_N \times \DeltaNRes
 \\
 &=& \frac{\sigma_0}{\pi i}\int_{C_{\rm MP}[\text{Im}[N]>0]}dN ~\tau_0^{-(N-1)} \times \mathcal{L}_N \times \DeltaNRes.
\end{eqnarray}
In the second line, a factor of $2$ follows from the integrand being even with respect to $\text{Im}[N]$. {This follows from the fact that the original cross section in momentum space Eq.~(\ref{eq:appGenFactThem}) is a real function.}

Following Ref.~\cite{Bonvini:2014joa}, and the associated code \texttt{ResHiggs}, we choose the path,
\begin{equation}
 N(t) = c_{\rm MP} + (m_{\rm MP}-i)\log(t), \quad t \in (0,1),
\end{equation}
where $c_{\rm MP},~ m_{\rm MP}\in \mathbb{R}$ cannot be too large numerically without hitting machine precision limitations in 
$\tau_0^{-(N-1)} = \exp[-(N-1)\log\tau_0]$. 
We have checked $5 < c_{\rm MP} < 15$ and $m_{\rm MP} = c_{\rm MP}/10$ leaves the integral unchanged.
Making the change of variable to $t$, one has
\begin{eqnarray}
 \sigma^{Res.}(s) &=& \frac{\sigma_0}{\pi i} \int_0^1 \frac{dt}{t} (i-m) ~\tau_0^{-(N-1)} \times \mathcal{L}_N \times \DeltaNRes
 \\
 &=& \frac{\sigma_0}{\pi} \int_0^1  \frac{dt}{t} \text{Im}\left[ (i-m) ~e^{-(N-1)\log\tau_0} \times \mathcal{L}_N \times \DeltaNRes\right].
 \label{eq:appInvMellinRes}
\end{eqnarray}
In the last line, we use the fact that $\sigma^{\rm Res.}$ is a physical rate, i.e., positive-definite,
implying that the integrand must be purely imaginary to cancel the $1/i$.

\subsection{Matching Resummed and Fixed Ordered Expressions}
The resummation procedure is derived in the threshold limit to leading power in $(1-z)$ and therefore neglects sub-leading power corrections, e.g., hard, wide-angle radiation.
To make viable predictions at collider experiments, it is desirable to supplement resummed formulae with exact FO results beyond LO.
This is achievable by subtracting from the resummed expression the $\mathcal{O}(\as^k)$ radiation terms common to both calculations, 
to avoid double counting of soft radiation, and add back the full FO $\mathcal{O}(\as^k)$ result, which describes both soft and hard radiation.
A convenient way to isolate those common terms is to Taylor expand the resummed expression $\sigmaRes$ about $\alpha_s=0$.
Up to NLO, this is given by the first two terms of the expansion
\begin{eqnarray}
\label{eq:resTaylor}
\sigmaRes  &=&
\sum_{l=0}^\infty \frac{\alpha_s^l}{l!}\left[\frac{\partial^l}{\partial \alpha_s^l} \sigmaRes \right]_{\alpha_s=0} \\
&=& 	\sigmaRes \Big|_{\alpha_s=0} 
+  	\alpha_s 		~\left[ \partial_{\alpha_s}\sigmaRes 	\right]_{\alpha_s=0} 
+ 	\frac{\alpha_s^2}{2}	~\left[ \partial^2_{\alpha_s} \sigmaRes \right]_{\alpha_s=0} 
+ \mathcal{O}(\alpha_s^3)
\end{eqnarray}

As the Mellin- and inverse-Mellin operators commute with the $\partial_{\as}$ operator, the expansion holds in Mellin-space. 
Furthermore, as there is no explicit $\as$ dependence in $\sigma_0$ and $\mathcal{L}_N$ in Eq.~(\ref{eq:appNNLLMellin}),
the expansion of $\sigmaRes_N$ is simply proportional to the Taylor expansion of the exponentiated coefficient $\DeltaNRes$.
In Mellin space, the $\mathcal{O}(\as)$-subtracted resummed cross section for $pp\rightarrow \WR$ is then
\begin{eqnarray}
 \sigmaRes_N\Big\vert_{\as-\text{Subtracted}}
 = \sigma_0 ~\times~ \mathcal{L}_N ~\times~\left[\DeltaNRes-\DeltaNRes\bigg|_{\alpha_s=0}
 -\alpha_s \left[\partial_{\alpha_s}\DeltaNRes\right]_{\alpha_s=0} \right]
 \label{eq:appResSub}
\end{eqnarray}
For color-singlet $q\overline{q'}$ initial states, explicit calculation shows
\begin{equation}
\Delta_{(q\bar{q}')N}^{\rm Res.} \bigg|_{\alpha_s=0} = g_{00} 
\quad\text{and}\quad 
\partial_{\alpha_s} \; \Delta_{(q\bar{q}')N}^{\rm Res.} \bigg|_{\alpha_s=0}= (\mathcal{S}_1 + g_{01}) \; , 
\end{equation}
which correspond to terms in Eq.~(\ref{eq:appSudakov}) and can be found in~\cite{Bonvini:2010tp}. 
To NLO+NNLL accuracy, the physical inclusive $W'$ production cross section in $pp$ collisions is at last
\begin{eqnarray}
 \sigma^{\rm NLO+NNLL}(pp\rightarrow W' +X) &=& \sigma^{\rm NLO} + \sigma^{\rm NNLL}\Big\vert_{\as-\text{Subtracted}},
\end{eqnarray}
where the modified resummed term is obtained by inserting Eq.~(\ref{eq:appResSub}) into Eq.~(\ref{eq:appInvMellinRes}):
\begin{eqnarray}
\sigma^{\rm NNLL}\Big\vert_{\as-\text{Subtracted}} = 
\frac{\sigma_0}{\pi} \int_0^1  \frac{dt}{t} \text{Im}\Bigg[ &(i-m)& ~\tau_0^{-(N-1)} \times \mathcal{L}_N \times
\nonumber\\
&\times& \left(\Delta_{(q\bar{q}')N}^{\rm NNLL}-g_{00} - \alpha_s(\mathcal{S}_1+g_{01}) \right)
\Bigg].
\end{eqnarray}

\subsection{Parton Luminosities in Mellin Space}
The resummation formalism we exploit requires parton luminosities $\mathcal{L}_{q\bar{q}'}$ in Mellin space.
This introduces a technical difficulty as modern PDF sets are typically only available numerically.
It is possible, however, at a fixed factorization scale $\mu_f$, 
to approximate luminosities $\mathcal{L}_{ab}(\tau)$ [and individual PDFs $f_{a/p}(\xi)$]
using a basis of polynomials that can be Mellin-transformed analytically.
Here we use a basis of Chebyshev polynomials of the first kind $T_n(x)$, 
for which fast numerical algorithms exist for calculating the expansion coefficients, e.g.~\cite{gslLib}.
The implementation and optimization of the Chebyshev approximation procedure has been documented in Refs.~\cite{Bonvini:2010tp,Bonvini:2012sh}.
We briefly summarize for completeness how to obtain the approximated $\mathcal{L}_{(q\bar{q}')N}$.

We write a general Chebyshev polynomial of degree $n$ defined over the domain $x\in(-1,1)$ as
$$
T_n(x) = \sum_{m=0}^nT_{mn}x^n, \qquad T_{mn} \in \mathbbm{Z}.
$$
The function $F(u)$ over the domain $u\in(u_{\min},u_{\max})$ can then be approximated by the first $n_{ch}$ polynomials by the relationship
\begin{equation}
F(u)\approx-\frac{c_0}{2}+\sum\limits_{k=0}^{n_{ch}} c_k T_k(Au+B),
\label{eq:appCheb_approx1}
\end{equation}
with $A$ and $B$ given by
\begin{equation}
A=\frac{2}{u_{\max}-u_{\min}} \quad\text{and}\quad B  = -\frac{u_{\max}+u_{\min}}{u_{\max}-u_{\min}},
\end{equation}
and the $k$th Chebyshev coefficient $c_k$ by~\cite{gslLib}
\begin{eqnarray}
 c_k &=& \frac{2}{n_{ch}+1}\sum_{j=0}^{n_{ch}} ~\tilde{F}_j~ \times ~\cos\left(\frac{k\pi(j+\frac{1}{2})}{n_{ch} + 1}\right).
\end{eqnarray}
The $j$th moment of $F(u)$, i.e., $\tilde{F}_j$, is defined as
\begin{equation}
\tilde{F}_j =  F(y_j), \quad\text{with}\quad 
 y_j = \frac{1}{2}(u_{\max} - u_{\min})\cos\left(\frac{\pi(j+\frac{1}{2})}{n_{ch} + 1}\right) + \frac{(u_{\max} + u_{\min})}{2}.
\end{equation}

Such efficient algorithms allow us in principle to immediately obtain the luminosity $\mathcal{L}_{q \bar{q}'}(\tau,\mu_f)$ in Mellin space
by transforming Eq.~(\ref{eq:appCheb_approx1}) directly.
However, $\mathcal{L}(\tau)$ is generally poorly behaved across $\tau \in (0,1)$, particularly at the origin.
This is resolvable by approximating a regularized version of the luminosity and set
\begin{equation}
 F(u) = \tau(u) \mathcal{L}_{q\bar{q}'}(\tau(u),\mu_f), \quad\text{with}\quad \tau(u) = e^u \quad\text{for}\quad u\in(\log\tau_0,0).
\end{equation}
As defined in Eq.~(\ref{eq:foFactThm}), $\tau_0=M_{W'}^2/s$ is the threshold above which $pp\rightarrow W'$ is kinematically allowed to proceed.
After a wee bit of algebra, we obtain an expression for the Mellin transformed parton luminosities,
\begin{eqnarray}
\label{eq:LN_exp}
\mathcal{L}_{(q\bar{q}')N}
=\int_0^1 d\tau ~\tau^{N-1} ~\mathcal{L}_{q\bar{q}'}(\tau,\mu_f) 
=\int_0^1 d\tau(u) ~\tau^{N-2}(u) ~F(u)  
=\sum_{p=0}^{n_{ch}}\frac{\bar{c}_p}{(N-1)^{p+1}},
\end{eqnarray}
where we have defined
\begin{equation}
\bar{c}_p=\frac{2^p}{u_{min}^p}\sum\limits_{j=p}^{n_{ch}}\frac{j!}{(j-p)!}\tilde{c}_j \; , 
\quad\text{with}\quad
\tilde{c}_j = -\frac{c_0}{2}\delta_{j0}+\sum_{k=j}^{n_{ch}} c_k T_{kj}.
\label{eq:LN_cbar}
\end{equation}
Once one calculates the initial coefficients $c_i$, 
it is straightforward to approximate the Mellin transform of $\mathcal{L}_{q\bar{q}'}(\tau,\mu_f)$ by using Eqs.~(\ref{eq:LN_exp})-(\ref{eq:LN_cbar}).
However, for different $\mu_f$ choices, the function being approximated changes and therefore the coefficients $c_k$ need to be recomputed.
This should be taken into account if one intends to use a dynamic factorization scale.

\section{Modeling Manifest Left-Right Symmetric Model with FeynRules}\label{app:modelFile}
The most generic scalar potential of the LRSM consists of {18} parameters:
three mass scales  $\mu_{1,\dots,3}$, 
14 dimensionless couplings $\lambda_{1,\dots,4}$, $\rho_{1,\dots,4}$, $\alpha_{1,\dots,3}$, $\beta_{1,\dots,3}$,
and {one} CP violating phase $\delta_2$.
It is given by \cite{Deshpande:1990ip}, 
\begin{equation}
 V(\Phi,\Delta_L,\Delta_R) = V_{\mu} + V_{\Phi} + V_{\Delta} + V_{\Phi\Delta} + V_{\Phi\Delta_L\Delta_R},
 \label{eq:potential}
\end{equation}
where the scalar mass and self-coupling terms of the bidoublet $\Phi$  are, respectively,
\begin{eqnarray}
 V_{\mu} &=& -\mu_1^2 \Tr{\Phi^\dagger\Phi}
 - \mu_2^2\Tr{\Phi^\dagger\tilde{\Phi} + \tilde{\Phi}^\dagger\Phi}
 - \mu_3^2\Tr{\Delta_L^\dagger\Delta_L + \Delta_R^\dagger\Delta_R},
 \\
 V_{\Phi} &=& 
  \lambda_1\left[\Tr{\Phi^\dagger\Phi}\right]^2 +
  \lambda_2\left[\Tr{\Phi^\dagger\tilde{\Phi}}\right]^2 + \lambda_2\left[\Tr{\tilde{\Phi}^\dagger\Phi}\right]^2 
    \nonumber\\   &+&
  \lambda_3\Tr{\Phi^\dagger\tilde{\Phi}}\Tr{\tilde{\Phi}^\dagger\Phi} + 
  \lambda_4\Tr{\Phi^\dagger\Phi}\Tr{\Phi^\dagger\tilde{\Phi} + \tilde{\Phi}^\dagger\Phi},
\end{eqnarray}
The $\Delta_{L,R}$ self- and cross couplings are:
\begin{eqnarray}
 V_{\Delta} &=& 
 \rho_1\left[\Tr{\Delta_L^\dagger\Delta_L}\right]^2 + 
 \rho_1\left[\Tr{\Delta_R^\dagger\Delta_R}\right]^2 +
 \rho_3\Tr{\Delta_L^\dagger\Delta_L}\Tr{\Delta_R^\dagger\Delta_R}
  \nonumber\\ &+& 
 \rho_2\Tr{\Delta_L\Delta_L}\Tr{\Delta_L^\dagger\Delta_L^\dagger} + \rho_2\Tr{\Delta_R\Delta_R}\Tr{\Delta_R^\dagger\Delta_R^\dagger}
 \nonumber\\ &+&
 \rho_4\Tr{\Delta_L\Delta_L}\Tr{\Delta_R^\dagger\Delta_R^\dagger} + \rho_4\Tr{\Delta_L^\dagger\Delta_L^\dagger}\Tr{\Delta_R\Delta_R}.
\end{eqnarray}
The $\Phi-\Delta_L$ and $\Phi-\Delta_R$ couplings are
\begin{eqnarray}
 V_{\Phi\Delta} &=& 
 \alpha_1\Tr{\Phi^\dagger\Phi}\Tr{\Delta_L^\dagger\Delta_L + \Delta_R^\dagger\Delta_R} +
 \alpha_3 \Tr{\Phi\Phi^\dagger\Delta_L\Delta_L^\dagger + \phi^\dagger\phi\Delta_R\Delta_R^\dagger} 
 \nonumber\\ &+&
 \left\{
 \alpha_2 e^{i\delta_2}\Tr{\Phi^\dagger\tilde{\Phi}}\Tr{\Delta_L^\dagger\Delta_L} + 
 \alpha_2 e^{i\delta_2}\Tr{\tilde{\Phi}^\dagger\Phi}\Tr{\Delta_R^\dagger\Delta_R} + \text{H.c.}\right\},
\end{eqnarray}
with $\delta_2 =0$ making CP conservation explicit, and the $\Phi-\Delta_L-\Delta_R$ couplings are
\begin{eqnarray}
 V_{\Phi\Delta_L\Delta_R} &=& 
 \beta_1\Tr{\Phi^\dagger\Delta_L^\dagger\Phi\Delta_R + \Delta_R^\dagger\Phi^\dagger\Delta_L\Phi} +
 \beta_2\Tr{\Phi^\dagger\Delta_L^\dagger\tilde{\Phi}\Delta_R + \Delta_R^\dagger\tilde{\Phi}^\dagger\Delta_L\Phi}
 \nonumber\\ &+&
 \beta_3\Tr{\tilde{\Phi}^\dagger\Delta_L^\dagger\Phi\Delta_R + \Delta_R^\dagger\Phi^\dagger\Delta_L\tilde{\Phi}}.
 \label{eq:triScalarCoup}
\end{eqnarray}

After LR and EWSB, there exists 10 physical scalars:
four neutral, CP even states $H_{0,\dots,3}^0$, including one at $\mh\approx125\GeV$;
two neutral CP odd states $A_{0,1}^0$;
two  states singly charged  under $U(1)_{\rm EM}$ $H_{1,2}^\pm$;
and two  doubly charged states $\delta_{L,R}^{\pm\pm}$.
Subscripts do not indicate a mass ordering.
The mass spectrum in the vev limit of Eq.~(\ref{eq:vevHierarchy}) is given by \cite{Deshpande:1990ip,Duka:1999uc}: 
\begin{eqnarray} 
 \mh^2 \approx (125\GeV)^2 &\approx& 2\kp^2\left(\lambda_1 + 4\frac{k_1^2k_2^2}{\kp^4}(2\lambda_2+\lambda_3) + 4\lambda_4\frac{k_1k_2}{\kp^2}\right).
 \nonumber\\
M^2_{H^0_1}=M^2_{A^0_1}\approx\alpha_3 \frac{v^2_R}{2} \frac{k^2_{+}}{k^2_{-}},	&\quad& 
M^2_{H^0_3}=M^2_{A^0_2}\approx(\rho_3-2\rho_1) \frac{v^2_R}{2},			 \quad 
M^2_{H^0_2}\approx2 \rho_1 v^2_R,
\nonumber\\
M^2_{H^{\pm}_1}\approx(\rho_3-2\rho_1) \frac{v^2_R}{2} + \alpha_3 \frac{k^2_{-}}{4},	&\quad& 
M^2_{\delta^{\pm \pm}_L}\approx(\rho_3-2\rho_1) \frac{v^2_R}{2} + \alpha_3 \frac{k^2_{-}}{2},
\nonumber\\
M^2_{H^{\pm}_2}\approx\alpha_3 \frac{v^2_R}{2} \frac{k^2_{+}}{k^2_{-}}+\alpha_3 \frac{k^2_{-}}{4},	&\quad&
M^2_{\delta^{\pm \pm}_R}\approx2 \rho_2v^2_R+\alpha_3 \frac{k^2_{-}}{2},
\label{eq:higgsMasses}
\end{eqnarray}
where $k_\pm$ is defined in Eq.~(\ref{eq:ewvevDef}).

With choice assumptions, the potential can be configured such that the theory is consistent with experimental limits 
and features new gauge states accessible by the LHC or VLHC. 
Accordingly, the Manifest LRSM FeynRules model of~\cite{Roitgrund:2014zka} can be set to simulate this region of the MLRSM parameter space. 
We now discuss this configuration.

\subsection{Phenomenological Constraints on LRSM Scalar Potential}
Explicit CP conservation and minimization conditions of the potential give rise to the so-called vev Seesaw relationship~\cite{Deshpande:1990ip}:
\begin{equation}
\vL = \frac{\beta_2 k_1^2  + \beta_1 k_1 k_2 + \beta_3 k_2^2}{(2\rho_1 - \rho_3)\vR},
\label{eq:vevSeesaw}
\end{equation}
implying inherently small $\vL$ for $2\rho_1 \neq \rho_3$ and $\vR^2\gg~ k_1, k_2$. 
Though it is natural for all $\rho_i$ to be comparable in magnitude, 
it is contrived to expect a fine cancellation, particularly after radiative corrections.
Consistent with $\rho/T$-parameter measurements~\cite{Chen:2005jx,Chen:2008jg}, we choose
\begin{equation}
 \vL = 0 \quad\Longleftrightarrow\quad \beta_{1,\dots,3} = 0.
\end{equation}
This may also be achievable if Eq.~(\ref{eq:potential})  respects an approximate custodial symmetry.

Neglecting terms $\mathcal{O}(k_+^2/\vR^2)$, the minimization conditions also imply~\cite{Deshpande:1990ip}
\begin{eqnarray}
 \frac{\mu_1^2}{\vR^2} = \frac{\alpha_1}{2} - \frac{\alpha_3}{2}\left(\frac{k_2^2}{\km^2}\right), \qquad
 \frac{\mu_2^2}{\vR^2} = \frac{\alpha_1}{2} + \frac{\alpha_3}{4}\left(\frac{k_1k_2}{\km^2}\right), \qquad
 \frac{\mu_3^2}{\vR^2} = \rho_1.
\end{eqnarray}
As argued, one expects on naturalness grounds
\begin{equation}
 \alpha_{2,3} \sim \mathcal{O}(\alpha_1) \quad\text{and}\quad \rho_{2,3}\sim \mathcal{O}(\rho_1).
\end{equation}
Dropping terms relatively suppressed by 
$(k_2/\km)\sim (m_b/m_t)\sim10^{-2}$ [see~Eq.~(\ref{eq:bidoubletvevs})] gives
\begin{eqnarray}
 \frac{\mu_1^2}{\vR^2} \approx \frac{\mu_2^2}{\vR^2}  \approx \frac{\alpha_1}{2}, 
 \qquad \frac{\mu_3^2}{\vR^2} = \rho_1,
\end{eqnarray}
suggesting that LRSB is inherently at the mass scale of the scalar potential assuming 
\begin{equation}
 \alpha_1 \sim \mathcal{O}(1) \quad\text{and}\quad \rho_1 \sim \mathcal{O}(1).
 \label{eq:naturalCoup}
\end{equation}

In terms of $\MWR$ and $g$, Eq.~(\ref{eq:naturalCoup}) and 
positivity of squared masses for (physical) scalars imply several mass and coupling relationships:
\begin{eqnarray}
 \frac{\mH^2}{\MWR^2}, \frac{\mA^2}{\MWR^2}, \frac{\mHHp^2}{\MWR^2}	\approx \frac{\alpha_3}{g^2} > 1,
 &\quad&
 \frac{\mHH^2}{\MWR^2}		\approx  \frac{4\rho_1}{g^2} > 1, 
 \nonumber\\
 \frac{\mHHH^2}{\MWR^2}, \frac{\mAA^2}{\MWR^2}, \frac{\mHp^2}{\MWR^2}, \frac{\mHppL^2}{\MWR^2}	\approx \frac{(\rho_3 - 2\rho_1)}{g^2} > 0,
  &\quad&
 \frac{\mHppR^2}{\MWR^2}	\approx  \frac{4\rho_2}{g^2} > 1.
 \label{eq:higgsMassRels}
\end{eqnarray}
Imposing the strong requirement on Eq.~(\ref{eq:higgsMassRels}) to universally comply with bounds on FCNH, i.e., 
$m_{\rm FCNH}$ in Eq.~(\ref{eq:fcnhBound}), implies 
\begin{eqnarray}
 \rho_{1,2,4} > \frac{g^2}{4}\left(\frac{m_{\rm FCNH}}{\MWR}\right)^2,
 &\quad&
 \rho_{3} > g^2\left(\frac{m_{\rm FCNH}}{\MWR}\right)^2 + 2\rho_1 \sim 6\rho_1,
 \\
 \alpha_{1,\dots,3} > g^2\left(\frac{m_{\rm FCNH}}{\MWR}\right)^2,
 &\quad&
 \mu_{1,2}^2 > (m_{\rm FCNH})^2, \quad \mu_3^2 > \frac{1}{2}(m_{\rm FCNH})^2.
\end{eqnarray}

\begin{figure}[!t]
\begin{center}
\includegraphics[scale=1,width=.7\textwidth]{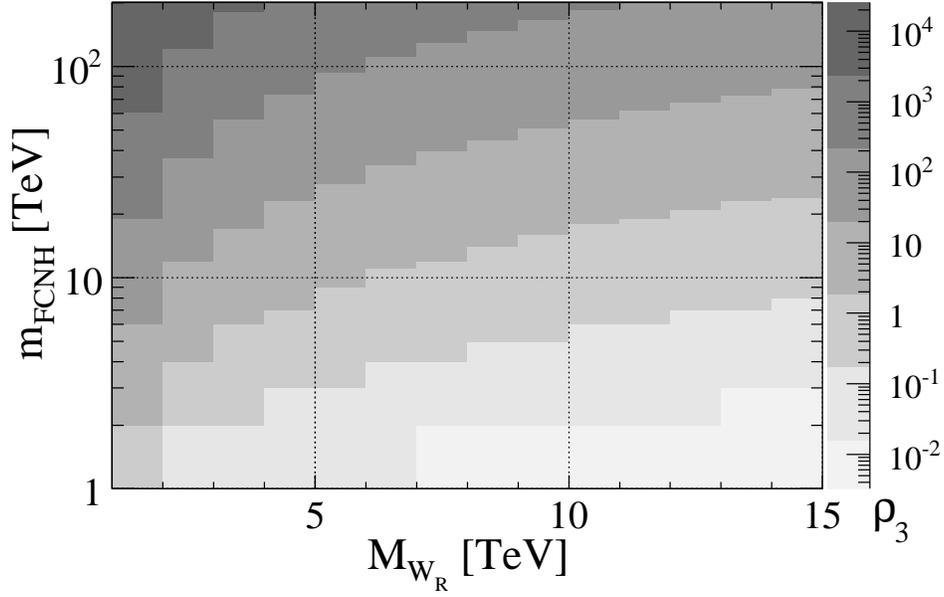}
\end{center}
\caption{Scalar triplet coupling $\rho_3$ contours for given $\MWR$ and $m_{\rm FCNH}$.} 
\label{fig:lrsm_rho3Contours}
\label{fig:rho3Limits}
\end{figure}

Several observations can be made from these relations:
First is that for $\MWR \lesssim 6.5\TeV$, one has $\rho_{1,2,4} > 1.$
Thus, discovery of a $\WR$ at the LHC would indicate a strongly coupled triplet sector.
Second is that a small hierarchy among the $\rho_i$ may exist. 
Requiring both $H_2^0$ and $H_3^0$ be heavier than $m_{\rm FCNH}$ suggests $\rho_3 \gtrsim 6\rho_1$.
Fig.~\ref{fig:rho3Limits} plots the values of $\rho_3$ for given $\MWR$ and $m_{\rm FCNH}$,
and shows, for example, $\rho_3<1$ and $m_{\rm FCNH}\sim15~(20)\TeV$ require $\MWR\gtrsim10~(12)\TeV$.
If $H_3^0$ and $A_2^0$ are largely responsible for neutral flavor transitions, 
then $\rho_{1,3}$ can be reduced while keeping their differences fixed. 
We do not apply theoretical prejudices against strongly coupled systems and treat this as a consistent prediction. 
A more detailed discussion on the perturbativity of the scalar sector can be found in~\cite{Maiezza:2016bzp}.

An ambiguity arises for the bidoublet self-couplings $\lambda_{1,3,4}$ as the self-coupling of the SM-like Higgs is unconstrained. 
Using Eq.~(\ref{eq:higgsMasses}), we take without impacting our study
\begin{equation}
 \lambda_1 \approx \frac{\mh^2}{2\kp^2}, \quad \lambda_{2,3}=0.
\end{equation}

\subsection{Configuration of LRSM FeynRules File}\label{app:UFO}
We implement our configuration of the scalar potential and choice for quark and lepton mixing
into one FR restriction file that can be invoked when generating UFOs for the Manifest LRSM v1.1.6-MIX model file by~\cite{Roitgrund:2014zka}.
See~\cite{Christensen:2008py} for instructions.
Internal parameters, e.g., $v_R$ and SM inputs of Eq.~(\ref{eq:smInputs}), can be modified via MG5 input parameter cards.
The restriction file, \texttt{lrsmLHCRestrictions.rst}, is available 
from the source directory for the arXiv preprint version of this report~\cite{modelFiles}.
It contains the following parameter identifications:
\begin{quote}\texttt{\noindent
(* Turn off CKM mixing *)\\
	s12 -> 0,\\
	s23 -> 0,\\
	s13 -> 0,\\
\hfill
(* Turn off light neutrino mixing and set PMNS to diagonal *)\\
	sL13 -> 0,\\
	sL23 -> 0,\\
	sL13 -> 0,\\
\hfill
(* Turn off off-diagonal heavy/light neutrino mixing [V,X in Eq.(A.11) of 0901.3589] *)\\
	VKe  -> 0, \\
	VKmu -> 0, \\
	VKta -> 0,\\
\hfill
(* Make mixing in LRSM manifest: all +1. Quasi-manifest: at least one -1 *)\\
	Wl11 -> 1,\\
	Wl22 -> 1,\\
	Wl33 -> 1,\\
	WU11 -> 1,\\
	WU22 -> 1,\\
	WU33 -> 1,\\
	WD11 -> 1,\\
	WD22 -> 1,\\
	WD33 -> 1,\\
\hfill
(* LH vev *)\\
	vL 	->  0,\\
\hfill
(* Quark masses and Yukawas *)\\	
        MU 	-> 0,\\
        MD  	-> 0,\\
	MC 	-> 0,\\
	MS  	-> 0,\\
\hfill
(* Lepton masses and Yukawas *)\\
	Me 	-> 0,\\
	Mmu 	-> 0,\\
	Mta 	-> 0,\\
	MN1 	-> 0,\\
	MN2 	-> 0,\\
	MN3 	-> 0
}
\end{quote}


\begin{thebibliography}{99}

\bibitem{Weinberg:1979sa} 
  S.~Weinberg,
  \textit{Baryon and Lepton Nonconserving Processes,}
  Phys.\ Rev.\ Lett.\  {\bf 43}, 1566 (1979).
  
  \bibitem{Wilczek:1979hc} 
  F.~Wilczek and A.~Zee,
  \textit{Operator Analysis of Nucleon Decay,}
  Phys.\ Rev.\ Lett.\  {\bf 43}, 1571 (1979).
  doi:10.1103/PhysRevLett.43.1571

\bibitem{Ma:1998dn} 
  E.~Ma,
  \textit{Pathways to naturally small neutrino masses,}
  Phys.\ Rev.\ Lett.\  {\bf 81}, 1171 (1998)
  doi:10.1103/PhysRevLett.81.1171
  [hep-ph/9805219].

\bibitem{Minkowski:1977sc} 
  P.~Minkowski,
  \textit{$\mu \to e\gamma$ at a Rate of One Out of $10^{9}$ Muon Decays?,}
  Phys.\ Lett.\ B {\bf 67}, 421 (1977).
  
\bibitem{Mohapatra:1979ia} 
  R.~N.~Mohapatra and G.~Senjanovic,
  \textit{Neutrino Mass and Spontaneous Parity Violation,}
  Phys.\ Rev.\ Lett.\  {\bf 44}, 912 (1980).
  
  \bibitem{Yanagida:1979as} 
  T.~Yanagida,
  \textit{Horizontal Symmetry And Masses Of Neutrinos,}
  Conf.\ Proc.\ C {\bf 7902131}, 95 (1979).
  
  \bibitem{GellMann:1980vs} 
  M.~Gell-Mann, P.~Ramond and R.~Slansky,
  \textit{Complex Spinors and Unified Theories,}
  Conf.\ Proc.\ C {\bf 790927}, 315 (1979)
  [arXiv:1306.4669 [hep-th]].
  
  \bibitem{Schechter:1980gr} 
  J.~Schechter and J.~W.~F.~Valle,
  \textit{Neutrino Masses in SU(2) x U(1) Theories,}
  Phys.\ Rev.\ D {\bf 22}, 2227 (1980).
  
  \bibitem{Shrock:1980ct} 
  R.~E.~Shrock,
  \textit{General Theory of Weak Leptonic and Semileptonic Decays. 1. 
  Leptonic Pseudoscalar Meson Decays, with Associated Tests For, and Bounds on, Neutrino Masses and Lepton Mixing,}
  Phys.\ Rev.\ D {\bf 24}, 1232 (1981).
        
  \bibitem{Magg:1980ut} 
  M.~Magg and C.~Wetterich,
  \textit{Neutrino Mass Problem and Gauge Hierarchy,}
  Phys.\ Lett.\ B {\bf 94}, 61 (1980).
  
  \bibitem{Cheng:1980qt} 
  T.~P.~Cheng and L.~F.~Li,
  \textit{Neutrino Masses, Mixings and Oscillations in SU(2) x U(1) Models of Electroweak Interactions,}
  Phys.\ Rev.\ D {\bf 22}, 2860 (1980).
  
  \bibitem{Lazarides:1980nt} 
  G.~Lazarides, Q.~Shafi and C.~Wetterich,
  \textit{Proton Lifetime and Fermion Masses in an SO(10) Model,}
  Nucl.\ Phys.\ B {\bf 181}, 287 (1981).
  
  \bibitem{Mohapatra:1980yp} 
  R.~N.~Mohapatra and G.~Senjanovic,
  \textit{Neutrino Masses and Mixings in Gauge Models with Spontaneous Parity Violation,}
  Phys.\ Rev.\ D {\bf 23}, 165 (1981).
   
  \bibitem{Foot:1988aq} 
  R.~Foot, H.~Lew, X.~G.~He and G.~C.~Joshi,
  \textit{Seesaw Neutrino Masses Induced by a Triplet of Leptons,}
  Z.\ Phys.\ C {\bf 44}, 441 (1989).
  
  \bibitem{Arkani-Hamed:2015vfh} 
  N.~Arkani-Hamed, T.~Han, M.~Mangano and L.~T.~Wang,
  \textit{Physics Opportunities of a 100 TeV Proton-Proton Collider,}
  arXiv:1511.06495 [hep-ph].
  
  \bibitem{Golling:2016gvc} 
  T.~Golling {\it et al.},
  \textit{Physics at a 100 TeV pp collider: beyond the Standard Model phenomena,}
  arXiv:1606.00947 [hep-ph].  
  
  \bibitem{Chen:2011de} 
  M.~C.~Chen and J.~Huang,
  \textit{TeV Scale Models of Neutrino Masses and Their Phenomenology,}
  Mod.\ Phys.\ Lett.\ A {\bf 26}, 1147 (2011)
  doi:10.1142/S0217732311035985
  [arXiv:1105.3188 [hep-ph]].
  
  \bibitem{Pati:1974yy} 
  J.~C.~Pati and A.~Salam,
  \textit{Lepton Number as the Fourth Color,}
  Phys.\ Rev.\ D {\bf 10}, 275 (1974)
  Erratum: [Phys.\ Rev.\ D {\bf 11}, 703 (1975)].
  doi:10.1103/PhysRevD.10.275, 10.1103/PhysRevD.11.703.2
  
  \bibitem{Mohapatra:1974gc} 
  R.~N.~Mohapatra and J.~C.~Pati,
  \textit{A Natural Left-Right Symmetry,}
  Phys.\ Rev.\ D {\bf 11}, 2558 (1975).
  doi:10.1103/PhysRevD.11.2558

  \bibitem{Senjanovic:1975rk} 
  G.~Senjanovic and R.~N.~Mohapatra,
  \textit{Exact Left-Right Symmetry and Spontaneous Violation of Parity,}
  Phys.\ Rev.\ D {\bf 12}, 1502 (1975).
  doi:10.1103/PhysRevD.12.1502

  \bibitem{Duka:1999uc} 
  P.~Duka, J.~Gluza and M.~Zralek,
  \textit{Quantization and renormalization of the manifest left-right symmetric model of electroweak interactions,}
  Annals Phys.\  {\bf 280}, 336 (2000)
  doi:10.1006/aphy.1999.5988
  [hep-ph/9910279].
  
  
  
  
  \bibitem{Chen:2005jx} 
  M.~C.~Chen, S.~Dawson and T.~Krupovnickas,
  \textit{Constraining new models with precision electroweak data,}
  Int.\ J.\ Mod.\ Phys.\ A {\bf 21}, 4045 (2006)
  doi:10.1142/S0217751X0603388X
  [hep-ph/0504286].
  
  \bibitem{Zhang:2007fn} 
  Y.~Zhang, H.~An, X.~Ji and R.~N.~Mohapatra,
  \textit{Right-handed quark mixings in minimal left-right symmetric model with general CP violation,}
  Phys.\ Rev.\ D {\bf 76}, 091301 (2007)
  doi:10.1103/PhysRevD.76.091301
  [arXiv:0704.1662 [hep-ph]].

  \bibitem{Zhang:2007da} 
  Y.~Zhang, H.~An, X.~Ji and R.~N.~Mohapatra,
  \textit{General CP Violation in Minimal Left-Right Symmetric Model and Constraints on the Right-Handed Scale,}
  Nucl.\ Phys.\ B {\bf 802}, 247 (2008)
  doi:10.1016/j.nuclphysb.2008.05.019
  [arXiv:0712.4218 [hep-ph]].
  
  \bibitem{Chen:2008jg} 
  M.~C.~Chen, S.~Dawson and C.~B.~Jackson,
  \textit{Higgs Triplets, Decoupling, and Precision Measurements,}
  Phys.\ Rev.\ D {\bf 78}, 093001 (2008)
  doi:10.1103/PhysRevD.78.093001
  [arXiv:0809.4185 [hep-ph]].
  
  \bibitem{Tello:2010am} 
  V.~Tello, M.~Nemevsek, F.~Nesti, G.~Senjanovic and F.~Vissani,
  \textit{Left-Right Symmetry: from LHC to Neutrinoless Double Beta Decay,}
  Phys.\ Rev.\ Lett.\  {\bf 106}, 151801 (2011)
  [arXiv:1011.3522 [hep-ph]].
  
  \bibitem{Nemevsek:2011aa} 
  M.~Nemevsek, F.~Nesti, G.~Senjanovic and V.~Tello,
  \textit{Neutrinoless Double Beta Decay: Low Left-Right Symmetry Scale?,}
  arXiv:1112.3061 [hep-ph].

  \bibitem{Deppisch:2012nb} 
  F.~F.~Deppisch, M.~Hirsch and H.~Pas,
  \textit{Neutrinoless Double Beta Decay and Physics Beyond the Standard Model,}
  J.\ Phys.\ G {\bf 39}, 124007 (2012)
  doi:10.1088/0954-3899/39/12/124007
  [arXiv:1208.0727 [hep-ph]].
  
  \bibitem{Chakrabortty:2012mh} 
  J.~Chakrabortty, H.~Z.~Devi, S.~Goswami and S.~Patra,
  \textit{Neutrinoless double-$\beta$ decay in TeV scale Left-Right symmetric models,}
  JHEP {\bf 1208}, 008 (2012)
  doi:10.1007/JHEP08(2012)008
  [arXiv:1204.2527 [hep-ph]].

  \bibitem{Nemevsek:2012iq} 
  M.~Nemevsek, G.~Senjanovic and V.~Tello,
  \textit{Connecting Dirac and Majorana Neutrino Mass Matrices in the Minimal Left-Right Symmetric Model,}
  Phys.\ Rev.\ Lett.\  {\bf 110}, no. 15, 151802 (2013)
  doi:10.1103/PhysRevLett.110.151802
  [arXiv:1211.2837 [hep-ph]].

  \bibitem{Dev:2013vxa} 
  P.~S.~Bhupal Dev, S.~Goswami, M.~Mitra and W.~Rodejohann,
  \textit{Constraining Neutrino Mass from Neutrinoless Double Beta Decay,}
  Phys.\ Rev.\ D {\bf 88}, 091301 (2013)
  doi:10.1103/PhysRevD.88.091301
  [arXiv:1305.0056 [hep-ph]];

  \bibitem{Awasthi:2016kbk} 
  R.~L.~Awasthi, A.~Dasgupta and M.~Mitra,
  \textit{Limiting the Effective Mass and New Physics Parameters from $0\nu\beta\beta$,}
  arXiv:1607.03835 [hep-ph].
 
  \bibitem{Barry:2013xxa} 
  J.~Barry and W.~Rodejohann,
  \textit{Lepton number and flavour violation in TeV-scale left-right symmetric theories with large left-right mixing,}
  JHEP {\bf 1309}, 153 (2013)
  doi:10.1007/JHEP09(2013)153
  [arXiv:1303.6324 [hep-ph]].

  \bibitem{Maiezza:2014ala} 
  A.~Maiezza and M.~Nemevsek,
  \textit{Strong P invariance, neutron electric dipole moment, and minimal left-right parity at LHC,}
  Phys.\ Rev.\ D {\bf 90}, no. 9, 095002 (2014)
  doi:10.1103/PhysRevD.90.095002
  [arXiv:1407.3678 [hep-ph]].
  
  \bibitem{Bertolini:2014sua} 
  S.~Bertolini, A.~Maiezza and F.~Nesti,
  \textit{Present and Future K and B Meson Mixing Constraints on TeV Scale Left-Right Symmetry,}
  Phys.\ Rev.\ D {\bf 89}, no. 9, 095028 (2014)
  doi:10.1103/PhysRevD.89.095028
  [arXiv:1403.7112 [hep-ph]].  
   
  \bibitem{Awasthi:2015ota} 
  R.~L.~Awasthi, P.~S.~B.~Dev and M.~Mitra,
  \textit{Implications of the Diboson Excess for Neutrinoless Double Beta Decay and Lepton Flavor Violation in TeV Scale Left Right Symmetric Model,}
  Phys.\ Rev.\ D {\bf 93}, no. 1, 011701 (2016)
  doi:10.1103/PhysRevD.93.011701
  [arXiv:1509.05387 [hep-ph]].

  \bibitem{Keung:1983uu} 
  W.~Y.~Keung and G.~Senjanovic,
  \textit{Majorana Neutrinos and the Production of the Right-handed Charged Gauge Boson,}
  Phys.\ Rev.\ Lett.\  {\bf 50}, 1427 (1983).
  doi:10.1103/PhysRevLett.50.1427

  \bibitem{Ferrari:2000sp} 
  A.~Ferrari, J.~Collot, M.~L.~Andrieux, B.~Belhorma, P.~de Saintignon, J.~Y.~Hostachy, P.~Martin and M.~Wielers,
  \textit{Sensitivity study for new gauge bosons and right-handed Majorana neutrinos in $p p$ collisions at $s$ = 14-TeV,}
  Phys.\ Rev.\ D {\bf 62}, 013001 (2000).
  doi:10.1103/PhysRevD.62.013001
  
  \bibitem{Frank:2010cj} 
  M.~Frank, A.~Hayreter and I.~Turan,
  \textit{Production and Decays of $W_R$ bosons at the LHC,}
  Phys.\ Rev.\ D {\bf 83}, 035001 (2011)
  doi:10.1103/PhysRevD.83.035001
  [arXiv:1010.5809 [hep-ph]].
  
  \bibitem{Das:2012ii} 
  S.~P.~Das, F.~F.~Deppisch, O.~Kittel and J.~W.~F.~Valle,
  \textit{Heavy Neutrinos and Lepton Flavour Violation in Left-Right Symmetric Models at the LHC,}
  Phys.\ Rev.\ D {\bf 86}, 055006 (2012)
  doi:10.1103/PhysRevD.86.055006
  [arXiv:1206.0256 [hep-ph]].

  \bibitem{Han:2012vk} 
  T.~Han, I.~Lewis, R.~Ruiz and Z.~g.~Si,
  \textit{Lepton Number Violation and $W^\prime$ Chiral Couplings at the LHC,}
  Phys.\ Rev.\ D {\bf 87}, no. 3, 035011 (2013)
  doi:10.1103/PhysRevD.87.035011,   [arXiv:1211.6447 [hep-ph]]. 
  
  \bibitem{Chen:2013fna}
  C.~Y.~Chen, P.~S.~B.~Dev and R.~N.~Mohapatra,
  \textit{Probing Heavy-Light Neutrino Mixing in Left-Right Seesaw Models at the LHC,}
  Phys.\ Rev.\ D {\bf 88}, 033014 (2013)
  doi:10.1103/PhysRevD.88.033014
  [arXiv:1306.2342 [hep-ph]].
  
  \bibitem{Dev:2014iva} 
  P.~S.~Bhupal Dev, C.~H.~Lee and R.~N.~Mohapatra,
  \textit{Leptogenesis Constraints on the Mass of Right-handed Gauge Bosons,}
  Phys.\ Rev.\ D {\bf 90}, no. 9, 095012 (2014)
  doi:10.1103/PhysRevD.90.095012
  [arXiv:1408.2820 [hep-ph]].
  
    \bibitem{Vasquez:2014mxa} 
  J.~C.~Vasquez,
  \textit{Right-handed lepton mixings at the LHC,}
  JHEP {\bf 1605}, 176 (2016)
  doi:10.1007/JHEP05(2016)176
  [arXiv:1411.5824 [hep-ph]].
  
  \bibitem{Maiezza:2015lza} 
  A.~Maiezza, M.~Nemevsek and F.~Nesti,
  \textit{Lepton Number Violation in Higgs Decay at LHC,}
  Phys.\ Rev.\ Lett.\  {\bf 115}, 081802 (2015)
  doi:10.1103/PhysRevLett.115.081802
  [arXiv:1503.06834 [hep-ph]].
  
  \bibitem{Gluza:2015goa} 
  J.~Gluza and T.~Jeli\'nski,
  \textit{Heavy neutrinos and the pp→lljj CMS data,}
  Phys.\ Lett.\ B {\bf 748}, 125 (2015)
  doi:10.1016/j.physletb.2015.06.077
  [arXiv:1504.05568 [hep-ph]].
  
  \bibitem{Chakrabortty:2016wkl} 
  J.~Chakrabortty, J.~Gluza, T.~Jeli\'	nski and T.~Srivastava,
  \textit{Theoretical constraints on masses of heavy particles in Left-Right Symmetric Models,}
  Phys.\ Lett.\ B {\bf 759}, 361 (2016)
  doi:10.1016/j.physletb.2016.05.092
  [arXiv:1604.06987 [hep-ph]].
  
  
   \bibitem{Dev:2015kca} 
  P.~S.~B.~Dev, D.~Kim and R.~N.~Mohapatra,
  \textit{Disambiguating Seesaw Models using Invariant Mass Variables at Hadron Colliders,}
  JHEP {\bf 1601}, 118 (2016)
  doi:10.1007/JHEP01(2016)118
  [arXiv:1510.04328 [hep-ph]].

  \bibitem{Ng:2015hba} 
  J.~N.~Ng, A.~de la Puente and B.~W.~P.~Pan,
  \textit{Search for Heavy Right-Handed Neutrinos at the LHC and Beyond in the Same-Sign Same-Flavor Leptons Final State,}
  JHEP {\bf 1512}, 172 (2015)
  doi:10.1007/JHEP12(2015)172
  [arXiv:1505.01934 [hep-ph]].
 
 \bibitem{Mondal:2015zba} 
  S.~Mondal and S.~K.~Rai,
  \textit{Polarized window for left-right symmetry and a right-handed neutrino at the Large Hadron-Electron Collider,}
  Phys.\ Rev.\ D {\bf 93}, no. 1, 011702 (2016)
  doi:10.1103/PhysRevD.93.011702
  [arXiv:1510.08632 [hep-ph]].
 
  \bibitem{Lindner:2016lxq} 
  M.~Lindner, F.~S.~Queiroz, W.~Rodejohann and C.~E.~Yaguna,
  \textit{Left-Right Symmetry and Lepton Number Violation at the Large Hadron Electron Collider,}
  JHEP {\bf 1606}, 140 (2016)
  doi:10.1007/JHEP06(2016)140
  [arXiv:1604.08596 [hep-ph]].
   
  \bibitem{Maiezza:2016bzp} 
  A.~Maiezza, M.~Nemevsek and F.~Nesti,
  \textit{Perturbativity and mass scales of Left-Right Higgs bosons,}
  arXiv:1603.00360 [hep-ph].
 
  \bibitem{FileviezPerez:2016erl} 
  P.~Fileviez Perez, C.~Murgui and S.~Ohmer,
  \textit{Simple Left-Right Theory: Lepton Number Violation at the LHC,}
  arXiv:1607.00246 [hep-ph].
  
  \bibitem{Binosi:2003yf} 
  D.~Binosi and L.~Theussl,
  \textit{JaxoDraw: A Graphical user interface for drawing Feynman diagrams,}
  Comput.\ Phys.\ Commun.\  {\bf 161}, 76 (2004)
  [hep-ph/0309015].
  
  
  \bibitem{Aad:2015xaa} 
  G.~Aad {\it et al.} [ATLAS Collaboration],
  \textit{Search for heavy Majorana neutrinos with the ATLAS detector in pp collisions at $ \sqrt{s}=8 $ TeV,}
  JHEP {\bf 1507}, 162 (2015)
  doi:10.1007/JHEP07(2015)162
  [arXiv:1506.06020 [hep-ex]].

  \bibitem{Khachatryan:2014dka} 
  V.~Khachatryan {\it et al.} [CMS Collaboration],
  \textit{Search for heavy neutrinos and $\mathrm {W}$ bosons with right-handed couplings in proton-proton collisions at $\sqrt{s} = 8\,\text {TeV} $,}
  Eur.\ Phys.\ J.\ C {\bf 74}, no. 11, 3149 (2014)
  doi:10.1140/epjc/s10052-014-3149-z
  [arXiv:1407.3683 [hep-ex]].
 
  \bibitem{Patra:2015bga} 
  S.~Patra, F.~S.~Queiroz and W.~Rodejohann,
  \textit{Stringent Dilepton Bounds on Left-Right Models using LHC data,}
  Phys.\ Lett.\ B {\bf 752}, 186 (2016)
  doi:10.1016/j.physletb.2015.11.009
  [arXiv:1506.03456 [hep-ph]].

  \bibitem{Lindner:2016lpp} 
  M.~Lindner, F.~S.~Queiroz and W.~Rodejohann,
  \textit{Dilepton bounds on left-right symmetry at the LHC run II and neutrinoless double beta decay,}
  arXiv:1604.07419 [hep-ph].
 
  \bibitem{ATLAS:2015nsi} 
  G.~Aad {\it et al.} [ATLAS Collaboration],
  \textit{Search for new phenomena in dijet mass and angular distributions from $pp$ collisions at $\sqrt{s}=$ 13 TeV with the ATLAS detector,}
  Phys.\ Lett.\ B {\bf 754}, 302 (2016)
  doi:10.1016/j.physletb.2016.01.032
  [arXiv:1512.01530 [hep-ex]].
 
  \bibitem{Khachatryan:2015dcf} 
  V.~Khachatryan {\it et al.} [CMS Collaboration],
  \textit{Search for narrow resonances decaying to dijets in proton-proton collisions at $\sqrt(s) =$ 13 TeV,}
  Phys.\ Rev.\ Lett.\  {\bf 116}, no. 7, 071801 (2016)
  doi:10.1103/PhysRevLett.116.071801
  [arXiv:1512.01224 [hep-ex]].
  
  \bibitem{Jezo:2014wra} 
  T.~Jezo, M.~Klasen, D.~R.~Lamprea, F.~Lyonnet and I.~Schienbein,
  \textit{NLO+NLL limits on $W'$ and $Z'$ gauge boson masses in general extensions of the Standard Model,}
  JHEP {\bf 1412}, 092 (2014)
  doi:10.1007/JHEP12(2014)092
  [arXiv:1410.4692 [hep-ph]]. 
  
  \bibitem{CMS:2015kjy} 
  CMS Collaboration [CMS Collaboration],
  \textit{Search for SSM W' production,  in the lepton+MET final state at a center-of-mass energy of 13 TeV,}
  CMS-PAS-EXO-15-006.
  
  \bibitem{Kang:2015uoc} 
  Z.~Kang, P.~Ko and J.~Li,
  \textit{New Avenues to Heavy Right-handed Neutrinos with Pair Production at Hadronic Colliders,}
  Phys.\ Rev.\ D {\bf 93}, no. 7, 075037 (2016)
  doi:10.1103/PhysRevD.93.075037
  [arXiv:1512.08373 [hep-ph]].
  
  \bibitem{Kaplan:2008ie} 
  D.~E.~Kaplan, K.~Rehermann, M.~D.~Schwartz and B.~Tweedie,
  \textit{Top Tagging: A Method for Identifying Boosted Hadronically Decaying Top Quarks,}
  Phys.\ Rev.\ Lett.\  {\bf 101}, 142001 (2008)
  doi:10.1103/PhysRevLett.101.142001
  [arXiv:0806.0848 [hep-ph]].
  
  \bibitem{Plehn:2009rk} 
  T.~Plehn, G.~P.~Salam and M.~Spannowsky,
  \textit{Fat Jets for a Light Higgs,}
  Phys.\ Rev.\ Lett.\  {\bf 104}, 111801 (2010)
  doi:10.1103/PhysRevLett.104.111801
  [arXiv:0910.5472 [hep-ph]].
  
  \bibitem{Plehn:2010st} 
  T.~Plehn, M.~Spannowsky, M.~Takeuchi and D.~Zerwas,
  \textit{Stop Reconstruction with Tagged Tops,}
  JHEP {\bf 1010}, 078 (2010)
  doi:10.1007/JHEP10(2010)078
  [arXiv:1006.2833 [hep-ph]].
  
  \bibitem{Soper:2012pb} 
  D.~E.~Soper and M.~Spannowsky,
  \textit{Finding top quarks with shower deconstruction,}
  Phys.\ Rev.\ D {\bf 87}, 054012 (2013)
  doi:10.1103/PhysRevD.87.054012
  [arXiv:1211.3140 [hep-ph]].
  
  \bibitem{Schaetzel:2013vka} 
  S.~Schaetzel and M.~Spannowsky,
  \textit{Tagging highly boosted top quarks,}
  Phys.\ Rev.\ D {\bf 89}, no. 1, 014007 (2014)
  doi:10.1103/PhysRevD.89.014007
  [arXiv:1308.0540 [hep-ph]].
  
  \bibitem{Bonvini:2015ira} 
  M.~Bonvini {\it et al.},
  \textit{Parton distributions with threshold resummation,}
  JHEP {\bf 1509}, 191 (2015)
  doi:10.1007/JHEP09(2015)191
  [arXiv:1507.01006 [hep-ph]].
  
  \bibitem{Beenakker:2015rna} 
  W.~Beenakker, C.~Borschensky, M.~Kr\"amer, A.~Kulesza, E.~Laenen, S.~Marzani and J.~Rojo,
  \textit{NLO+NLL squark and gluino production cross-sections with threshold-improved parton distributions,}
  Eur.\ Phys.\ J.\ C {\bf 76}, no. 2, 53 (2016)
  doi:10.1140/epjc/s10052-016-3892-4
  [arXiv:1510.00375 [hep-ph]].
  
    \bibitem{Sullivan:2002jt} 
  Z.~Sullivan,
  \textit{Fully differential $W^\prime$ production and decay at next-to-leading order in QCD,}
  Phys.\ Rev.\ D {\bf 66}, 075011 (2002)
  doi:10.1103/PhysRevD.66.075011
  [hep-ph/0207290].
  
  \bibitem{Gavin:2012sy} 
  R.~Gavin, Y.~Li, F.~Petriello and S.~Quackenbush,
  \textit{W Physics at the LHC with FEWZ 2.1,}
  Comput.\ Phys.\ Commun.\  {\bf 184}, 208 (2013)
  doi:10.1016/j.cpc.2012.09.005
  [arXiv:1201.5896 [hep-ph]].
  
  
  \bibitem{Roitgrund:2014zka} 
  A.~Roitgrund, G.~Eilam and S.~Bar-Shalom,
  \textit{Implementation of the left-right symmetric model in FeynRules/CalcHep,}
  arXiv:1401.3345 [hep-ph].
  
 \bibitem{Korner:1992zk} 
  J.~G.~Korner, A.~Pilaftsis and K.~Schilcher,
  \textit{Leptonic CP asymmetries in flavor changing H0 decays,}
  Phys.\ Rev.\ D {\bf 47}, 1080 (1993)
  doi:10.1103/PhysRevD.47.1080
  [hep-ph/9301289].

\bibitem{Grimus:2000vj}
  W.~Grimus and L.~Lavoura,
  \textit{The seesaw mechanism at arbitrary order: Disentangling the small scale from the large scale},
  JHEP {\bf 0011}, 042 (2000)
  [arXiv:hep-ph/0008179].

 \bibitem{Atre:2009rg} 
  A.~Atre, T.~Han, S.~Pascoli and B.~Zhang,
  \textit{The Search for Heavy Majorana Neutrinos,}
  JHEP {\bf 0905}, 030 (2009)
  doi:10.1088/1126-6708/2009/05/030
  [arXiv:0901.3589 [hep-ph]].  
  
  \bibitem{KamLAND-Zen:2016pfg} 
  A.~Gando {\it et al.} [KamLAND-Zen Collaboration],
  \textit{Search for Majorana Neutrinos near the Inverted Mass Hierarchy region with KamLAND-Zen,}
  arXiv:1605.02889 [hep-ex].
  
  \bibitem{Zhang:2007fn} 
  Y.~Zhang, H.~An, X.~Ji and R.~N.~Mohapatra,
  Phys.\ Rev.\ D {\bf 76}, 091301 (2007)
  doi:10.1103/PhysRevD.76.091301
  [arXiv:0704.1662 [hep-ph]].
  
  \bibitem{Zhang:2007da} 
  Y.~Zhang, H.~An, X.~Ji and R.~N.~Mohapatra,
  Nucl.\ Phys.\ B {\bf 802}, 247 (2008)
  doi:10.1016/j.nuclphysb.2008.05.019
  [arXiv:0712.4218 [hep-ph]].
  
    \bibitem{Maiezza:2010ic} 
  A.~Maiezza, M.~Nemevsek, F.~Nesti and G.~Senjanovic,
  Phys.\ Rev.\ D {\bf 82}, 055022 (2010)
  doi:10.1103/PhysRevD.82.055022
  [arXiv:1005.5160 [hep-ph]].
  
  \bibitem{Senjanovic:2014pva} 
  G.~Senjanović and V.~Tello,
  Phys.\ Rev.\ Lett.\  {\bf 114}, no. 7, 071801 (2015)
  doi:10.1103/PhysRevLett.114.071801
  [arXiv:1408.3835 [hep-ph]].
  
  \bibitem{Senjanovic:2015yea} 
  G.~Senjanović and V.~Tello,
  Phys.\ Rev.\ D {\bf 94}, no. 9, 095023 (2016)
  doi:10.1103/PhysRevD.94.095023
  [arXiv:1502.05704 [hep-ph]].
    
  \bibitem{Agashe:2014kda} 
  K.~A.~Olive {\it et al.} [Particle Data Group Collaboration],
  \textit{Review of Particle Physics,}
  Chin.\ Phys.\ C {\bf 38}, 090001 (2014).
  doi:10.1088/1674-1137/38/9/090001
  
  \bibitem{Buckley:2014ana} 
  A.~Buckley, J.~Ferrando, S.~Lloyd, K.~Nordstr\"om, B.~Page, M.~Rüfenacht, M.~Schönherr and G.~Watt,
  \textit{LHAPDF6: parton density access in the LHC precision era,}
  Eur.\ Phys.\ J.\ C {\bf 75}, 132 (2015)
  doi:10.1140/epjc/s10052-015-3318-8
  [arXiv:1412.7420 [hep-ph]].


  \bibitem{Ball:2014uwa} 
  R.~D.~Ball {\it et al.} [NNPDF Collaboration],
  \textit{Parton distributions for the LHC Run II,}
  JHEP {\bf 1504}, 040 (2015)
  doi:10.1007/JHEP04(2015)040
  [arXiv:1410.8849 [hep-ph]].
  
    \bibitem{Hahn:2004fe} 
  T.~Hahn,
  \textit{CUBA: A Library for multidimensional numerical integration,}
  Comput.\ Phys.\ Commun.\  {\bf 168}, 78 (2005)
  doi:10.1016/j.cpc.2005.01.010
  [hep-ph/0404043].
  
  \bibitem{Fabricius:1981sx} 
  K.~Fabricius, I.~Schmitt, G.~Kramer and G.~Schierholz,
  \textit{Higher Order Perturbative QCD Calculation of Jet Cross-Sections in e+ e- Annihilation,}
  Z.\ Phys.\ C {\bf 11}, 315 (1981).
  
  \bibitem{Kramer:1986mc} 
  G.~Kramer and B.~Lampe,
  \textit{Jet Cross-Sections in e+ e- Annihilation,}
  Fortsch.\ Phys.\  {\bf 37}, 161 (1989).
  
  \bibitem{Baer:1989jg} 
  H.~Baer, J.~Ohnemus and J.~F.~Owens,
  \textit{A Next-To-Leading Logarithm Calculation of Jet Photoproduction,}
  Phys.\ Rev.\ D {\bf 40}, 2844 (1989).
  
  \bibitem{Harris:2001sx} 
  B.~W.~Harris and J.~F.~Owens,
  \textit{The Two cutoff phase space slicing method,}
  Phys.\ Rev.\ D {\bf 65}, 094032 (2002)
  [hep-ph/0102128].

  \bibitem{Ruiz:2015zca} 
  R.~Ruiz,
  \textit{QCD Corrections to Pair Production of Type III Seesaw Leptons at Hadron Colliders,}
  JHEP {\bf 1512}, 165 (2015)
  doi:10.1007/JHEP12(2015)165
  [arXiv:1509.05416 [hep-ph]].
  
  \bibitem{Alwall:2014hca} 
  J.~Alwall {\it et al.},
  \textit{The automated computation of tree-level and next-to-leading order differential cross sections, 
  and their matching to parton shower simulations,}
  JHEP {\bf 1407}, 079 (2014)
  doi:10.1007/JHEP07(2014)079
  [arXiv:1405.0301 [hep-ph]].
  
  
  \bibitem{Sterman:1986aj} 
  G.~F.~Sterman,
  \textit{Summation of Large Corrections to Short Distance Hadronic Cross-Sections,}
  Nucl.\ Phys.\ B {\bf 281}, 310 (1987).
  doi:10.1016/0550-3213(87)90258-6
  
  \bibitem{Catani:1989ne} 
  S.~Catani and L.~Trentadue,
  \textit{Resummation of the QCD Perturbative Series for Hard Processes,}
  Nucl.\ Phys.\ B {\bf 327}, 323 (1989).
  doi:10.1016/0550-3213(89)90273-3
  
  \bibitem{Catani:1990rp} 
  S.~Catani and L.~Trentadue,
  \textit{Comment on QCD exponentiation at large x,}
  Nucl.\ Phys.\ B {\bf 353}, 183 (1991).
  doi:10.1016/0550-3213(91)90506-S
  

    \bibitem{Catani:1996yz} 
  S.~Catani, M.~L.~Mangano, P.~Nason and L.~Trentadue,
  \textit{The Resummation of soft gluons in hadronic collisions,}
  Nucl.\ Phys.\ B {\bf 478}, 273 (1996)
  doi:10.1016/0550-3213(96)00399-9
  [hep-ph/9604351].


  \bibitem{Bonvini:2010tp} 
  M.~Bonvini, S.~Forte and G.~Ridolfi,
  \textit{Soft gluon resummation of Drell-Yan rapidity distributions: Theory and phenomenology,}
  Nucl.\ Phys.\ B {\bf 847}, 93 (2011)
  doi:10.1016/j.nuclphysb.2011.01.023
  [arXiv:1009.5691 [hep-ph]].
 
  \bibitem{Catani:2003zt} 
  S.~Catani, D.~de Florian, M.~Grazzini and P.~Nason,
  \textit{Soft gluon resummation for Higgs boson production at hadron colliders,}
  JHEP {\bf 0307}, 028 (2003)
  doi:10.1088/1126-6708/2003/07/028
  [hep-ph/0306211].
  
  \bibitem{Bonvini:2012sh} 
  M.~Bonvini,
  \textit{Resummation of soft and hard gluon radiation in perturbative QCD,}
  arXiv:1212.0480 [hep-ph].
  
  \bibitem{Appell:1988ie} 
  D.~Appell, G.~F.~Sterman and P.~B.~Mackenzie,
  Nucl.\ Phys.\ B {\bf 309}, 259 (1988).
  doi:10.1016/0550-3213(88)90082-X
  
  \bibitem{Alloul:2013bka} 
  A.~Alloul, N.~D.~Christensen, C.~Degrande, C.~Duhr and B.~Fuks,
  \textit{FeynRules  2.0 - A complete toolbox for tree-level phenomenology,}
  Comput.\ Phys.\ Commun.\  {\bf 185}, 2250 (2014)
  doi:10.1016/j.cpc.2014.04.012
  [arXiv:1310.1921 [hep-ph]].
  
  \bibitem{Christensen:2008py} 
  N.~D.~Christensen and C.~Duhr,
  \textit{FeynRules - Feynman rules made easy,}
  Comput.\ Phys.\ Commun.\  {\bf 180}, 1614 (2009)
  doi:10.1016/j.cpc.2009.02.018
  [arXiv:0806.4194 [hep-ph]].
  
    \bibitem{Degrande:2011ua} 
  C.~Degrande, C.~Duhr, B.~Fuks, D.~Grellscheid, O.~Mattelaer and T.~Reiter,
  \textit{UFO - The Universal FeynRules Output,}
  Comput.\ Phys.\ Commun.\  {\bf 183}, 1201 (2012)
  doi:10.1016/j.cpc.2012.01.022
  [arXiv:1108.2040 [hep-ph]].
  
 \bibitem{Sjostrand:2014zea} 
  T.~Sj\"ostrand {\it et al.},
  \textit{An Introduction to PYTHIA 8.2,}
  Comput.\ Phys.\ Commun.\  {\bf 191}, 159 (2015)
  doi:10.1016/j.cpc.2015.01.024
  [arXiv:1410.3012 [hep-ph]]. 
    
  \bibitem{Cacciari:2005hq} 
  M.~Cacciari and G.~P.~Salam,
  \textit{Dispelling the $N^{3}$ myth for the $k_t$ jet-finder,}
  Phys.\ Lett.\ B {\bf 641}, 57 (2006)
  doi:10.1016/j.physletb.2006.08.037
  [hep-ph/0512210].

 \bibitem{Cacciari:2011ma} 
  M.~Cacciari, G.~P.~Salam and G.~Soyez,
  \textit{FastJet User Manual,}
  Eur.\ Phys.\ J.\ C {\bf 72}, 1896 (2012)
  doi:10.1140/epjc/s10052-012-1896-2
  [arXiv:1111.6097 [hep-ph]]. 

  \bibitem{Dokshitzer:1997in} 
  Y.~L.~Dokshitzer, G.~D.~Leder, S.~Moretti and B.~R.~Webber,
  \textit{Better jet clustering algorithms,}
  JHEP {\bf 9708}, 001 (1997)
  doi:10.1088/1126-6708/1997/08/001
  [hep-ph/9707323].
  
  \bibitem{Wobisch:1998wt} 
  M.~Wobisch and T.~Wengler,
  \textit{Hadronization corrections to jet cross-sections in deep inelastic scattering,}
  In *Hamburg 1998/1999, Monte Carlo generators for HERA physics* 270-279
  [hep-ph/9907280].  

  
  \bibitem{Alva:2014gxa} 
  D.~Alva, T.~Han and R.~Ruiz,
  \textit{Heavy Majorana neutrinos from $W\gamma$ fusion at hadron colliders,}
  JHEP {\bf 1502}, 072 (2015)
  doi:10.1007/JHEP02(2015)072
  [arXiv:1411.7305 [hep-ph]].
  
 \bibitem{Aad:2009wy} 
  G.~Aad {\it et al.} [ATLAS Collaboration],
  \textit{Expected Performance of the ATLAS Experiment - Detector, Trigger and Physics,}
  arXiv:0901.0512 [hep-ex]. 

    \bibitem{Khachatryan:2016olu} 
  V.~Khachatryan {\it et al.} [CMS Collaboration],
  \textit{Search for heavy Majorana neutrinos in e+/- e+/- plus jets and e+/- mu+/- 
  plus jets events in proton-proton collisions at sqrt(s) = 8 TeV,}
  arXiv:1603.02248 [hep-ex].
  
 
  
  \bibitem{ATL-PHYS-PUB-2013-004}
\textit{Performance assumptions for an upgraded ATLAS detector at a High-Luminosity LHC}
  Tech.\ Rep.\ ATL-PHYS-PUB-2013-004, CERN, Geneva, (2013).

\bibitem{Rehermann:2010vq} 
  K.~Rehermann and B.~Tweedie,
  \textit{Efficient Identification of Boosted Semileptonic Top Quarks at the LHC,}
  JHEP {\bf 1103}, 059 (2011)
  doi:10.1007/JHEP03(2011)059
  [arXiv:1007.2221 [hep-ph]].
  
  
 \bibitem{Martin:2009iq} 
  A.~D.~Martin, W.~J.~Stirling, R.~S.~Thorne and G.~Watt,
  \textit{Parton distributions for the LHC,}
  Eur.\ Phys.\ J.\ C {\bf 63}, 189 (2009)
  doi:10.1140/epjc/s10052-009-1072-5
  [arXiv:0901.0002 [hep-ph]].
  
  
\bibitem{Collins:1981uk} 
  J.~C.~Collins and D.~E.~Soper,
  \textit{Back-To-Back Jets in QCD,}
  Nucl.\ Phys.\ B {\bf 193}, 381 (1981)
  Erratum: [Nucl.\ Phys.\ B {\bf 213}, 545 (1983)].
  doi:10.1016/0550-3213(81)90339-4
  
\bibitem{Collins:1981va} 
  J.~C.~Collins and D.~E.~Soper,
  \textit{Back-To-Back Jets: Fourier Transform from B to K-Transverse,}
  Nucl.\ Phys.\ B {\bf 197}, 446 (1982).
  doi:10.1016/0550-3213(82)90453-9
  
\bibitem{Collins:1984kg} 
  J.~C.~Collins, D.~E.~Soper and G.~F.~Sterman,
  \textit{Transverse Momentum Distribution in Drell-Yan Pair and W and Z Boson Production,}
  Nucl.\ Phys.\ B {\bf 250}, 199 (1985).
  doi:10.1016/0550-3213(85)90479-1

  
\bibitem{Li:1998is} 
  H.~n.~Li,
  \textit{Unification of the k(T) and threshold resummations,}
  Phys.\ Lett.\ B {\bf 454}, 328 (1999)
  doi:10.1016/S0370-2693(99)00350-0
  [hep-ph/9812363].
  
\bibitem{Laenen:2000de}
  E.~Laenen, G.~F.~Sterman and W.~Vogelsang,
  \textit{Higher order QCD corrections in prompt photon production,}
  Phys.\ Rev.\ Lett.\  {\bf 84} (2000) 4296
  doi:10.1103/PhysRevLett.84.4296
  [hep-ph/0002078].
  
\bibitem{Laenen:2000ij}
  E.~Laenen, G.~F.~Sterman and W.~Vogelsang,
  \textit{Recoil and threshold corrections in short distance cross-sections,}
  Phys.\ Rev.\ D {\bf 63} (2001) 114018
  doi:10.1103/PhysRevD.63.114018
  [hep-ph/0010080].    

  
  \bibitem{Bozzi:2007qr} 
  G.~Bozzi, B.~Fuks and M.~Klasen,
  \textit{Threshold Resummation for Slepton-Pair Production at Hadron Colliders,}
  Nucl.\ Phys.\ B {\bf 777}, 157 (2007)
  doi:10.1016/j.nuclphysb.2007.03.052
  [hep-ph/0701202].  
  
    \bibitem{Forte:2002ni} 
  S.~Forte and G.~Ridolfi,
  \textit{Renormalization group approach to soft gluon resummation,}
  Nucl.\ Phys.\ B {\bf 650}, 229 (2003)
  doi:10.1016/S0550-3213(02)01034-9
  [hep-ph/0209154].
  
 \bibitem{Luisoni:2015xha} 
  G.~Luisoni and S.~Marzani,
  \textit{QCD resummation for hadronic final states,}
  J.\ Phys.\ G {\bf 42}, no. 10, 103101 (2015)
  doi:10.1088/0954-3899/42/10/103101
  [arXiv:1505.04084 [hep-ph]].  
  
\bibitem{Bonvini:2014joa} 
  M.~Bonvini and S.~Marzani,
  \textit{Resummed Higgs cross section at N$^{3}$LL,}
  JHEP {\bf 1409}, 007 (2014)
  doi:10.1007/JHEP09(2014)007
  [arXiv:1405.3654 [hep-ph]].
  
\bibitem{gslLib}
  M. Galassi et al, 
  GNU Scientific Library Reference Manual (3rd Ed.), 
  ISBN 0954612078
  
 \bibitem{Deshpande:1990ip} 
  N.~G.~Deshpande, J.~F.~Gunion, B.~Kayser and F.~I.~Olness,
  \textit{Left-right symmetric electroweak models with triplet Higgs,}
  Phys.\ Rev.\ D {\bf 44}, 837 (1991).
  doi:10.1103/PhysRevD.44.837  
  
  \bibitem{modelFiles}
  \href{http://arxiv.org/abs/1607.03504}{http://arxiv.org/abs/1607.03504}  
  

  
\end{thebibliography}
\end{document}